\documentclass[a4paper,11pt]{article}
\pdfoutput=1 
\usepackage{jheppub} 
\usepackage{cleveref}
\usepackage{lipsum}
\usepackage{amsmath}
\usepackage{slashed}
\usepackage{subcaption}

\usepackage{xcolor}
\usepackage[normalem]{ulem}

\usepackage{graphicx}
\usepackage{tikz}
\usepackage{tikz-feynman} 
\tikzfeynmanset{
  every vertex/.style={dot,minimum size=0.6mm},
  small/.style={scale=0.95}
}
\usepackage{float}
\usepackage[utf8]{inputenc}

\title{Minimalist Seesaw-Scotogenic Mechanism as a Source for Neutrino Physics}  

\author[a]{Eduardo Peinado}
\author[b]{Abdel P\'erez-Lorenzana}
\author[b]{Isangel Toledo}

\affiliation[a]{Instituto de F\'isica, Universidad Nacional Aut\' onoma de M\'exico, A.P. 20-364, Ciudad de M\'exico 01000, M\'exico.}
\affiliation[b]{Departamento de F\'{\i}sica, Centro de Investigaci\'on y de Estudios Avanzados del I.P.N. \\
    Apartado Postal 14-740, 07000, Ciudad de M\'exico, M\'exico.}

\emailAdd{epeinado@fisica.unam.mx}
\emailAdd{abdel.perez@cinvestav.mx}
\emailAdd{isangel.toledo@cinvestav.mx}

\date{\today}

\abstract{We identify the smallest discrete flavor symmetry and minimal field content that can simultaneously account for neutrino masses and the dark sector in the context of the discrete dark matter. Minimality is achieved with a $D_4$ symmetry by adding three right-handed neutrinos and three scalar doublets. Within this framework, tree-level and one-loop contributions yield the observed solar--atmospheric mass splitings as well as the  observed neutrino mixings. Furthermore, by including CP-violating phases, the model successfully accommodates both normal and inverted ordering. Spontaneous $D_4$ breaking leaves a residual $\mathbb{Z}_2$ symmetry that stabilizes the dark matter candidate, with a predicted mass within the $50$--$200$ GeV range, consistent with LUX-ZEPLIN limits.}

\arxivnumber{}

\makeatletter

\begin{document}

\maketitle
\flushbottom

\section{\label{sec:introduction} Introduction}
The Standard Model (SM) of particle physics provides a precise description of fundamental particles and their interactions governed by the strong and electroweak forces. However, neutrino masses are not considered in its formulation, which leaves out any understanding of neutrino oscillation experiments that establish their existence \cite{BILENKY1978225}. The framework of neutrino oscillations, dependent on neutrino masses and lepton mixing, has proven successful, driving continuous theoretical and experimental advancements. In recent years, novel frameworks based on non-holomorphic modular symmetries~\cite{feruglio2017neutrinomassesmodularforms,nomura2024nonholomorphicmodularmathcala4symmetric,nomura2025radiativeneutrinomassmodel} have been developed to provide a mechanism for generating neutrino masses. In parallel, various proposals have been developed to explain the observed patterns of neutrino masses and mixing via invoking non-Abelian discrete flavor symmetries~\cite{tanimoto1999neutrinomassesmixingsflavor,Ishimori:2010au,Grimus_2003,Bonilla_2020}. These approaches include models for neutrino masses generated at tree level, such as the seesaw mechanism, and at the one-loop level, such as the Zee and Scotogenic models; for a review see, for instance~\cite{RevModPhys.59.671,_vila_2022,Morisi_2012}.

On the other hand, observations of large-scale structure formation and the shape of the power spectrum of cosmic microwave background (CMB) anisotropies conclusively establish the existence of dark matter. These observations have also fixed its current abundance. The Planck data and CMB anisotropies yield the following value for the relic density~\cite{2020}:
\begin{align}
\label{eq:relicdensity}
\Omega_c h^2=0.1200\pm0.0012
\end{align}

However, the Standard Cosmological Model does not restrict the spin or mass of the dark matter candidate, which spans a viable spectrum of many orders of magnitude, leading to candidates ranging from ultralight axions to Weakly Interacting Massive Particles (WIMPs). Due to the above, it is not possible to uniquely identify the specific mechanism responsible for the observed relic density, and thus, the thermal history and strength of interactions with the primordial plasma remain model-dependent features that must be explored in theoretical extensions.

These considerations motivate extensions of the SM in which neutrino masses are generated either at the radiative level or through a combination of tree-level and radiative mechanisms, while accounting for dark matter~\cite{Bonilla_2023}. In such scenarios, models based on discrete flavor symmetries can address both sectors simultaneously, with the breaking of the flavor symmetry playing a role in the origin of dark matter and neutrino mass generation~\cite{PhysRevD.82.116003,Boucenna:2011tj}. There, the breaking of the underlying flavor symmetry, should leave a residual $\mathbb{Z}_2$ symmetry, under which dark fields are odd, ensuring the stability of the dark sector, with the lightest of these fields constituting a dark matter candidate.

The scotogenic model~\cite{Ma_2006} is the best example of this class of SM extensions, in which a discrete $\mathbb{Z}_2$ symmetry stabilizes the dark sector at the loop level. This $\mathbb{Z}_2$ symmetry need not be imposed ad hoc; it can be obtained by imposing a larger non-Abelian discrete symmetry. In this work, we employ this mechanism - combined with a type-I Seesaw, to address the issue of identifying and exploring the minimal discrete symmetry that allows a realization of the scotogenic model with the minimal field content extention over the SM. As we argue, a non-Abelian $D_4$ flavor symmetry turns out to be the appropriate one. 

The theoretical base of this framework was established in~\cite{Bonilla_2023} by one of the authors of the present paper. Here, we present a new possibility to explain the origin of neutrino mass hierarchy and the production of DM, guided by the principle of minimality: we require the lowest discrete group order and the minimal matter content necessary to remain consistent with the experimental data. We consider both normal and inverted neutrino mass orderings, in agreement with current neutrino oscillation global fit data.~\cite{de_Salas_2021, Esteban_2024}. In this realization, through the spontaneous breaking of $D_4$ to $\mathbb{Z}_2$, the DM is stabilized, and neutrino masses are generated at both tree and one-loop levels. To validate the model, numerical constraints from Electroweak Precision Observables and Lepton Flavor Violation (LFV) rates are incorporated, accounting for general charged-lepton mixing and the seesaw-scotogenic interplay, as well as dark sector phenomenology to identify the parameter space regions consistent with both neutrino data and cosmological observations.

For completeness, we present in Appendices A and B a thorough model-building search exploring several discrete symmetries ($S_3, D_4, Q_6, D_6$) and alternative representation assignments. We explicitly demonstrate why simpler configurations (such as those without the singlet neutrino $N_s$, or with different singlet representations) fail due to structural features, such as uncompensated zeroes in the mixing matrix or a rank-1 mass matrix constraint.

This paper is organized as follows. In Sec.~\ref{sec:TheModel}, we introduce the model by searching for the lowest group order and field content that accommodate both neutrino masses and mixings and the scotogenic model. We write down the scalar potential, discuss the stability of the DM candidate under the Non-Abelian $D_4$ symmetry, and detail the flavor symmetry breakdown and the scalar spectrum. Subsequently, we write down the Yukawa Lagrangian and detail the Yukawa matrices for the active and dark sectors. In Sec.~\ref{Sec:NeutrinoMasses}, we calculate all contributions to the total neutrino mass matrix, both the active and dark sector contributions. In Sec.~\ref{Sec:Constraints}, we identify theoretical and experimental constraints on the scalar spectrum, the charged-lepton sector, namely LFV limits, the neutrino oscillation data, and DM observables. In Sec.~\ref{Sec:Numerical_Results}, we present the results of the numerical scan, addressing the mixing angle correlation in the general CP-violating case and the effective Majorana mass, as well as the region where the correct DM relic density is obtained, showing that it remains consistent for both mass hierarchies while satisfying the Direct Detection limits. Finally, in Sec.~\ref{sec:conclusions}, we conclude on the model viability by showing that it provides a plausible DM candidate while simultaneously generating the observed neutrino masses and mixings, all while remaining well within current experimental limits on oblique parameters, LFV, and cosmology.

\section{The Model}\label{sec:TheModel}
As stated above, in this work, we aim to construct a minimalist model that offers predictions for the neutrinos and dark matter sectors. The proposed model is based on two principles: using the flavor symmetry group of the lowest possible order and introducing the minimum number of new fields to the SM necessary to generate neutrino masses and mixing. We explored various non-abelian discrete symmetry groups $\left(S_3, D_4, Q_6, D_6\right)$, an implementation of a 4HDM model, and different representations for the matter fields. All of these scenarios are described in more detail in Appendix~\ref{appB}. From this analysis, we determined that the model satisfying the imposed conditions is the one based on the $D_4$ group~\cite{Ishimori:2010au}, extended by two scalar doublets and three right-handed neutrinos, that we shall discuss hereafter.

Therefore, we propose an extension of the SM with a $D_4$ flavor symmetry, with the field content minimally extended by two $SU(2)_L$ doublet scalar fields in the doublet representation of $D_4$, $\Phi_D=\left(\phi_1,\phi_2\right)$, two right-handed neutrinos also as an $D_4$ doublet~\cite{Boucenna_2012,MELONI2011281}, $N_D=\left(N_1,N_2\right)$, and one right-handed neutrino $N_s$ which is a $D_4$ singlet. Meanwhile, the scalar doublet already in the SM matter content, $\Phi_s$, is assigned to a singlet representation of $D_4$.

On the other hand, fields associated with both, left-handed $\left(L_i\right)$ and right-handed leptons $\left(\ell_i\right)$, where $i=1,2,3$, are assigned to the $D_4$ singlet representations $1_1, 1_3,$ and $ 1_3$, respectively. This assignment naturally leads to a charged lepton mass matrix exhibiting a $\mu-\tau$ texture~\cite{MELONI2011281,Blum_2008,Adulpravitchai_2009}. The relevant quantum numbers for the fields in the model are summarized in Tab.~\ref{tab:tab1}. In this section, we study the scalar potential, the structure of the Yukawa sector constrained by the flavor symmetry, flavor symmetry breaking, and the scalar field spectrum.

\begin{table}[h!]
\centering
\begin{tabular}{|l|l|l|l|l|l|l|l|l|l|l|}
\hline
        & $L_1$ & $L_2$ & $L_3$ & $\ell_1$ & $\ell_2$ & $\ell_3$ & $\Phi_s$ & $\Phi_D$ & $N_D$ & $N_s$\\ \hline
$D_4$   & $1_1$ & $1_3$ & $1_3$ & $1_1$    & $1_3$    & $1_3$    & $1_1$    & $2$      & $2$   & $1_3$\\ \hline
$SU(2)$ & $2$   & $2$   & $2$   & $1$      & $1$      & $1$      & $2$      & $2$      & $1$   & $1$\\ \hline
\end{tabular}
\caption{Relevant particle content and quantum numbers of the model}
\label{tab:tab1}
\end{table}

\subsection{Scalar Potential and Stability of the Dark Matter Candidate}

The properties of $D_4$ symmetry representations are described in the Appendix~\ref{appA}. However, it is worth mentioning that throughout this paper, we work in the real basis. Based on these properties under $D_4$, the most general invariant scalar potential under $SU\left(2\right)_L\times U\left(1\right)_Y\times D_4$ can be written as:
\begin{align}
    \label{eq:Vtermsinv}
    V&=\mu_s^2\Phi_s^\dagger\Phi_s+\mu_{d}^2\left[\Phi_D^\dagger\Phi_D\right]_{1_1}+\lambda_1\left(\Phi_s^\dagger\Phi_s\right)^2+\sum_{i=1}^4\lambda_{i+1}\left(\left[\Phi_D^\dagger\Phi_D\right]_{1_i}\right)^2\nonumber\\
    +&\lambda_6\left[\Phi_D^\dagger\Phi_D\right]_{1_1}\left(\Phi_s^\dagger\Phi_s\right)+\lambda_7\left[\left[\Phi_s^\dagger\Phi_D\right]_2\left[\Phi_D^\dagger\Phi_s\right]_2\right]_{1_1}+\left[\lambda_8\left(\left[\Phi_s^\dagger\Phi_D\right]_2\left[\Phi_s^\dagger\Phi_D\right]_2\right)_{1_1}+h.c\right]\nonumber\\
    +&\lambda_9\left[\Phi_D^\dagger\Phi_D^\dagger\right]_{1_1}\left[\Phi_D\Phi_D\right]_{1_1}+\lambda_{10}\left[\Phi_D^\dagger\Phi_D^\dagger\right]_{1_2}\left[\Phi_D\Phi_D\right]_{1_2}+\lambda_{11}\left[\Phi_D^\dagger\Phi_D^\dagger\right]_{1_3}\left[\Phi_D\Phi_D\right]_{1_3}\\
    +&\lambda_{12}\left[\Phi_D^\dagger\Phi_D^\dagger\right]_{1_4}\left[\Phi_D\Phi_D\right]_{1_4},\nonumber
\end{align}
where the subscript in the parentheses denotes the $D_4$ contraction of two doublets. We have studied the minimization of potential $V$ solving the equation $\partial V/\partial v_i=0$ where $v_i$ are the vevs of the fields $\Phi_s,\phi_1$ and $\phi_2$. A minimum local value allowed for the potential $V$ is
\begin{align}
    \label{eq:setvevs}
    v_s\neq0 ,& & v_1^2-v_2^2=&0.
\end{align}

Following the above condition, we adopt the configuration $v_1=v_2=v_d$. Then, the vevs structure takes the following form
\begin{align}
    \label{eq:vevstructure}
    \langle{\Phi_s\rangle}=\frac{v_s}{\sqrt{2}},& & \langle{\phi_1\rangle}&=\frac{v_d}{\sqrt{2}}, & \langle{\phi_2\rangle}&=\frac{v_d}{\sqrt{2}} .
\end{align}

Equivalently, we can write it as follows within the $D_4$ representations,
\begin{align}
    \label{eq:vevsfin}
    \langle{\Phi_s\rangle}=\frac{v_s}{\sqrt{2}},& & \langle{\Phi_D\rangle}\equiv\begin{pmatrix} \langle{\phi_1\rangle}\\ \langle{\phi_2\rangle} \end{pmatrix}=\begin{pmatrix}
        1\\
        1
    \end{pmatrix}\frac{v_d}{\sqrt{2}}.
\end{align}

Thus,  after electroweak symmetry breaking, we introduce the following explicit field expansion in flavour basis
\begin{gather}
  \Phi_s=\begin{pmatrix}
    \sigma_s^{\prime+}\\
    \displaystyle\frac{v_s+\rho_s^\prime+i\eta_s^\prime}{\sqrt{2}}
  \end{pmatrix},
  \qquad \qquad 
  \phi_1=\begin{pmatrix}
    \psi_1^{\prime+}\\
   \displaystyle\frac{v_d+\zeta_1^\prime+i\omega_1^\prime}{\sqrt{2}}
  \end{pmatrix}
  \label{eq:dobletes1}, \nonumber\\
  \phi_2=\begin{pmatrix}
    \psi_2^{\prime+}\\
    \displaystyle\frac{v_d+\zeta_2^\prime+i\omega_2^\prime}{\sqrt{2}}
  \end{pmatrix}.
  \label{eq:doblete2}
\end{gather}
\subsubsection{Flavor Symmetry Breakdown}

We can now study the mechanism responsible for DM stability in our model. First, we should notice that there is a $\mathbb{Z}_2$ subgroup of $D_4$ that leaves the $\langle \Phi_D \rangle$ vev invariant, which is generated by the group generator $B\equiv a^3b$\footnote{See Appendix~\ref{appA} for the notation regarding group generators.}:
{ 
\renewcommand{\arraystretch}{1.5} 
\begin{align}
    \label{eq:Bgenerator}
    B\left(\langle \Phi_D \rangle\right) = \begin{pmatrix}
        0 & 1\\
        1 & 0 
    \end{pmatrix}
    \begin{pmatrix}
        \displaystyle\frac{v_d}{\sqrt{2}} \\[10pt] 
        \displaystyle\frac{v_d}{\sqrt{2}}
    \end{pmatrix}=\langle \Phi_D \rangle.
\end{align}
} 

By performing a unitary transformation $U$ on the group representation basis (Appendix \ref{appA}), the generator $B$ can be cast in a diagonal form:
\begin{align}
    \label{eq:Bz2}
    \hat{B}=\begin{pmatrix}
        1 & 0\\
        0 & -1
    \end{pmatrix}.
\end{align}

 This transformation allows us to decompose the subspace spanned by the $D_4$ doublet into even and odd eigenstates under the remnant $\mathbb{Z}_2$ symmetry~\cite{MELONI2011281}. In this new bases, the scalar fields are expressed as:
\begin{align}
    \label{eq:cambiodebase}
    \varphi_1=&\frac{\phi_1+\phi_2}{\sqrt{2}}, & \varphi_2=&\frac{\phi_1-\phi_2}{\sqrt{2}},
\end{align}
which, according to Eq.~\eqref{eq:Bz2}, transform under this $\mathbb{Z}_2$ as follows:

\begin{align}
    \label{eq:scalarrotation}
    \varphi_1&\longrightarrow\varphi_1, & \varphi_2&\longrightarrow-\varphi_2,
\end{align}
In this basis, the corresponding vevs are
\begin{align}
    \label{eq:vevsrot}
    \langle \varphi_1 \rangle=&v_d, & \langle \varphi_2 \rangle=0. 
\end{align}

Since the transformation $U$ acts on the $D_4$ doublet representation, it must be applied consistently to all fields transforming under this representation. Thus, for the right-handed neutrino doublet $N_D$, we define the rotated fermionic state components as $\chi_1$ and $\chi_2$ in an analogous manner to Eq.~\eqref{eq:cambiodebase}. In this way, considering the criteria of dark matter stability~\cite{PhysRevD.82.116003}, we define dark fields as those that are odd under the residual $\mathbb{Z}_2$, which, for our model, are given by:

\begin{align}
    \label{eq:iddark}
    \mathbb{Z}_2: \varphi_2&\longrightarrow-\varphi_2\nonumber\\
    \chi_2&\longrightarrow-\chi_2
\end{align}

Finally, to facilitate the subsequent identification of the scalar spectrum, we explicitly parameterize the rotated fields~$\varphi_1$ and $\varphi_2$, defined in Eq.~\eqref{eq:cambiodebase}, in terms of their CP-even, CP-odd, and charged components:

\begin{align}
	\label{eq:fieldsrot}
	\varphi_{1} = &\begin{pmatrix} \sigma_{1}^{\prime+} \\
	 \frac{\sqrt{2}v_{d} + \rho_{1}^{\prime} + i\eta_{1}^{\prime}}{\sqrt{2}} \end{pmatrix}, &
\varphi_{2} =& \begin{pmatrix} \sigma^{\prime+} \\ 
\frac{\rho^{\prime} + i\eta^{\prime}}{\sqrt{2}} \end{pmatrix},
\end{align}
where these new states, although expressed in the rotated basis defined by Eq.~\eqref{eq:cambiodebase}, still correspond to flavor-basis components. Specifically, they are linear combinations of the original fields introduced in Eq.~\eqref{eq:doblete2}, such that $\rho_1^\prime=(\zeta_1^\prime+\zeta_2^\prime)/\sqrt{2}$ and $\rho^\prime=(\zeta_1^\prime-\zeta_2^\prime)/\sqrt{2}$, with analogous relations for the CP-odd components $\eta_1^\prime$, $\eta^\prime$, and the charged components $\sigma_1^\prime$, $\sigma^\prime$.

\subsection{Scalar Spectrum}

To analyze the scalar spectrum of the model, it is convenient to define the mass eigenstate basis for the full scalar fields as follows:

\begin{equation}
   \left. 
   \begin{aligned}
        \lbrace\sigma_s^{\prime +},\sigma_1^{\prime+},\sigma^{\prime+}\rbrace &\rightarrow \lbrace G_0^+,\sigma_1^+,\sigma_2^+\rbrace, \\
        \lbrace\rho^\prime_s,\rho_1^\prime,\rho^\prime\rbrace&\rightarrow\lbrace  h_s,h_1,h_2\rbrace ,\\
        \lbrace\eta_s^\prime,\eta_1^\prime,\eta^\prime\rbrace&\rightarrow\lbrace G_\eta,A_1,A_2\rbrace.
   \end{aligned}
   \right\} \textbf{Mass Eigenstates}
\end{equation}
In this basis, we identify the active scalar physical Higgs field observed at the LHC as $h_s$, while $h_1$ is a heavier active scalar. Finally, the fields with the subscript $2$ constitute the odd scalar sector and do not mix with the even scalar sector due to the preserved $\mathbb{Z}_2$ symmetry.

The expressions for the scalar mass matrices are written in Appendix~\ref{appA}. The pseudo-scalar mass matrix ($M_\eta^2$) and charged scalar mass matrix ($M_c^2$) each have a vanishing eigenvalue corresponding to the neutral $(G_\eta)$ and charged $(G_0^+)$ Goldstone bosons, respectively.

Since the mixing is restricted to the active sector, only this sector requires diagonalization. We identify the physical mass eigenstates through the following orthogonal transformations:

\begin{align}
    \label{eq:Rotpsueo}
    D_{G_\eta,A_1}^2=O_{G_\eta,A_1}M^2_{\eta_s^{\prime}\eta_{1}^\prime}O_{G_\eta,A_1}^T,\\
    D_{G_0^{+},\sigma_1^+}=O_{G_0^{+},\sigma_1^+}M^2_{\sigma_s^{\prime +}\sigma_{1}^{\prime +}}O_{G_0^{+},\sigma_1^+}^T,
\end{align}
where we define
\begin{align}
    O_{G_\eta,A_1}=O_{G_0^{+},\sigma_1^+}=\begin{pmatrix}
        \cos{\beta} & -\sin{\beta}\\
         \sin{\beta} & \cos{\beta}
    \end{pmatrix},
\end{align}
and the mixing angle $\beta$ is determined by:
\begin{align}
    \tan{2\beta}=\frac{v_d v_s}{v_d^2-\frac{v_s^2}{4}}.
\end{align}
Here, $M^2_{\eta_s^{\prime}\eta_{1}^\prime}$ and $M^2_{\sigma_s^{\prime +}\sigma_{1}^{\prime +}}$ are the $2\times 2$ mass sub-matrices for the pseudo-scalar and charged scalar sectors, respectively.

Similarly, for the CP-even neutral sector, the mixing between the active scalars $\rho_s^\prime$ and $\rho_1^\prime$ is parametrized by the angle $\alpha$, as follows:
\begin{align}
    \label{eq:diagonalneutralscalar}
    D_{h_s,h_1}^2=O_{h_s,h_1}M^2_{\rho_s^\prime,\rho_1^\prime}O_{h_s,h_1}^T,
\end{align}
where the orthogonal matrix is defined as
\begin{align}
    \label{eq:mixingneutral}
    O_{h_s,h_1}=\begin{pmatrix}
        \cos{\alpha} & -\sin{\alpha}\\
         \sin{\alpha} & \cos{\alpha}
    \end{pmatrix},
\end{align}
and the mixing angle $\alpha$ is determined by:
\begin{align}
    \label{eq:tanalpha}
    \tan{2\alpha}=\frac{v_dv_s\left(\lambda_6+\lambda_7+2\lambda_8\right)}{v_d^2\left(\lambda_2+\lambda_4+\lambda_9+\lambda_{11}\right)-v_s^2\lambda_1}.
\end{align}

\subsection{The Yukawa Sector}
The lepton sector Yukawa Lagrangian of the model in the new basis defined previously for $U$ is the following
\begin{align}
    \label{eq:yukawalagrangian1}
    \mathcal{L}_Y=&y_1^\ell\overline{L_1}\Phi_s\ell_1+y_2^\ell\overline{L_2}\Phi_s\ell_2+y_3^\ell\overline{L_3}\Phi_s\ell_3+y_4^\ell\overline{L_2}\Phi_s\ell_3+y_5^\ell\overline{L_3}\Phi_s\ell_2\nonumber\\
    +&y_1^\nu\overline{L_1}\left(\Tilde{\varphi}_D \chi_D\right)_{1_1}+y_2^\nu\overline{L_2}\left(\Tilde{\varphi}_D \chi_D\right)_{1_3}+y_3^\nu\overline{L_3}\left(\Tilde{\varphi}_D \chi_D\right)_{1_3}+M\left(\overline{\chi_D^C}\chi_D\right)_{1_1}\nonumber\\
    +& y_4^\nu \overline{L_2}\Tilde{\phi}_s N_s+ y_5^\nu \overline{L_3}\Tilde{\phi}_s N_s+M_s \overline{N}_s^c N_s+h.c,
\end{align}
where $\Phi_i=i\tau_2\phi_i^\dagger$ and the Yukawa couplings may be complex, introducing CP-violating phases in the lepton sector, with a set of physical CP-violating phases remaining after field redefinitions. This equation includes the Majorana mass term for the heavy right-handed (RH) neutrinos. Notice that due to the flavor symmetry, the two RH-Neutrinos, $\chi_D = (\chi_1,\chi_2)$, are degenerated with mass $M$. The $D_4$ symmetry forces the charged lepton matrix not to be diagonal, where, considering the quantum numbers of the model, as shown in Tab.~\ref{tab:tab1}, we have

\begin{align}
    M_l=\frac{v_s}{\sqrt{2}}\begin{pmatrix}
        y_1^\ell & 0 & 0\\
        0 & y_2^\ell & y_4^\ell\\
        0 & y_5^\ell & y_3^\ell
    \end{pmatrix}.
\end{align}

From Eq.\eqref{eq:yukawalagrangian1}, the Yukawa coupling matrices of the $\varphi_D$ and $\Phi_s$ fields with the neutrinos are
\begin{align}
    \label{eq:yukawacouplin}
    Y^{\varphi_1}=& \begin{pmatrix}
        y_1^\nu & 0 & 0\\
        y_2^\nu & 0 & 0\\
        y_3^\nu & 0 & 0
    \end{pmatrix}, & Y^{\varphi_2}=&\begin{pmatrix}
        0 & y_1^\nu & 0\\
        0 & -y_2^\nu & 0\\
        0 & -y_3^\nu & 0
    \end{pmatrix},
    &  Y^{\Phi_s}=&\begin{pmatrix}
        0 & 0 & 0\\
        0 & 0 & y_4^\nu\\
        0 & 0 & y_5^\nu
    \end{pmatrix}
\end{align}
Notice that there are five parameters $y_1^\nu,y_2^\nu,y_3^\nu,y_4^\nu,y_5^\nu$, governing the Yukawa interactions of neutrinos with the three scalar doublets, $\Phi_s,\varphi_1$, and $\varphi_2$. Observe that the RH-neutrinos, $\chi_D$, interact with $\varphi_1$ and $\varphi_2$ through the same Yukawa coupling due to the implementation of the $D_4$ flavor symmetry. This is an important point, since the neutrino masses will be generated through both a type-I seesaw and a scotogenic mechanism. Furthermore, it is worth mentioning that the Yukawa coupling $Y^{\Phi_s}$, arising from the interaction between $N_s$ and $\Phi_s$, contributes exclusively to the type-I seesaw sector. This contribution is essential for obtaining the correct order of magnitude for the mixing angle $\sin^2{\theta_{13}}$ (See Appendix \ref{appB}) \cite{Grimus_2003}. 

This approach offers a clear advantage. In previous works that include a scoto-seesaw mechanism not based on non-Abelian flavor symmetries, the dominance between the two mechanisms typically depends on assumed hierarchies among different Yukawa couplings \cite{Mandal_2021,Aranda_2019,Rojas_2019,Barreiros_2021,Barreiros_2022}. In contrast, models based on such symmetries, such as those with an $A_4$ flavor structure \cite{Bonilla_2023} or the $D_4$ model presented here, provide a framework in which the interplay between the mechanism naturally arises from the underlying symmetry.

\section{Neutrino Masses} \label{Sec:NeutrinoMasses}
In this model, neutrino mass generation arises from contributions from both the active and dark sectors \cite{Bonilla_2023}, as show below
\begin{align}
    \label{eq:neutrinomass}
    \left(m_\nu\right)_{\alpha\beta}=\left(m_\nu^{active}\right)_{\alpha\beta}+\left(m_\nu^{dark}\right)_{\alpha\beta}.
\end{align}
It is important to identify which fields contribute to which sectors. Specifically, the fields contributing to the dark sector are those odd under the $\mathbb{Z}_2$ symmetry, which we identify from Eq.\eqref{eq:iddark} as $\varphi_2$ and $\chi_2$. All other fields, $N_s,\chi_1,\varphi_1$, $\Phi_s$ and $Z$ are part of the active sector. We analyze the contributions from both sectors separately. 
\subsection{Active Fields Contributions}
Taking into account the Eqs.\eqref{eq:vevsfin}, \eqref{eq:vevsrot}, \eqref{eq:yukawalagrangian1} and \eqref{eq:yukawacouplin} the Dirac neutrino mass matrix, $m_D$ and the heavy Majorana neutrino mass matrix $M_R$ are given respectively by
\begin{align}
    \label{eq:massmatrices}
    m_D=&\begin{pmatrix}
        y_1^\nu v_d & 0 & 0\\
        y_2^\nu v_d & 0 & y_4^\nu v_s/\sqrt{2}\\
        y_3^\nu v_d & 0 & y_5^\nu v_s/\sqrt{2}
    \end{pmatrix} , & M_R=&\begin{pmatrix}
        M & 0 & 0\\
        0 & M & 0\\
        0 & 0  & M_s
    \end{pmatrix}.
\end{align}

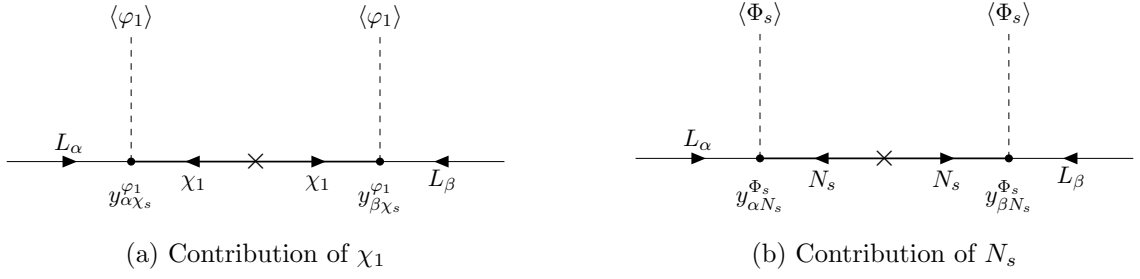
\begin{figure}[h!]
	\centering
	
	\begin{subfigure}[b]{0.45\textwidth}
		\centering
		\resizebox{\textwidth}{!}{ 
			\begin{tikzpicture}
				\begin{feynman}
					
					\vertex (a) at (-4, 0);
					\vertex (v1) at (-2, 0);
					\vertex (m) at (0, 0);
					\vertex (v2) at (2, 0);
					\vertex (b) at (4, 0);
					\vertex (h1) at (-2, 2);
					\vertex (h2) at (2, 2);

					\diagram*{
						
						(a) -- [fermion, edge label=$L_\alpha$] (v1),
						
						(b) -- [fermion, edge label=$L_\beta$] (v2),

						(v1) -- [scalar] (h1),
						(v2) -- [scalar] (h2)
					};

					\node[above=0pt of h1] {$\langle \varphi_1 \rangle$};
					\node[above=0pt of h2] {$\langle \varphi_1 \rangle$};

					\draw[fermion, thick] (m) -- (v1);
					
					\draw[fermion, thick] (m) -- (v2);
					
					\node at (m) {\Large $\times$};
					\node at (-1.0, -0.35) {$\chi_1$};
					\node at ( 1.0, -0.35) {$\chi_1$};

					\filldraw[black] (v1) circle (1.6pt) node[below=6pt] {$y_{\alpha \chi_s}^{\varphi_1}$};
					\filldraw[black] (v2) circle (1.6pt) node[below=6pt] {$y_{\beta \chi_s}^{\varphi_1}$};
				\end{feynman}
			\end{tikzpicture}
		}
		\caption{Contribution of $\chi_1$}
		\label{fig:seesaw_chi}
	\end{subfigure}
	\hfill 
	\begin{subfigure}[b]{0.45\textwidth}
		\centering
		\resizebox{\textwidth}{!}{ 
			\begin{tikzpicture}
				\begin{feynman}
					\vertex (a) at (-4, 0);
					\vertex (v1) at (-2, 0);
					\vertex (m) at (0, 0);
					\vertex (v2) at (2, 0);
					\vertex (b) at (4, 0);
					\vertex (h1) at (-2, 2);
					\vertex (h2) at (2, 2);

					\diagram*{
						
						(a) -- [fermion, edge label=$L_\alpha$] (v1),
						
						(b) -- [fermion, edge label=$L_\beta$] (v2),

						(v1) -- [scalar] (h1),
						(v2) -- [scalar] (h2)
					};

					\node[above=0pt of h1] {$\langle \Phi_s \rangle$};
					\node[above=0pt of h2] {$\langle \Phi_s \rangle$};

					\draw[fermion, thick] (m) -- (v1);
					
					\draw[fermion, thick] (m) -- (v2);
					
					\node at (m) {\Large $\times$};
					\node at (-1.0, -0.35) {$N_s$};
					\node at ( 1.0, -0.35) {$N_s$};

					\filldraw[black] (v1) circle (1.6pt) node[below=6pt] {$y_{\alpha N_s}^{\Phi_s}$};
					\filldraw[black] (v2) circle (1.6pt) node[below=6pt] {$y_{\beta N_s}^{\Phi_s}$};
				\end{feynman}
			\end{tikzpicture}
		}
		\caption{Contribution of $N_s$}
		\label{fig:seesaw_N}
	\end{subfigure}
	
	\caption{Feynman diagrams of tree-level type-I seesaw mediated by~(\subref{fig:seesaw_chi}) $\chi_1$ and~(\subref{fig:seesaw_N}) $N_s$. Only the active fields partake in this mass mechanism. The contribution from these diagrams generates a rank-2 mass matrix for light neutrinos.}
	\label{fig:full_seesaw}
\end{figure}

\begin{figure}[t!]
	\centering
	\begin{subfigure}[b]{0.48\textwidth}
		\centering
		\resizebox{\textwidth}{!}{ 
			\begin{tikzpicture}
				\begin{feynman}
					
					\vertex (a) at (-5, 0);  
					\vertex (b) at (5, 0);   
					\vertex (v1) at (-4, 0);
					\vertex (v2) at (4, 0);  
					\vertex (f1) at (-2, 0); 
					\vertex (m) at (0, 0);   
					\vertex (f2) at (2, 0);  
					\vertex (e1) at (-2, 2.5); 
					\vertex (e2) at (2, 2.5);

					\diagram*{
						
						(a) -- [fermion, edge label=$L_\alpha$] (v1),
						(v1) -- [fermion] (f1),

						(b) -- [fermion, edge label=$L_\beta$] (v2),
						(v2) -- [fermion] (f2),

						(m) -- [fermion, thick] (f1),
						(m) -- [fermion, thick] (f2)
					};

					\draw[boson, thick] (v1) .. controls (-4, 5.5) and (4, 5.5) .. (v2)
					node[pos=0.5, above=4pt] {$Z$};

					\diagram*{
						(f1) -- [scalar] (e1),
						(f2) -- [scalar] (e2)
					};

					\node[above=0pt of e1] {$\langle \varphi_1 \rangle$};
					\node[above=0pt of e2] {$\langle \varphi_1 \rangle$};
					
					\node at (m) {\Large $\times$};
					\node[below=4pt] at (-1.0, 0) {$\chi_1$};      
					\node[below=4pt] at ( 1.0, 0) {$\chi_1$};      
					
					\filldraw[black] (f1) circle (1.6pt) node[below=6pt] {$y_{\alpha \chi_s}^{\varphi_1}$};
					\filldraw[black] (f2) circle (1.6pt) node[below=6pt] {$y_{\beta \chi_s}^{\varphi_1}$};
					
				\end{feynman}
			\end{tikzpicture}
		}
		\caption{1-loop $Z$-boson correction to the mass term mediated by $\chi_1$.}
		\label{fig:zloop_chi}
	\end{subfigure}
	\hfill 
	\begin{subfigure}[b]{0.48\textwidth}
		\centering
		\resizebox{\textwidth}{!}{ 
			\begin{tikzpicture}
				\begin{feynman}
					\vertex (a) at (-5, 0);  
					\vertex (b) at (5, 0);   
					\vertex (v1) at (-4, 0); 
					\vertex (v2) at (4, 0);  
					\vertex (f1) at (-2, 0); 
					\vertex (m) at (0, 0);   
					\vertex (f2) at (2, 0);  
					\vertex (e1) at (-2, 2.5); 
					\vertex (e2) at (2, 2.5);

					\diagram*{
						
						(a) -- [fermion, edge label=$L_\alpha$] (v1),
						(v1) -- [fermion] (f1),

						(b) -- [fermion, edge label=$L_\beta$] (v2),
						(v2) -- [fermion] (f2),

						(m) -- [fermion, thick] (f1),
						(m) -- [fermion, thick] (f2)
					};

					\draw[boson, thick] (v1) .. controls (-4, 5.5) and (4, 5.5) .. (v2)
					node[pos=0.5, above=4pt] {$Z$};

					\diagram*{
						(f1) -- [scalar] (e1),
						(f2) -- [scalar] (e2)
					};

					\node[above=0pt of e1] {$\langle \Phi_s \rangle$};
					\node[above=0pt of e2] {$\langle \Phi_s \rangle$};
					
					\node at (m) {\Large $\times$};
					\node[below=4pt] at (-1.0, 0) {$N_s$};      
					\node[below=4pt] at ( 1.0, 0) {$N_s$};      
					
					\filldraw[black] (f1) circle (1.6pt) node[below=6pt] {$y_{\alpha N_s}^{\Phi_s}$};
					\filldraw[black] (f2) circle (1.6pt) node[below=6pt] {$y_{\beta N_s}^{\Phi_s}$};
					
				\end{feynman}
			\end{tikzpicture}
		}
		\caption{1-loop $Z$-boson correction to the mass term mediated by $N_s$.}
		\label{fig:zloop_N}
	\end{subfigure}
	
	\caption{Corrections from the active fields, $N_s$, $\chi_1$, $\Phi_s$, $\varphi_1$, and the $Z$-boson, mediated by (\subref{fig:zloop_chi}) $\chi_1$ and (\subref{fig:zloop_N}) $N_s$. These corrections are proportional to the tree-level seesaw mechanism.}
	\label{fig:zloop_correction}
\end{figure}
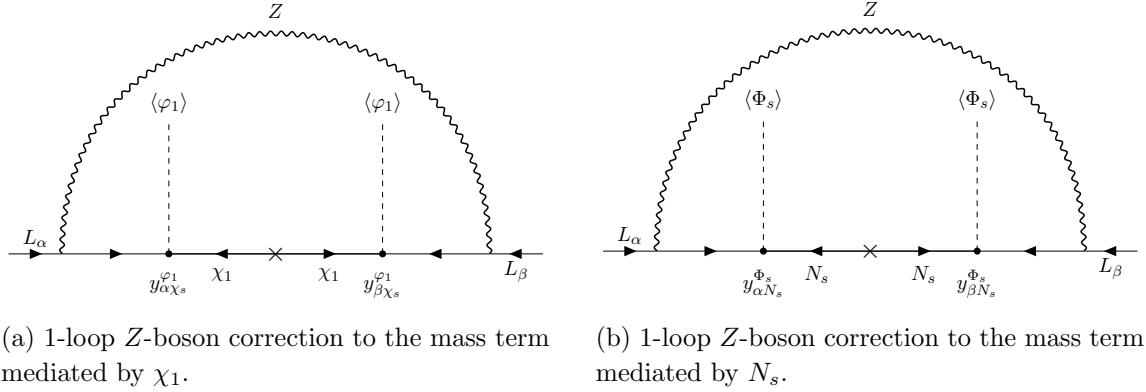

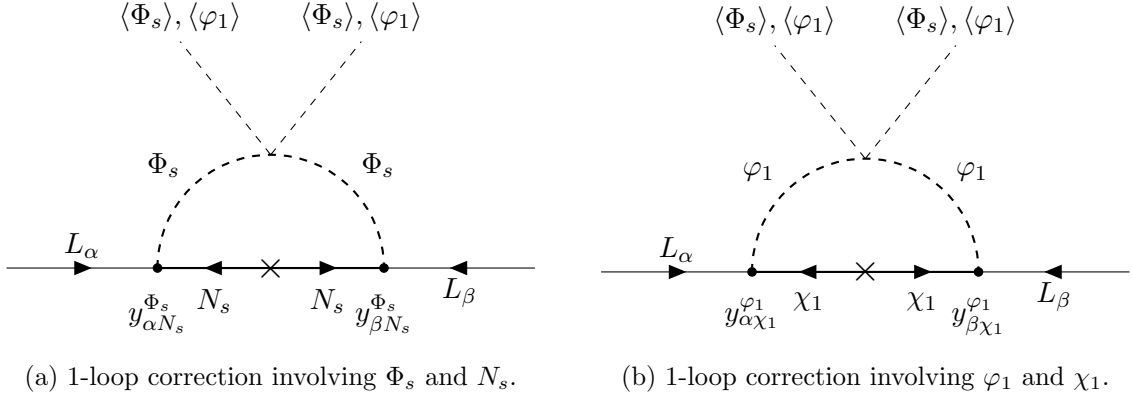
\begin{figure}[t!]
  \centering

\begin{subfigure}[b]{0.48\textwidth}
	\centering
	\resizebox{\textwidth}{!}{ 
		\begin{tikzpicture}
			\begin{feynman}
				
				\vertex (a) at (-3.5, 0);  
				\vertex (b) at (3.5, 0);   
				\vertex (v1) at (-1.5, 0); 
				\vertex (v2) at (1.5, 0);  
				\vertex (m) at (0, 0);

				\vertex (top) at (0, 1.5); 
				\vertex (vev1) at (-1.2, 3.0); 
				\vertex (vev2) at (1.2, 3.0);

				\diagram*{
					
					(a) -- [fermion, edge label=$L_\alpha$] (v1),

					(b) -- [fermion, edge label=$L_\beta$] (v2),

					(m) -- [fermion, thick] (v1), 
					(m) -- [fermion, thick] (v2) 
				};

				\draw[scalar, thick] (v1) arc [start angle=180, end angle=0, radius=1.5]
				node[pos=0.25, above left] {$\Phi_s$}  
				node[pos=0.75, above right] {$\Phi_s$};

				\draw[scalar, dashed] (top) -- (vev1) node[above] {$\langle \Phi_s \rangle, \langle \varphi_1 \rangle$};
				\draw[scalar, dashed] (top) -- (vev2) node[above] {$\langle \Phi_s \rangle, \langle \varphi_1 \rangle$};

				\node at (m) {\Large $\times$};
				\node[below=4pt] at (-0.75, 0) {$N_s$};
				\node[below=4pt] at ( 0.75, 0) {$N_s$};

				\filldraw[black] (v1) circle (1.6pt) node[below=6pt] {$y_{\alpha N_s}^{\Phi_s}$};
				\filldraw[black] (v2) circle (1.6pt) node[below=6pt] {$y_{\beta N_s}^{\Phi_s}$};
				
			\end{feynman}
		\end{tikzpicture}
	}
	\caption{1-loop correction involving $\Phi_s$ and $N_s$.}
	\label{fig:loopactive_Ns}
\end{subfigure}
  \hfill 
 \begin{subfigure}[b]{0.48\textwidth}
 	\centering
 	\resizebox{\textwidth}{!}{ 
 		\begin{tikzpicture}
 			\begin{feynman}
 				
 				\vertex (a) at (-3.5, 0); 
 				\vertex (b) at (3.5, 0);  
 				\vertex (v1) at (-1.5, 0); 
 				\vertex (v2) at (1.5, 0);  
 				\vertex (m) at (0, 0);     
 				\vertex (top) at (0, 1.5); 
 				\vertex (vev1) at (-1.2, 3.0); 
 				\vertex (vev2) at (1.2, 3.0);

 				\diagram*{
 					
 					(a) -- [fermion, edge label=$L_\alpha$] (v1),
 					(b) -- [fermion, edge label=$L_\beta$] (v2),

 					(m) -- [fermion, thick] (v1),

 					(m) -- [fermion, thick] (v2)
 				};

 				\draw[scalar, thick] (v1) arc [start angle=180, end angle=0, radius=1.5]
 				node[pos=0.25, above left] {$\varphi_1$} 
 				node[pos=0.75, above right] {$\varphi_1$};

 				\draw[scalar, dashed] (top) -- (vev1) node[above] {$\langle \Phi_s \rangle, \langle \varphi_1 \rangle$};
 				\draw[scalar, dashed] (top) -- (vev2) node[above] {$\langle \Phi_s \rangle, \langle \varphi_1 \rangle$};

 				\node at (m) {\Large $\times$};
 				\node[below=4pt] at (-0.75, 0) {$\chi_1$};
 				\node[below=4pt] at ( 0.75, 0) {$\chi_1$};

 				\filldraw[black] (v1) circle (1.6pt) node[below=6pt] {$y_{\alpha \chi_1}^{\varphi_1}$};
 				\filldraw[black] (v2) circle (1.6pt) node[below=6pt] {$y_{\beta \chi_1}^{\varphi_1}$};
 				
 			\end{feynman}
 		\end{tikzpicture}
 	}
 	\caption{1-loop correction involving $\varphi_1$ and $\chi_1$.}
 	\label{fig:loopactive_Chi1}
 \end{subfigure}

  \caption{1-loop corrections from the active fields involving (\subref{fig:loopactive_Ns}) $\Phi_s$ and $N_s$, and (\subref{fig:loopactive_Chi1}) $\varphi_1$ and $\chi_1$. These corrections are proportional to the tree-level seesaw mechanism.}
  \label{fig:loopactive_vshape}
\end{figure}

The fields in the $\chi_D$ doublet are mass degenerate. The fields participating in the type-I seesaw mechanism are $\mathbb{Z}_2$ even states, namely $\chi_1$ and $N_s$, whose tree-level topologies are depicted in Fig.~\ref{fig:seesaw_chi} and Fig.~\ref{fig:seesaw_N}, respectively. Their contribution to the neutrino masses is

\begin{align}
    \label{eq:seesaw}
    m_\nu^{tree} = & \underbrace{-\frac{v_d^2}{M}\begin{pmatrix}
         y_1^\nu y_1^\nu &  y_1^\nu y_2^\nu &  y_1^\nu y_3^\nu\\
         y_1^\nu y_2^\nu &  y_2^\nu y_2^\nu &  y_2^\nu y_3^\nu\\
         y_1^\nu y_3^\nu &  y_2^\nu y_3^\nu &  y_3^\nu y_3^\nu
    \end{pmatrix}}_{\left[m_\nu^{tree}\right]_{\chi_1}}
    \underbrace{-\frac{v_s^2}{2M_s}\begin{pmatrix}
         0 & 0 & 0 \\
         0 & y_4^{\nu} y_4^{\nu} & y_4^{\nu}y_5^{\nu}\\
         0 & y_5^{\nu}y_4^{\nu} & y_5^{\nu}y_5^{\nu}
    \end{pmatrix}}_{\left[m_\nu^{tree}\right]_{N_s}}
\end{align}
which is rank-2, meaning that only two neutrinos are massive at tree-level. The mass matrix $m_\nu^{active}$ contains the type-I seesaw contribution, as well as 1-loop corrections from the $Z$ boson and the active neutral scalar fields \cite{Aristizabal_Sierra_2011}, corresponding to the Feynman diagrams shown in Fig.~\ref{fig:zloop_correction} and Fig.~\ref{fig:loopactive_vshape} respectively. These contributions can be written as follows:
\begin{align}
    \label{eq:masstotact}
    m_\nu^{active}\equiv\mathcal{M}^{tree}+\mathcal{M}^{1-loop}=m_\nu^{tree}+m_{\nu,\chi_1}^{1-loop}+m_{\nu,N_s}^{1-loop}+m_{\nu,Z}^{1-loop}.
\end{align}

The 1-loop corrections mediated by the $Z$ boson preserve the flavor structure associated with each heavy neutrino separately. In our case, since $M_R$ is diagonal, there is no mixing between the heavy fermions $\chi_1$ and $N_s$. Consequently, the $Z$-loop contributions remain aligned with the corresponding tree-level matrix structures, $Y^{\varphi_1}Y^{\varphi_1^T}$ and $Y^{\Phi_s}Y^{\Phi_s^T}$. Therefore, they do not generate additional independent flavor structures beyond those already present at the tree-level. As a consequence, the rank of the active neutrino mass matrix remains equal to its tree-level value (rank-2). The same conclusion holds for the loop-induced seesaw contributions $m_{\nu,\chi_1}^{1-loop}$ and $m_{\nu,N_s}^{1-loop}$, which are also proportional to their respective tree-level structures and hence do not increase the rank of the active neutrino mass matrix either.

From the Yukawa Lagrangian in Eq.\eqref{eq:yukawalagrangian1}, we derive the Yukawa coupling matrices for the different scalars in the flavor basis. Within the active sector, the only non-vanishing coupling matrices are those connecting the $N_s$ fermion to the $\rho_s^\prime$ and $\eta_s^\prime$ scalars, and the $\chi_1$ fermion to the $\rho_1^\prime$ and $\eta_1^\prime$ scalars, namely
\begin{align}
    \label{eq:yuwakasflavor}
    Y^{\rho_1^\prime}=&Y^{\varphi_1} & Y^{\eta_1^\prime}=iY^{\varphi_1},\\
    Y^{\rho_s^\prime}=&Y^{\Phi_s} & Y^{\eta_s^\prime}=iY^{\Phi_s}.
\end{align}

Consequently, the one-loop contributions involving the active neutral scalar fields yield the following mass terms:
\begin{align}
    \left(m^{1-loop}_{\nu,\chi_1}\right)_{\alpha\beta} = & -\frac{1}{32\pi^2}\sum_a Y^a_{\alpha,\chi_1} M\left(\frac{m_a^2}{M^2-m_a^2}\ln{\left(\frac{m_a^2}{M^2}\right)}\right) Y^a_{\beta,\chi_1}, 
    \qquad  a=h_s,h_1,A_1 \label{eq:mnuchi1} \\
    \left(m^{1-loop}_{\nu,N_s}\right)_{\alpha\beta} = & -\frac{1}{32\pi^2}\sum_a Y^a_{\alpha,N_s} M_s\left(\frac{m_a^2}{M_s^2-m_a^2}\ln{\left(\frac{m_a^2}{M_s^2}\right)}\right) Y^a_{\beta,N_s} \label{eq:mnuNs}
\end{align}

Note that the Yukawa coupling matrices in Eqs. \eqref{eq:mnuchi1} and \eqref{eq:mnuNs} are written in the scalar mass-eigenstate basis. These are obtained using the basis-change matrices defined in Eqs.\eqref{eq:Rotpsueo} and \eqref{eq:diagonalneutralscalar}, which are given by:
\begin{align}
    Y^{a}_{\alpha,i} = & \left(O_{h_s,h_1}\right)^a_k Y^{k}_{\alpha,i}, 
    && \text{for } k = \rho_s', \rho_1', \ a = h_s, h_1 \label{eq:Y_a_up} \\
    Y_{\alpha,i}^{a} = & \left(O_{G_\eta, A_1}\right)^a_k Y^k_{\alpha,i}, 
    && \text{for } a = A_1, \ k = \eta_s', \eta_1' \label{eq:Y_a_down}
\end{align}

where $i=\chi_1,N_s$.

The mass matrix due to the 1-loop correction involving the $Z$-boson \cite{Aristizabal_Sierra_2011} is given by

\begin{align}
    \left(m_{\nu,Z}^{1-loop}\right)_{\alpha\beta}=-\frac{3m_Z^2}{16\pi^2v^2}\ln{\left(\frac{m_Z^2}{M^2}\right)}\left[m_\nu^{tree}\right]_{\chi_1}-\frac{3m_Z^2}{16\pi^2v^2}\ln{\left(\frac{m_Z^2}{M_s^2}\right)}\left[m_\nu^{tree}\right]_{N_s}.
\end{align}

As previously discussed, all 1-loop contributions from the active sector yield a rank-2 mass matrix, resulting in two massive neutrinos. On the other hand, the dark fields generate only one light neutrino mass via the scotogenic mechanism.

\subsection{Dark Field Contributions}
\begin{figure}[t!]
  \centering
  \begin{tikzpicture}
    \begin{feynman}
      
      \vertex (a) at (-3, 0);  
      \vertex (b) at (3, 0);   
      \vertex (v1) at (-1.5, 0); 
      \vertex (v2) at (1.5, 0);  
      \vertex (m) at (0, 0);

      \vertex (top) at (0, 1.5); 
      
      \vertex (vev1) at (-1.0, 2.5); 
      \vertex (vev2) at (1.0, 2.5);

      \diagram*{
        (a) -- [fermion, edge label=$L_\alpha$] (v1),
        (v2) -- [anti fermion, edge label=$L_\beta$] (b)
      };

      \draw[fermion, thick] (m) -- (v1);
      \draw[fermion, thick] (m) -- (v2);

      \node at (m) {\Large $\times$}; 
      \node at (-0.75, -0.35) {$\chi_2$};
      \node at ( 0.75, -0.35) {$\chi_2$};

      \draw[scalar, thick] (v1) arc [start angle=180, end angle=0, radius=1.5]
        node[pos=0.15, left] {$\varphi_2$}   
        node[pos=0.85, right] {$\varphi_2$};

      \draw[scalar, dashed] (top) -- (vev1);
      \draw[scalar, dashed] (top) -- (vev2);

      \node[above=1pt] at (vev1) {\small $\langle \Phi_s \rangle, \langle \varphi_1 \rangle$};
      \node[above=1pt] at (vev2) {\small $\langle \Phi_s \rangle, \langle \varphi_1 \rangle$};

      \filldraw[black] (v1) circle (1.6pt);
      \filldraw[black] (v2) circle (1.6pt);

    \end{feynman}
  \end{tikzpicture}
  \caption{Feynman diagram of the scotogenic mechanism. The dark fermion $\chi_2$ and the scalar $\varphi_2$ partake in this mass mechanism. The contribution from this diagram generates a rank-1 mass matrix for light neutrinos.}
  \label{fig:scotogenicdiag}
\end{figure}
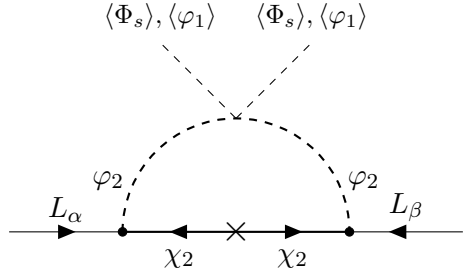

As illustrated in Fig.~\ref{fig:scotogenicdiag}, the contribution from the dark fields to the neutrino mass matrix arises from the scotogenic mechanism~\cite{Ma_2006,Escribano:2020iqq}:

\begin{align}
    \label{eq:massdarkcontribution}
    m_\nu^{\text{dark}}=m_{\nu,\chi_2}^{1-\text{loop}}.
\end{align}
The Yukawa coupling matrices for the dark sector are
\begin{equation}
\label{eq:yukawasdark}
    Y^{\rho_{D}^\prime} = Y^{\varphi_2}, \qquad Y^{\eta_{D}^\prime} = iY^{\varphi_2}.
\end{equation}
From this, the contribution to the neutrino mass matrix is
\begin{align}
     \label{eq:masaoscura}
    \left(m_\nu^{\text{dark}}\right)_{\alpha\beta}=&-\frac{1}{32\pi^2}\sum_{a}Y^a_{\alpha,\chi_2}M\left[\frac{m_a^2}{M^2-m_a^2}\ln{\frac{m_a^2}{M^2}}\right]Y^a_{\beta,\chi_2} & a=&h_2,A_2,
\end{align}
This dark mass matrix is rank-1. When combined with $m_\nu^{active}$, it allows us to explain both normal and inverted mass ordering scenarios. We will comment further on this in the following sections.
\section{Theoretical and Phenomenological Constraints}\label{Sec:Constraints}
In this section, we list the phenomenological and theoretical constraints to be considered in the study of the model's parameter space.
\subsection{General Constraints}
\paragraph{Theoretical requirements.}
Perturbativity of the theory requires the scalar and Yukawa couplings to satisfy
\begin{align}
    \label{eq:parampertur}
    \lvert \lambda_a\rvert,\lvert y_i^\nu\rvert^2, \lvert y_i\rvert^2&\leq4\pi; &a=1,\dots,12,\qquad i,j=1,\dots5
\end{align}

We also impose positivity of the Hessian and the bounded-from-below conditions for the potential \cite{Kannike_2012, Maniatis_2015,Degee_2013}.
\paragraph{Electroweak Precision Constraints.}
The new fields added to the SM Lagrangian modify the electroweak oblique parameters at the 1-loop level \cite{Peskin:1991sw}. We calculate the contribution to $S,T$ and $U$ following \cite{Grimus_2008,Grimus_2008_2,Hern_ndez_2015,Hern_ndez_2024} and compare them against the constraints from global fits to electroweak precision data \cite{ParticleDataGroup:2024cfk}. The allowed ranges we consider are:
\begin{align}
    \label{eq:obliqueparameters}
    &S=-0.04\pm0.01,\nonumber\\
    &T=0.01\pm0.12,\\
    &U=-0.01\pm0.09.\nonumber
\end{align}
\subsection{Lepton Flavor Violation}
 
In this model, LFV originates from different sources. The first source is the Yukawa interaction $Y^{\varphi_2}$, which induces LFV by generating neutrino masses via the Scotogenic mechanism \cite{Toma_2014}. The second source is the type-I seesaw mechanism, which also contributes to LFV \cite{Grimus_2002}, due to the non-diagonal flavor structure of the Yukawa matrices $Y^{\varphi_1}$ and $Y^{\phi_s}$. Finally, the charged lepton sector exhibits a non-diagonal mass matrix with a $\mu-\tau$ texture \cite{Sher:1991km}, arising from the flavor assignment. By restricting the interactions to a single scalar doublet, this assignment prevents the appearance of Higgs-mediated flavor changing at tree level.

All sources of LFV are subject to stringent experimental constraints from radiative decays $\ell_i\to\ell_j\gamma$, three-body decays, and $\mu-e$ conversion processes. In our analysis, we apply rigorous limits on these rare processes, specifically focusing on the $\ell_i\to \ell_j\gamma$ and $\ell_i\to3\ell_j$ channels. We use these channels to constrain the scalar spectrum and Yukawa couplings. Consequently, we restrict our analysis to the regions of the parameter space that are in full accordance with these limits, ensuring that all presented points satisfy these experimental constraints.

The relevance of each specific process depends on the specific source of LFV. In particular, due to the $\mu-\tau$ texture in the charged lepton mass matrix, the $\tau-\mu$ process channel is significantly enhanced. On the other hand, contributions from the Scotogenic sector and the type-I seesaw can induce all transitions and must therefore be considered simultaneously. The expressions for all processes were calculated following \cite{Sun:2025jmx,Redigolo:2024ztw,Lavoura2003,Toma_2014} and compared with the constraints from \cite{ParticleDataGroup:2024cfk,megiicollaboration2024searchmutoegammadataset,PhysRevLett.104.021802}. The current experimental bounds are summarized in Table \ref{tab:lfv_limits}.\\

\begin{table}[h!]
	\centering
	
	\renewcommand{\arraystretch}{1.5}
	
	\begin{tabular}{|l|c|}
		\hline
		\textbf{LFV observables} & \textbf{Experimental limits} \\
		\hline
		$\text{BR}(\mu \to e\gamma)$ & $\leq 4.2 \times 10^{-13}$ \\
		\hline
		$\text{BR}(\tau \to e\gamma)$ & $\leq 3.3 \times 10^{-8}$ \\
		\hline
		$\text{BR}(\tau \to \mu\gamma)$ & $\leq 4.4 \times 10^{-8}$ \\
		\hline
		$\text{BR}(\mu^- \to e^-e^+e^-)$ & $\leq 1.0 \times 10^{-12}$ \\
		\hline
		$\text{BR}(\tau^- \to e^-e^+e^-)$ & $\leq 1.4 \times 10^{-8}$ \\
		\hline
		$\text{BR}(\tau^- \to e^-\mu^+\mu^-)$ & $\leq 1.6 \times 10^{-8}$ \\
		\hline
		$\text{BR}(\tau^- \to \mu^-\mu^+\mu^-)$ & $\leq 1.6 \times 10^{-8}$ \\
		\hline
	\end{tabular}
	\caption{Summary of experimental limits on LFV processes.}
	\label{tab:lfv_limits}
\end{table}
\subsection{Neutrino Oscillations}\label{NeutrinoOscillationSection}
In our model, we do not have a flavor-diagonal mass matrix for charged leptons. Therefore, the matrix $U_\ell$ is the matrix that diagonalizes the charged lepton squared mass matrix $M_\ell^\dagger M_\ell$. The neutrino masses are obtained after the diagonalization of the mass matrix in Eq.~\eqref{eq:neutrinomass}. Similarly, $U_\nu$ is the matrix that diagonalizes the neutrino squared mass matrix $m_\nu^\dagger m_\nu$.

Therefore the lepton mixing matrix is given by $U_{PMNS}\equiv U_L\left(\theta_{12},\theta_{13},\theta_{23},\delta_{CP}\right)=U_\ell^\dagger U_\nu$, where $\theta_{ij}$ are the mixing angles and $\delta_{CP}$ is the Dirac CP-violating phase. These angles, $\theta_{ij}$, are determined by neutrino oscillation experiments. The global fits of neutrino oscillation parameters provide the best-fit values and their $3\sigma$ intervals \cite{de_Salas_2021,deSalas:2018bym,Gariazzo:2018pei,Esteban_2020}. The values for both normal and inverted neutrino mass orderings are given in Table \ref{tab:neutrino_params_3sigma}.

\begin{table}[H]
\centering
\renewcommand{\arraystretch}{1.3}

\begin{tabular}{|c|c|c|}
\hline
\textbf{Parameter} & \textbf{NO ($3\sigma$)} & \textbf{IO ($3\sigma$)} \\
\hline
$\Delta m^2_{21}\,[10^{-5}\text{ eV}^2]$ & $6.94-8.14$ & $6.94-8.14$ \\
\hline
$|\Delta m^2_{31}|\,[10^{-3}\text{ eV}^2]$ & $2.47-2.63$ & $2.37-2.53$ \\
\hline
$\sin^2\theta_{12}/10^{-1}$ & $2.71-3.69$ & $2.71-3.69$ \\
\hline
$\sin^2\theta_{23}/10^{-1}$ & $4.34-6.10$ & $4.33-6.08$ \\
\hline
$\sin^2\theta_{13}/10^{-2}$ & $2.000-2.405$ & $2.018-2.424$ \\
\hline
$\delta_{CP}/\pi$ & $0.71-1.99$ & $1.11-1.96$ \\
\hline
\end{tabular}

\caption{Neutrino oscillation parameters at $3\sigma$ for normal ordering (NO) and inverted ordering (IO).}
\label{tab:neutrino_params_3sigma}
\end{table}

\subsection{Dark Matter Constraints}
We consider a cold dark matter scenario with a WIMP like DM particle. The relic density for the dark matter candidate must fulfill the cosmological limits derived by Planck satellite data~\cite{2020}: $\Omega h^2=0.12\pm0.0036$ (at $3\sigma$). Furthermore, we consider the data analysis by LEP II for IDM, which excludes the mass region defined by the intersection of the conditions  \cite{Belyaev_2018,Lundstr_m_2009}: 
\begin{align} 
    \label{eq:massregionexcluded} 
    m_{\rho_2}&<80\text{ GeV}, & m_{\eta_2}<&100\text{ GeV}, & &\text{and} & \Delta m=&m_{\eta_2}-m_{\rho_2} >8\text{ GeV}. 
\end{align} 

Searches at LEP for exotic charged particles (such as charginos or inert Higgs bosons) impose the following bounds on the charged Higgs mass \cite{Castro_Alvaredo_2004,Pierce_2007}: 
\begin{align} 
    \label{eq:higgscharged} 
    m_{\sigma_2^{\pm}}\gtrsim 70-90\text{ GeV} 
\end{align} 

We also impose constraints from gauge boson width measurements, which are crucial for the low-mass region. The model should respect the precise measurements of the $W$ and $Z$ widths, which lead to the following lower limits on the odd-scalar masses:
\begin{equation}
	\label{eq:massconstrainWZ}
	\begin{split}
		m_{\sigma_2^{+}} + m_{\rho_2} &> m_{W^+}, \qquad m_{\sigma_2^{+}} + m_{\eta_2} > m_{W^+}, \\
		m_{\eta_2} + m_{\rho_2} &> m_{Z}, \qquad 2m_{\sigma_2^{\pm}} > m_{Z}.
	\end{split}
\end{equation}

These conditions ensure that the decay channels $\Gamma\left(W^+\to \rho_2\sigma_2^+,\eta_2\sigma_2^+\right)$ and $\Gamma\bigl(Z\to \rho_2\eta_2,\sigma_2^+\sigma_2^-\bigr)$ are kinematically forbidden.

To ensure the phenomenological viability of the DM candidate, we employ the most recent limits for the Spin-Independent cross-section $\left(\sigma_{SI}\right)$ from direct detection experiments, including PandaX~\cite{Cui_2017}, LUX-ZEPLIN~\cite{Aalbers_2023} and XENONnT~\cite{Aprile_2023}, to delineate the phenomenologically viable regions. Consequently, points exceeding these limits are considered experimentally excluded.

\section{Numerical Results} \label{Sec:Numerical_Results}
We performed a numerical scan of the model's parameter space in order to investigate its ability to reproduce neutrino phenomenology and dark sector observables. In contrast to previous 3HDM studies, which restrict the analysis to a generic high-mass point in the inert scalar sector~\cite{Garcia-Cely_2016}, our approach explores the parameter space without prior mass restrictions. Present analysis shows the viable dark matter candidates across an unrestricted mass range that simultaneously satisfy current relic abundance constraints and severe limits from direct detection experiments. 

To perform our analysis, we used the following workflow: we first used SARAH~\cite{Staub_2014} to implement the model, which provides analytical expressions for the spectrum and couplings and generates a source code for micrOmegas~\cite{B_langer_2018}. We then performed the scan using a custom Python script (implementing a Monte Carlo method), which filters parameter points against theoretical and experimental constraints. The valid points were then passed to micrOmegas to compute the dark matter relic abundance and the Spin-Independent DM-nucleon Cross-section.

In our numerical scan, the dimensionless couplings $(\lambda_i,~ y_i^\ell,~ y_i^\nu)$ were explored within the bounds of Eq.~\eqref{eq:parampertur}, allowing for complex Yukawa couplings. Furthermore, for the dimensional parameters, the vevs were varied around the electroweak scale, while the heavy neutrino masses were scanned within the interval $[10^4, 10^8]~\text{GeV}$. This range was chosen to ensure that the lightest inert scalar is the lightest particle in the dark sector, thus establishing a scalar dark matter scenario.

In the following subsections, we emphasize some correlations between the lepton mixing parameters, the lightest neutrino mass, the neutrinoless double beta decay mass parameter, the relic density, and the spin-independent cross-section for the dark matter candidate.
\subsection{Neutrino Physics}

In a minimal scenario for the model, that is, if we take the Yukawa couplings to be real, the model is only compatible with NO for the neutrino masses, since this corresponds to a scenario consistent with lepton CP-conservation, viable results for IO are not obtainable.

On the other hand, when complex Yukawa couplings are allowed, enlarging the parameter space and introducing CP-violating phases~\cite{2025}, the model can accommodate both normal and inverted orderings. In this case, the parameter space becomes less constrained, but the neutrino oscillations observables remain consistent with experimental data within $3\sigma$.  In addition, the model accommodates the full experimentally allowed range of the Dirac CP-phase, without providing a significant restriction on this observable. In the following sections, we focus on dark matter and lightest neutrino mass results obtained from parameter points compatible with neutrino oscillation constraints at the $3\sigma$ level.

It is worth mentioning that the Type-I seesaw mechanism appears at tree level and, due to the $D_4$ symmetry and the presence of two active right-handed neutrinos, it generates only two non-zero neutrino masses. Complementing this, the scotogenic mechanism generates a rank-1 dark mass matrix, ensuring that the total mass matrix is rank-3. Therefore, an analysis of the parameter space shows that the scotogenic mechanism is essential not only to induce the mass of the lightest neutrino but also to naturally complement the active sector in accommodating the full neutrino mass spectrum and mixings.

\subsubsection{Lightest Neutrino Mass and $\nu0\beta\beta$ decay}

\begin{figure}[t!]
\centering
\includegraphics[width=\textwidth]{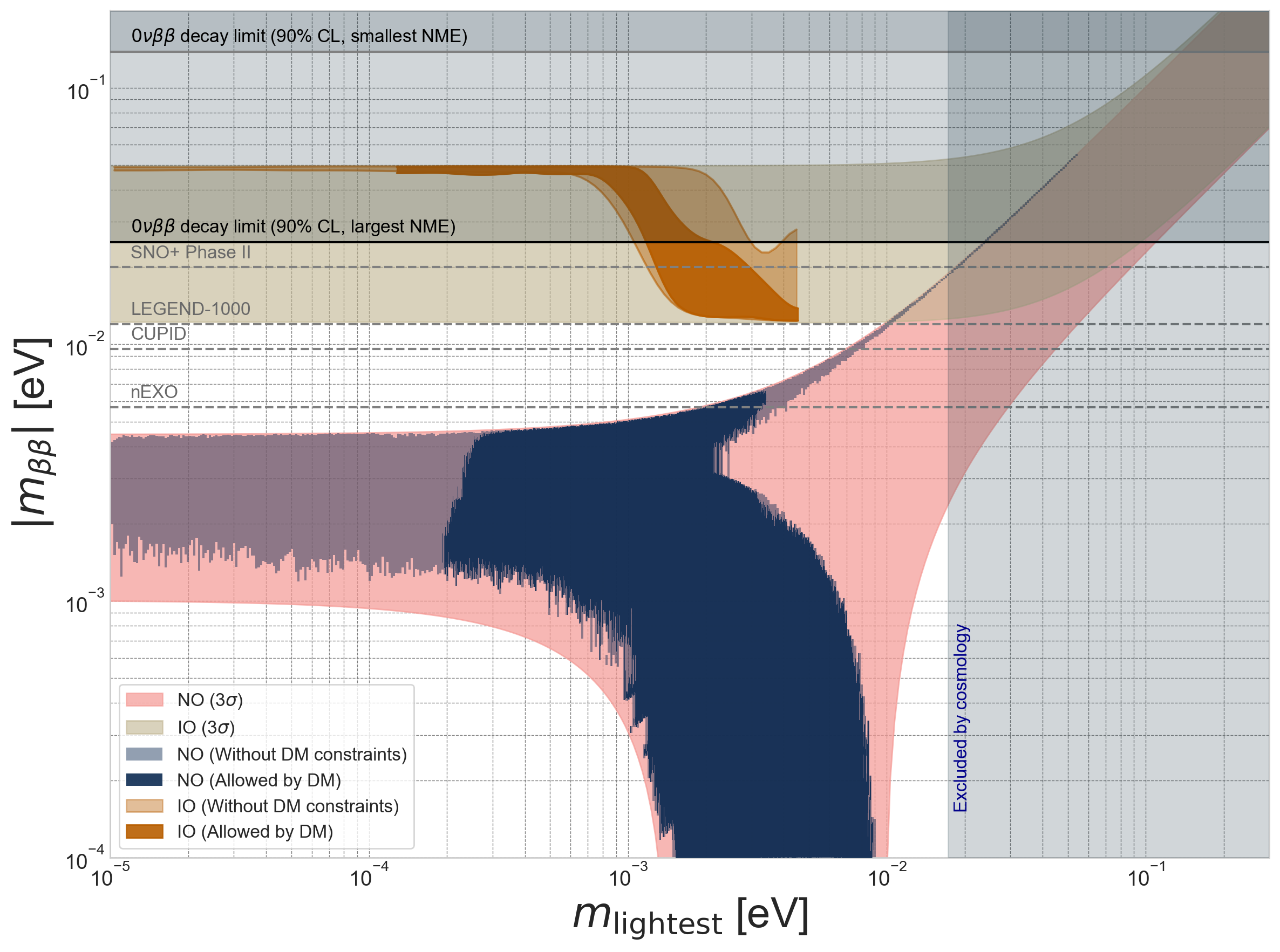} 
\caption{Viable model points in the $m_{\text{lightest}}$ vs $\langle m_{\beta\beta}\rangle$ plane. The beige and pink regions indicate the $3\sigma$ constraints from neutrino mixing parameters for the IO and NO scenarios, respectively. The light orange (IO) and light blue (NO) regions correspond to the model predictions before imposing the dark matter relic abundance and direct detection constraints, while the dark orange (IO) and dark blue (NO) regions represent the parameter space that remains after applying these constraints. The upper part of the plot also displays the $90\%$ CL exclusion limits from KamLAND-Zen, assuming the smallest and largest nuclear matrix elements}
\label{fig:neutrinoless}
\end{figure}

Given that neutrinos are Majorana particles in this framework, Lepton number violating processes are present, such as neutrinoless double-beta decay $\left(0\nu\beta\beta\right)$. The effective Majorana neutrino mass, $\langle m_{\beta\beta} \rangle$, is given by
\begin{align}
\label{eq:effectivemass}
\langle m_{\beta\beta} \rangle = \left| \sum_{i=1}^3 m_i U^2_{L_{ei}} \right|,
\end{align}
where $U_L$ denotes the lepton mixing matrix (see section \ref{NeutrinoOscillationSection}). It is important to remember that in our formalism, both the charged-lepton and neutrino mass matrices are non-diagonal. Thus, the contribution to the PMNS mixing matrix arises from both leptonic mass matrices.

Fig. \ref{fig:neutrinoless} illustrates our predictions for the $0\nu\beta\beta$, in which we display the effective mass parameter $\langle m_{\beta\beta}\rangle$ as a function of the lightest neutrino mass. The yellow region corresponds to the $3\sigma$ NO constraints from oscillation parameters, and the green region corresponds to the IO case, while the two grey horizontal bands represent the current experimental bounds on $\langle m_{\beta\beta}\rangle$ \cite{Abe_2023} for both the smallest and largest Nuclear Matrix Elements, together with the cosmological bound on $\sum m_\nu$ \cite{2020,PhysRevD.110.123537}. The projected sensitivities for future double-beta decay experiments (LEGEND \cite{legendcollaboration2021legend1000preconceptualdesignreport}, SNO+ Phase II \cite{Andringa_2016}, nEXO \cite{Adhikari_2021}, and CUPID \cite{alfonso2025sensitivitycupidexperiment0nubetabeta}) are displayed as horizontal dashed lines.

The viable range for the lightest neutrino mass in the NO scenario is $1.81\times10^{-4} \lesssim m_{\rm lightest} \lesssim 9.43\times10^{-3}$ eV, while the corresponding effective Majorana mass reaches a maximum value of $6.619\times10^{-3}$ eV. For the IO case, the lightest neutrino mass is found within the interval $1.259\times10^{-4} \lesssim m_{\rm lightest} \lesssim 4.61\times10^{-3}$ eV, whereas the effective Majorana mass spans the range $1.23\times10^{-2} \lesssim \langle m_{\beta\beta}\rangle \lesssim 4.952\times10^{-2}$ eV. As shown in Fig.~\ref{fig:neutrinoless}, the allowed parameter space for the IO scenario exhibits a pronounced narrowing. This behavior arises because the Majorana phases in the quasi-degenerate mass region collapse toward $\pi$, leading to destructive interference in the effective Majorana mass. It is important to emphasize that these results were obtained by considering only the parameter space consistent with the dark sector constraints of the model.

\subsection{Dark Matter}

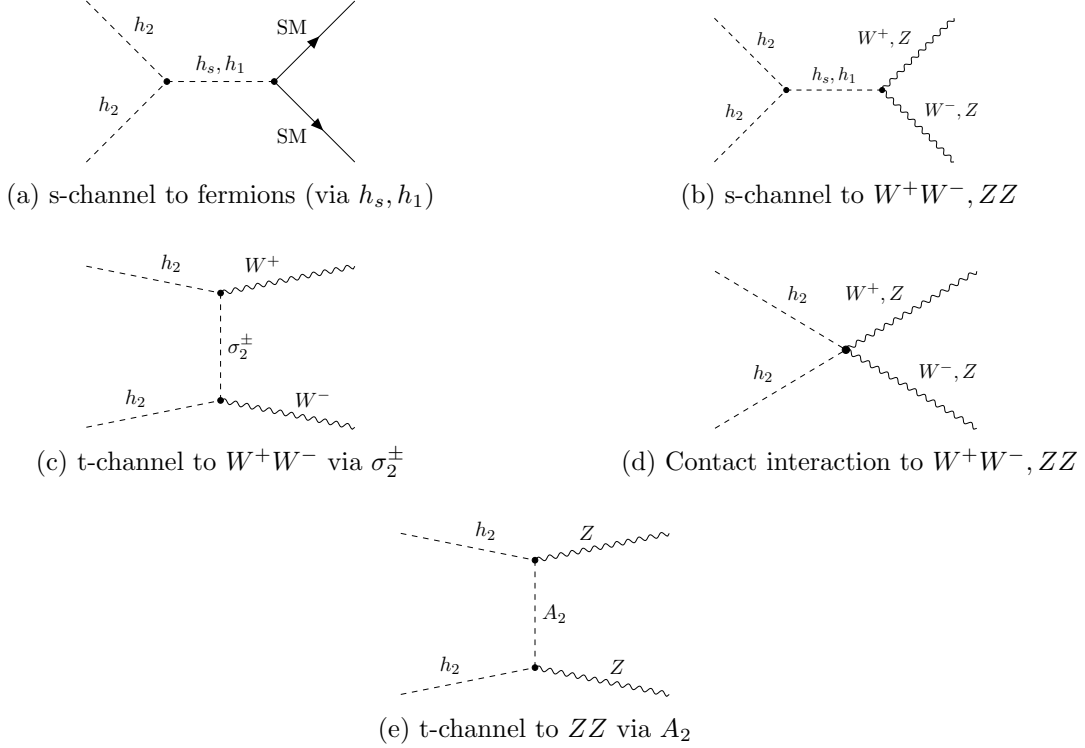
\begin{figure}[t!]
	\centering
	
	\begin{subfigure}[b]{0.45\textwidth}
		\centering
		\resizebox{0.55\textwidth}{!}{ 
			\begin{tikzpicture}
				\begin{feynman}
					\vertex (i1) at (-2.5, 1.5);
					\vertex (i2) at (-2.5, -1.5);
					\vertex (v1) at (-1, 0);
					\vertex (v2) at (1, 0);
					\vertex (f1) at (2.5, 1.5);
					\vertex (f2) at (2.5, -1.5);
					
					\diagram* {
						(i1) -- [scalar, edge label=$h_2$] (v1),
						(i2) -- [scalar, edge label=$h_2$] (v1),
						
						(v1) -- [scalar, edge label={$h_s, h_1$}] (v2),
						(v2) -- [fermion, edge label=SM] (f1),
						(f2) -- [anti fermion, edge label=SM] (v2),
					};
					\filldraw[black] (v1) circle (1.5pt);
					\filldraw[black] (v2) circle (1.5pt);
				\end{feynman}
			\end{tikzpicture}
		}
		\caption{s-channel to fermions (via $h_s, h_1$)}
		\label{fig:diag_s_fermions_combined}
	\end{subfigure}
	\hfill
	\begin{subfigure}[b]{0.45\textwidth}
		\centering
		\resizebox{0.55\textwidth}{!}{ 
			\begin{tikzpicture}
				\begin{feynman}
					\vertex (i1) at (-2.5, 1.5);
					\vertex (i2) at (-2.5, -1.5);
					\vertex (v1) at (-1, 0);
					\vertex (v2) at (1, 0);
					\vertex (f1) at (2.5, 1.5);
					\vertex (f2) at (2.5, -1.5);
					
					\diagram* {
						(i1) -- [scalar, edge label=$h_2$] (v1),
						(i2) -- [scalar, edge label=$h_2$] (v1),
						
						(v1) -- [scalar, edge label={$h_s, h_1$}] (v2),
						(v2) -- [boson, edge label={$W^+,Z$}] (f1),
						(v2) -- [boson, edge label={$W^-,Z$}] (f2),
					};
					\filldraw[black] (v1) circle (1.5pt);
					\filldraw[black] (v2) circle (1.5pt);
				\end{feynman}
			\end{tikzpicture}
		}
		\caption{s-channel to $W^+W^-,ZZ$ }
		\label{fig:diag_s_WW_combined}
	\end{subfigure}
	\par\bigskip 
	\begin{subfigure}[b]{0.45\textwidth}
		\centering
		\resizebox{0.55\textwidth}{!}{ 
			\begin{tikzpicture}
				\begin{feynman}
					\vertex (i1) at (-2.5, 1.5);
					\vertex (i2) at (-2.5, -1.5);
					\vertex (v1) at (0, 1.0);   
					\vertex (v2) at (0, -1.0);  
					\vertex (f1) at (2.5, 1.5);
					\vertex (f2) at (2.5, -1.5);
					
					\diagram* {
						(i1) -- [scalar, edge label=$h_2$] (v1),
						(i2) -- [scalar, edge label=$h_2$] (v2),
						(v1) -- [scalar, edge label=$\sigma_2^\pm$] (v2),
						(v1) -- [boson, edge label=$W^+$] (f1),
						(v2) -- [boson, edge label=$W^-$] (f2),
					};
					\filldraw[black] (v1) circle (1.5pt);
					\filldraw[black] (v2) circle (1.5pt);
				\end{feynman}
			\end{tikzpicture}
		}
		\caption{t-channel to $W^+W^-$ via $\sigma_2^\pm$}
		\label{fig:diag_t_charged}
	\end{subfigure}
	\hfill
	\begin{subfigure}[b]{0.45\textwidth}
		\centering
		\resizebox{0.55\textwidth}{!}{ 
			\begin{tikzpicture}
				\begin{feynman}
					\vertex (a) at (-2.5, 1.5);  
					\vertex (b) at (-2.5, -1.5); 
					\vertex (c) at (2.5, 1.5);    
					\vertex (d) at (2.5, -1.5);   
					\vertex (m) at (0, 0);        
					
					\diagram* {
						(a) -- [scalar, edge label=$h_2$] (m),
						(b) -- [scalar, edge label=$h_2$] (m),
						(m) -- [boson, edge label={$W^+,Z$}] (c),
						(m) -- [boson, edge label={$W^-,Z$}] (d),
					};
					\filldraw[black] (m) circle (2pt); 
				\end{feynman}
			\end{tikzpicture}
		}
		\caption{Contact interaction to $W^+W^-,ZZ$}
		\label{fig:diag_contact_WW}
	\end{subfigure}
    \par\bigskip 

    \begin{subfigure}[b]{0.45\textwidth}
		\centering
		\resizebox{0.55\textwidth}{!}{ 
			\begin{tikzpicture}
				\begin{feynman}
					\vertex (i1) at (-2.5, 1.5);
					\vertex (i2) at (-2.5, -1.5);
					\vertex (v1) at (0, 1.0);   
					\vertex (v2) at (0, -1.0);  
					\vertex (f1) at (2.5, 1.5);
					\vertex (f2) at (2.5, -1.5);
					
					\diagram* {
						(i1) -- [scalar, edge label=$h_2$] (v1),
						(i2) -- [scalar, edge label=$h_2$] (v2),
						(v1) -- [scalar, edge label=$A_2$] (v2),
						(v1) -- [boson, edge label=$Z$] (f1),
						(v2) -- [boson, edge label=$Z$] (f2),
					};
					\filldraw[black] (v1) circle (1.5pt);
					\filldraw[black] (v2) circle (1.5pt);
				\end{feynman}
			\end{tikzpicture}
		}
		\caption{t-channel to $ZZ$ via $A_2$}
		\label{fig:diag_t_Z}
	\end{subfigure}
    
	\caption{Main annihilation channels for CP-even dark scalar contributing to the DM annihilation cross-section: (a) s-channel annihilations to SM fermions via Higgs-portal mediated by $h_s$ or $h_1$, (b) s-channel annihilations to $W^{\pm}$ bosons mediated by $h_s$ or $h_1$, (c) t-channel annihilations to $W^{\pm}$ bosons mediated by dark charged scalar $\sigma_2^{\pm}$, (d) four-point contact interaction, and (e) t-channel annihilations to $Z$ mediated by dark pseudo-scalar $A_2$.}
	\label{fig:all_annihilation_channels}
\end{figure}


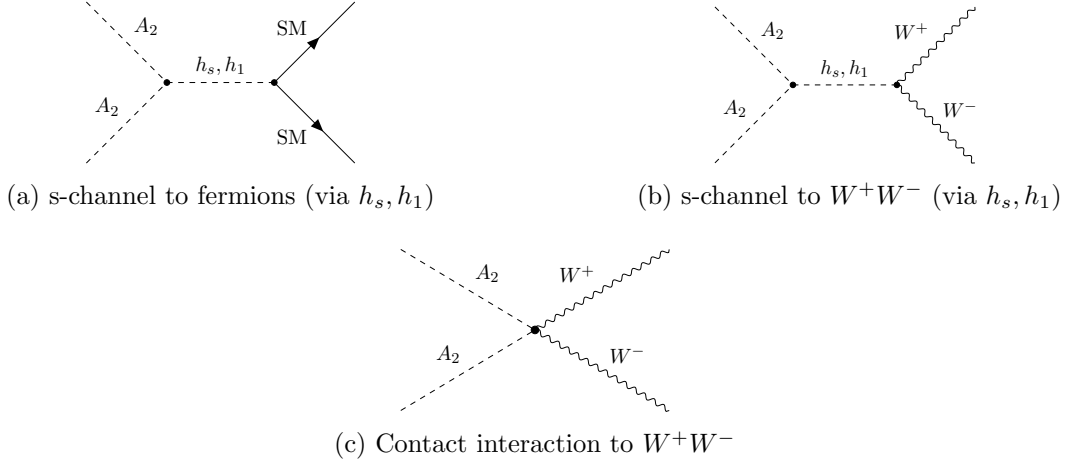
\begin{figure}[t!]
	\centering

	\begin{subfigure}[b]{0.45\textwidth}
		\centering
		\resizebox{0.55\textwidth}{!}{ 
			\begin{tikzpicture}
				\begin{feynman}
					\vertex (i1) at (-2.5, 1.5);
					\vertex (i2) at (-2.5, -1.5);
					\vertex (v1) at (-1, 0);
					\vertex (v2) at (1, 0);
					\vertex (f1) at (2.5, 1.5);
					\vertex (f2) at (2.5, -1.5);
					
					\diagram* {
						(i1) -- [scalar, edge label=$A_2$] (v1),
						(i2) -- [scalar, edge label=$A_2$] (v1),
						(v1) -- [scalar, edge label={$h_s, h_1$}] (v2),
						(v2) -- [fermion, edge label=SM] (f1),
						(f2) -- [anti fermion, edge label=SM] (v2),
					};
					\filldraw[black] (v1) circle (1.5pt);
					\filldraw[black] (v2) circle (1.5pt);
				\end{feynman}
			\end{tikzpicture}
		}
		\caption{s-channel to fermions (via $h_s, h_1$)}
		\label{fig:diag_s_A2A2_fermions}
	\end{subfigure}
	\hfill
	\begin{subfigure}[b]{0.45\textwidth}
		\centering
		\resizebox{0.55\textwidth}{!}{ 
			\begin{tikzpicture}
				\begin{feynman}
					\vertex (i1) at (-2.5, 1.5);
					\vertex (i2) at (-2.5, -1.5);
					\vertex (v1) at (-1, 0);
					\vertex (v2) at (1, 0);
					\vertex (f1) at (2.5, 1.5);
					\vertex (f2) at (2.5, -1.5);
					
					\diagram* {
						(i1) -- [scalar, edge label=$A_2$] (v1),
						(i2) -- [scalar, edge label=$A_2$] (v1),
						
						(v1) -- [scalar, edge label={$h_s, h_1$}] (v2),
						(v2) -- [boson, edge label=$W^+$] (f1),
						(v2) -- [boson, edge label=$W^-$] (f2), 
					};
					\filldraw[black] (v1) circle (1.5pt);
					\filldraw[black] (v2) circle (1.5pt);
				\end{feynman}
			\end{tikzpicture}
		}
		\caption{s-channel to $W^+W^-$ (via $h_s, h_1$)}
		\label{fig:diag_s_A2A2_WW}
	\end{subfigure}
	
	\par\bigskip 
    
	\begin{subfigure}[b]{0.45\textwidth}
		\centering
		\resizebox{0.55\textwidth}{!}{ 
			\begin{tikzpicture}
				\begin{feynman}
					\vertex (a) at (-2.5, 1.5);  
					\vertex (b) at (-2.5, -1.5); 
					\vertex (c) at (2.5, 1.5);    
					\vertex (d) at (2.5, -1.5);   
					\vertex (m) at (0, 0);        
					
					\diagram* {
						(a) -- [scalar, edge label=$A_2$] (m),
						(b) -- [scalar, edge label=$A_2$] (m),
						(m) -- [boson, edge label=$W^+$] (c),
						(m) -- [boson, edge label=$W^-$] (d),
					};
					\filldraw[black] (m) circle (2pt); 
				\end{feynman}
			\end{tikzpicture}
		}
		\caption{Contact interaction to $W^+W^-$}
		\label{fig:diag_4point_A2A2_WW}
	\end{subfigure}
	
	\caption{Main annihilation channels for CP-odd dark scalar: (a) s-channel annihilation to SM fermions via Higgs-portal mediated by $h_s$ or $h_1$, (b) s-channel annihilation to $W^{\pm}$ bosons mediated by $h_s$ or $h_1$, and (c) four-point contact interaction to $W^{\pm}$ bosons.}
	\label{fig:annihilation_A2A2_combined}
\end{figure}

\begin{figure}[t!]
  \centering

  \begin{subfigure}[b]{0.48\textwidth} 
    \centering
    \resizebox{0.55\textwidth}{!}{ 
    \begin{tikzpicture}
      \begin{feynman}
        
        \vertex (i1) at (-2.5, 1.5);
        \vertex (i2) at (-2.5, -1.5);
        \vertex (v1) at (-1, 0);
        \vertex (v2) at (1, 0);
        \vertex (f1) at (2.5, 1.5);
        \vertex (f2) at (2.5, -1.5);

        \diagram* {
          (i1) -- [scalar, edge label=$h_2$] (v1),
          (i2) -- [scalar, edge label=$A_2$] (v1),
          (v1) -- [scalar, edge label=$A_1$] (v2),
          (v2) -- [fermion, edge label=SM] (f1),
          (f2) -- [fermion, edge label=SM] (v2),
        };
        
        \filldraw[black] (v1) circle (1.5pt);
        \filldraw[black] (v2) circle (1.5pt);
      \end{feynman}
    \end{tikzpicture}
    }
    \caption{s-channel via $A_1$}
    \label{fig:s_channel_A1}
  \end{subfigure}
  \hfill 
  \begin{subfigure}[b]{0.48\textwidth}
    \centering
    \resizebox{0.55\textwidth}{!}{ 
    \begin{tikzpicture}
      \begin{feynman}
        
        \vertex (i1) at (-2.5, 1.5);
        \vertex (i2) at (-2.5, -1.5);
        \vertex (v1) at (-1, 0);
        \vertex (v2) at (1, 0);
        \vertex (f1) at (2.5, 1.5);
        \vertex (f2) at (2.5, -1.5);

        \diagram* {
          (i1) -- [scalar, edge label=$h_2$] (v1),
          (i2) -- [scalar, edge label=$A_2$] (v1),
          (v1) -- [boson, edge label=$Z$] (v2),
          (v2) -- [fermion, edge label=SM] (f1),
          (f2) -- [fermion, edge label=SM] (v2),
        };
        
        \filldraw[black] (v1) circle (1.5pt);
        \filldraw[black] (v2) circle (1.5pt);
      \end{feynman}
    \end{tikzpicture}
    }
    \caption{s-channel via $Z$ boson}
    \label{fig:s_channel_Z}
  \end{subfigure}

  \caption{Main coannihilation channels for $h_2 A_2\to SM+SM$ in s-channel mediated by CP-odd scalar, $A_1$, or $Z$ boson.}
  \label{fig:schannel_h2A2}
\end{figure}
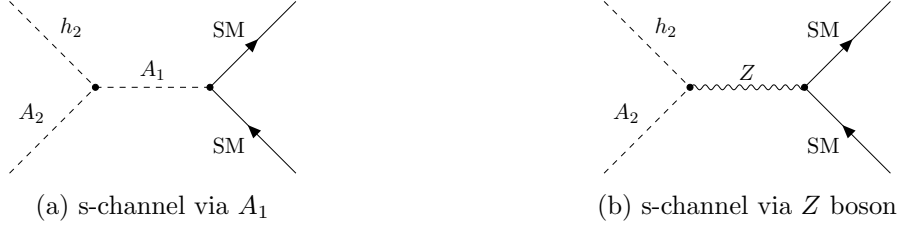

\begin{figure}[t!]
\centering
\includegraphics[width=0.95\textwidth]{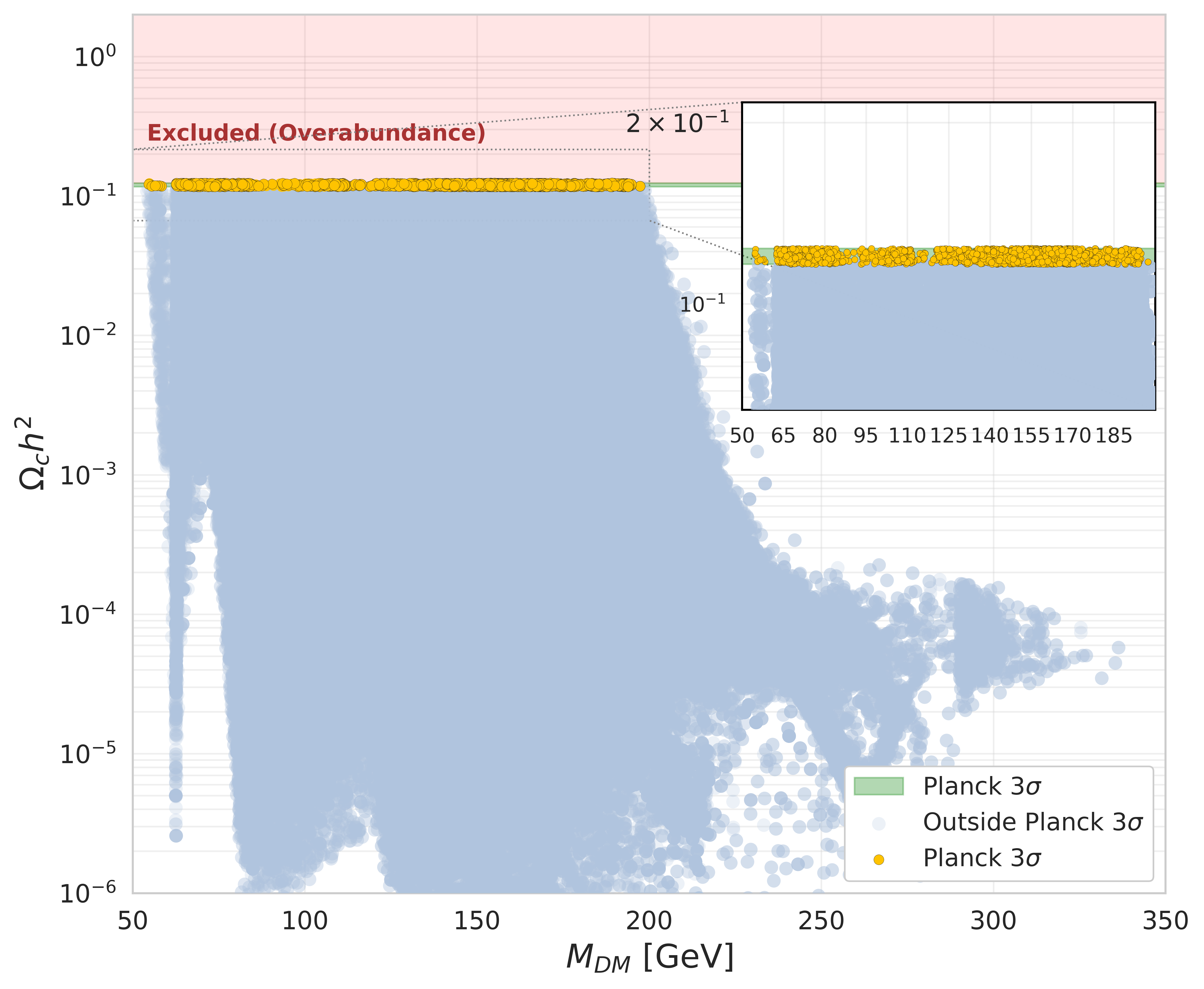}
\caption{Relic density as a function of DM candidate mass, $M_{DM}$. The horizontal green band corresponds to the value of the relic density observed by Planck within a $3\sigma$ range. Black points represent solutions with the spin-independent DM-nucleon cross-section, $\sigma_{SI}$, larger than the limit established by the LZ experiment. Yellow points represent solutions with $\sigma_{SI}$ lower than the limits established by the LZ experiment. We have taken into account that these limits vary with the mass scale.}
\label{fig:relicdensity}
\end{figure}

We consider thermally produced DM with a mass at the electroweak scale. The main annihilation channels contributing to the cross-section in this regime are shown in Figs.~\ref{fig:all_annihilation_channels}--\ref{fig:schannel_h2A2}. CP-even scalar DM that is thermally produced mainly through s-channel annihilations, with a contribution from t-channel into $W$ and $Z$ gauge bosons, through interactions with the dark sector parameterized by the scalar potential given in Eq.~\eqref{eq:Vtermsinv}, as illustrated in Fig.~\ref{fig:all_annihilation_channels}. In the case of the s-channel, the annihilation can occur through two active scalar mediators: the SM-like Higgs, denoted by $h_s$, or a heavy Higgs, $h_1$, leading to fermionic, $W^+ W^-$ or $ZZ$ final states, as shown in Figs.~\ref{fig:diag_s_fermions_combined} and \ref{fig:diag_s_WW_combined}.

Regarding the Higgs portal interactions, we emphasize that, due to the mixing in the active sector, the physical scalar states $h_s$ and $h_1$ are linear combinations of the active fields. Since only $\phi_s$ couples with the SM fermions, the state $h_s$, which we identify as the SM Higgs boson, inherits the dominant Yukawa interactions, while the heavier physical scalar $h_1$ couples to fermions only through its small mixing of $\phi_s$. Then,we find that the dark matter annihilation into a fermion final state is mainly mediated by $h_s$, in contrast to a gauge boson final states, where the heavy scalar $h_1$ becomes the dominant mediator,which is explicitly verified in our numerical analysis using micrOmegas.

Meanwhile, for the t-channel, shown in Fig.~\ref{fig:diag_t_charged}, the process proceeds via the exchange of a charged scalar from the dark sector, $\sigma_2^\pm$. Also, there is a contribution from the four-point contact interaction $h_2 h_2 W^+ W^-$ and $h_2 h_2 ZZ$, represented in Fig.~\ref{fig:diag_contact_WW}. Finally, there is a t-channel annihilations contribution via the exchange of a pseudo-scalar, from the dark sector $A_2$, to a final state $ZZ$, shown in Fig.~\ref{fig:diag_t_Z}. 

In the CP-odd case, the annihilation contributions arise exclusively from the s-channel, mediated by the scalar $h_s$ or $h_1$, leading to fermionic or $W^+ W^-$ final states, as shown in Figs. \ref{fig:diag_s_A2A2_fermions} and \ref{fig:diag_s_A2A2_WW}. Furthermore, the process can also occur through a four-point contact interaction, as shown in Fig \ref{fig:diag_4point_A2A2_WW}. In addition, co-annihilation processes between the CP-even and CP-odd states of the dark sector are possible via the s-channel, with the active CP-odd scalar, $A_1$, or the $Z$ boson exchanged, as shown in Fig. \ref{fig:schannel_h2A2}.

We find that, to reproduce the relic density observed by Planck, the freeze-out process is mainly dominated by annihilation into gauge bosons, with subdominant contributions to fermionic final states. Since the channels depending on the fermionic flavor structure have suppressed contributions to the thermal average, the normal and inverted ordering scenarios do not exhibit a significant difference in the dark matter annihilation mechanism. Furthermore, as shown in the $\langle m_{\beta\beta}\rangle$ vs $m_{\text{lightest}}$ plane (see Fig.~\ref{fig:neutrinoless}), the allowed parameter space for NO is significantly more expansive than the constrained band of the IO case. Therefore, in order to avoid redundancy and take advantage of the larger parameter space available in the NO scenario, we focus our discussion on dark matter phenomenology within this hierarchy.

\begin{figure}[t!]
	\centering
	\includegraphics[width=0.95\textwidth]{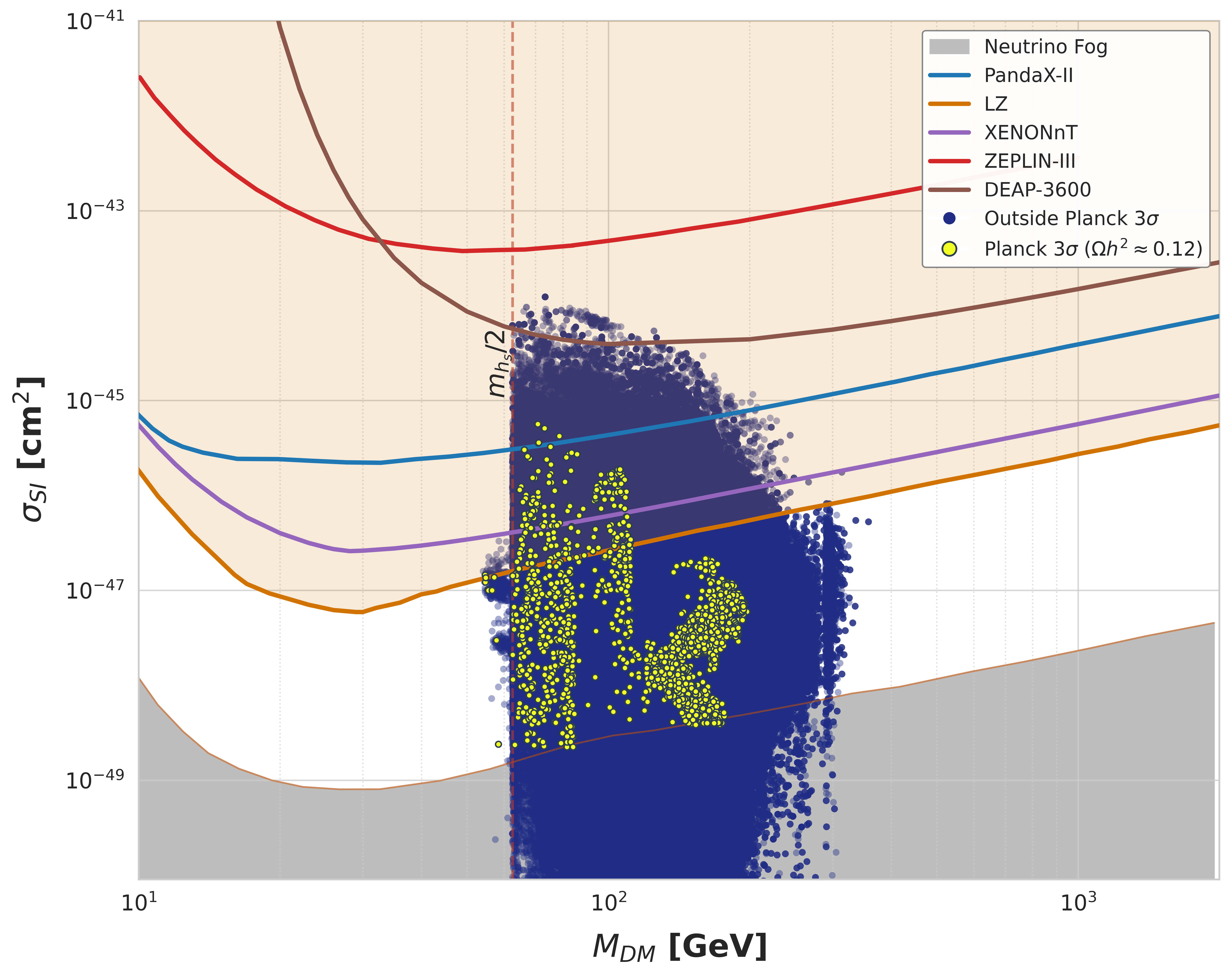}
	\caption{Direct detection in the model. All points satisfy neutrino oscillation parameters within $3\sigma$. The light gray points indicate where the relic abundance of DM is outside the Planck $3\sigma$ range. The yellow points are allowed by Planck to $3\sigma$. The dashed red line indicates half of the SM Higgs mass. It is worth noting that for masses $M_{DM}<M_{h_s}/2$, the invisible Higgs decay channel opens (Higgs-to-invisible decays). The green line represents the LZ-2022 constraint on spin-independent direct detection. The light orange region is the neutrino fog for Xenon.}
	\label{fig:directdetection}
\end{figure}

The numerical results for the dark sector are presented in Figs.~\ref{fig:relicdensity}--\ref{fig:directdetection}. First, we present the results for DM relic abundance obtained from parameter points consistent with the experimental limits described above. In Fig.~\ref{fig:relicdensity}, the value of the DM relic abundance predicted by the model is shown as a function of DM mass. The horizontal green band corresponds to the relic density values within the $3\sigma$ range observed by Planck.  In this plot, we project the constraints from direct detection (DD) to illustrate their impact: black points represent the solutions that are  excluded by the LZ upper limit, whereas yellow points represent the viable parameter space consistent with the current DD bounds.

As also seen in this figure, the pronounced dip around $65.5$ GeV corresponds to the Higgs resonance, which maximizes the annihilation cross-section and severely depletes the DM abundance. On the other hand,  the model's restricted viability region lies between $50\text{ GeV}-200\text{ GeV}$, where the processes in Figs.~\ref{fig:all_annihilation_channels}--\ref{fig:schannel_h2A2} yield the correct relic density.

An explicit visualization of the DD constraints projected in Fig.~\ref{fig:relicdensity} is provided in Fig.~\ref{fig:directdetection}, which shows the spin-independent DM-nucleon cross-section as a function of DM mass. The most stringent exclusion limit is imposed by the LZ experiment, represented by the solid green line. The light gray points represent data that are consistent with the production of at least a fraction of the DM relic abundance, while the yellow points are those that are within $3\sigma$ of the relic density observed by Planck.

In the region where $M_{DM}\leq M_{h_s}/2$, the invisible decay channel $h_s \to \text{DM} \text{ DM}$ opens kinematically. Collider searches by ATLAS and CMS impose a limit on the branching ratio of: $Br\left(h_s \to \text{inv} \right)<0.11$ at $95\%$ confidence level. This constraint directly translates into upper bounds on the potential couplings, restricting the parameter space allowed to the left of the dashed red line. Finally, in the mass region beyond the kinematic threshold, $M_{DM}\gtrsim M_{h_s}/2$, a portion of the parameter space exhibits a suppressed DD cross-section, rendering these points compatible with the LZ bound

\section{Discussion and Conclusions}\label{sec:conclusions}

In this work, we have studied a minimal extension of the SM based on a $D_4$ flavor symmetry in the lepton sector within a three-Higgs-doublet model, and three right-handed neutrinos, that with the incorporation of both a seesaw-type I mechanism and the scotogenic mechanism, are the source of neutrino mass generation with distinctive feature of this model is that the same Yukawa coupling governs both neutrino mass mechanisms. This structure is developed to naturally explore the origins of the hierarchy between the solar and atmospheric neutrino mass scales and, in the dark sector, our model provides a viable scalar dark matter candidate stabilized by the residual $\mathbb{Z}_2$ symmetry.

We begin our analysis by considering a possible minimal scenario with two right-handed neutrinos as a doublet under $D_4$, in which the total light neutrino mass matrix receives contributions from two rank-1 matrices, one from the active sector via the type-I seesaw mechanism, and the other from the dark sector at the one-loop level through the scotogenic mechanism. However, our analysis shows that this configuration is disfavored, as it fails to reproduce simultaneously the observed mass squared differences and mixing angles, in particular the reactor angle $\theta_{13}$.

In order to reproduce the reactor mixing, we needed to include a third right-handed neutrino as a singlet $1_3$. In this new scenario, the active sector generates a rank-2 mass matrix, while the dark sector correction contributes a rank-1 matrix, leading to a total rank-3 neutrino mass matrix, allowing us to successfully accommodate the neutrino oscillation data. On the other hand, given that the Lagrangian allows for complex Yukawa couplings, our model can naturally incorporate both normal and inverted orderings, as well as leptonic CP violation. However, our numerical analysis shows that $\delta_{\text{CP}}$ spans a wide range of values consistent with the latest neutrino global fits, whereas in other scoto-seesaw models, the lightest neutrino mass is tightly constrained by the solar mixing angle $\theta_{12}$~\cite{Bonilla_2023}, this work, implemented under the $D_4$ symmetry with a third heavy neutrino $N_s\sim1_3$, allows the lightest neutrino mass to vary over a wide range without exhibiting a strong correlation with $\theta_{12}$. In addition, the model structure leads to a global correlation between the reactor and atmospheric mixing angles in the CP-conserving limit, where Yukawa couplings are restricted to be real; conversely, if we allow the general scenario with complex Yukawas, this global correlation is lost, as the additional degrees of freedom provide greater flexibility to accommodate the experimental observables.

To assess the phenomenological viability, we performed an extensive numerical scan over the parameter space. We restricted the dimensionless couplings to conservative upper bounds to ensure perturbativity and theoretical stability, while also imposing the experimental bounds from electroweak precision observables and LFV. The numerical analysis demonstrates that the model successfully accommodates all current phenomenological limits. In particular, the LFV branching ratios remain below present bounds, with the dominant $\tau\to\mu\gamma$ channel yielding a branching ratio of order $\mathcal{O} \left(10^{-14}\right)$, while other LFV processes are significantly more suppressed.

About the dark matter, we identified a scalar dark matter candidate-the lightest neutral component of the inert doublet- that satisfies the relic density observed by Planck. Furthermore, taking into account the most recent exclusion limits from LUX-ZEPLIN, these limits yield a viable mass window of $50$--$200$ GeV for the dark matter candidate. It should be noted that all viable points in our numerical analysis simultaneously satisfy neutrino oscillation data, electroweak precision constraints, and LFV bounds, demonstrating that the existence of this viable DM region allows us to provide a consistent common origin for both neutrino physics and dark sector.

Finally, it is worth mentioning that our results show that there is no significant phenomenological distinction between the normal and inverted ordering scenarios regarding the relic density, as in both cases, the viability window remains identical. Furthermore, the dominance of annihilation channels during freeze-out is consistent: vector boson final states are the main contributors, with minor contributions from leptons. These results confirm that dark matter abundance is governed by flavor-independent interactions, indicating that dark sector physics is independent of the neutrino mass hierarchy.
\section*{Note Added}

During the final stages of this work, Ref.~\cite{Bonilla:2026jzk} appeared on arXiv, presenting a similar framework based on the $D_4$ discrete symmetry. We emphasize that our study extends and complements their findings in several crucial directions:
\begin{itemize}
 \item \textbf{Systematic model selection and no-go theorems:} While Ref.~\cite{Bonilla:2026jzk} directly constructs specific viable models for $S_3$ and $D_4$ by adding the singlet $N_S$, our study provides a systematic classification. In Appendix B, we analyze several alternatives (including $Q_6$ and $D_6$ groups, and three different lepton assignments in $D_4$ without $N_s$) and present the analytical reasons—such as the proportionality $m_\nu^{active} \propto m_\nu^{dark}$ or the propagation of structural zeroes—why they are phenomenologically excluded. This highlights the uniqueness of our chosen $D_4$ framework.
    \item We include physical CP-violating phases in the lepton sector, allowing us to fit the Dirac phase $\delta_{CP}$ and to reproduce both normal and inverted neutrino mass orderings, whereas Ref.~\cite{Bonilla:2026jzk} is restricted to the CP-conserving case and normal ordering.
    \item We conduct a full numerical scan of the parameter space to calculate the dark matter relic density and spin-independent cross-section. This allows us to identify a viable low-mass scalar dark matter window ($50\text{--}200$ GeV) compatible with the latest LUX-ZEPLIN limits, which was not explored in Ref.~\cite{Bonilla:2026jzk}.
    \item We perform a comprehensive analysis of lepton flavor violation constraints (such as $\ell_i \to \ell_j \gamma$ and three-body decays), ensuring the full experimental consistency of our benchmark scenarios.
\end{itemize}
\acknowledgments
This work was supported by DGAPA UNAM Grant No. PAPIIT-IN111625 and SECIHTI Grant CBF-2025-I-1589. E.P.R and A.P.L acknowledge the support of SNII (SECIHTI). I.T.M acknowledges support from SECIHTI and thanks IF-UNAM for its warm hospitality.
\appendix
\section{The $D_4$ Basis and its Representation Products} \label{appA}
$D_4$ group has two generators, $a\text{ and }b$, that satisfy $a^4=\mathbb{I},\text{ }b^2=\mathbb{I},\text{ and }aba=b$. In the real basis \cite{Ishimori:2010au,VIEN_2013}, these generators are defined as:
\begin{align}
    \label{eq:generatorD4}
      a&=\begin{pmatrix}
        0 & 1\\
        -1 & 0
    \end{pmatrix} & b&=\begin{pmatrix}
        1 & 0\\
        0 & -1
    \end{pmatrix}
\end{align}
This group consists of 8 elements and 5 conjugacy classes, which are:
\begin{align}
    \label{eq:conjugationclass}
     C_1&:\{e\}, & C_2&:\{a,a^3\}, & c_1^\prime&:\{a^2\}, & C_2^\prime&:\{b,a^2b\}, & c_2^{\prime\prime}&:\{ab,a^3b\}.
\end{align}
$D_4$ contains five irreducible representations (irreps): 4 singlets $\left(1_1,1_2,1_3,1_4\right)$ and one doublet, $2$. The multiplication rules for the singlets are:
\begin{align}
    \label{eq:productD4}
        1_i\otimes 1_i&=1_1 & 1_1\otimes 1_i&=1_i & 1_2\otimes 1_3&=1_4 & 1_2\otimes 1_4&=1_3 & 1_3\otimes 1_4&=1_2,
\end{align}
whereas the tensor product $2\otimes2$ decomposes into the four singlets: $2\otimes2=1_1\oplus1_2\oplus1_3\oplus1_4$. Thus, for two doublets $\left(a_1,a_2\right)^T$, $\left(b_1,b_2\right)^T\sim2$, the decomposition reads as follows
\begin{align}
    \label{eq:2x2descomposition}
        a_1b_1+a_2b_2&\sim1_1, & a_1b_1-a_2b_2&\sim1_2, & a_1b_2+a_2b_1&\sim1_3, & a_1b_2-a_2b_1&\sim1_4.
\end{align}

We have used these rules to write down the $D_4$ invariant scalar potential given in Eq.\eqref{eq:Vtermsinv}. Next, as already stated in the main text, we used the transformation 

\begin{align}
    \label{eq:Utransformation}
    U=\frac{1}{\sqrt{2}}\begin{pmatrix}
        1 & 1\\
        1 & -1
    \end{pmatrix}~,
\end{align}
in order to rotate the field basis, which yields the explicit scalar potential

\begin{align*}
     V&=\mu_s^2\left(\Phi_s^\dagger\Phi_s\right)+\mu_d^2\left[\left(\varphi_1^\dagger\varphi_1+\varphi_2^\dagger\varphi_2\right)\right]+\lambda_1\left(\Phi_s^\dagger\Phi_s\right)^2+\lambda_2\left(\varphi_1^\dagger\varphi_1+\varphi_2^\dagger\varphi_2\right)^2+\lambda_3\left(\varphi_1^\dagger\varphi_2+\varphi_2^\dagger\varphi_1\right)^2\nonumber\\
    +&\lambda_4\left(\varphi_1^\dagger\varphi_1-\varphi_2^\dagger\varphi_2\right)^2+\lambda_5\left(\varphi_2^\dagger\varphi_1-\varphi_1^\dagger\varphi_2\right)^2+\lambda_{6}\left(\varphi_1^\dagger\varphi_1+\varphi_2^\dagger\varphi_2\right)\left(\Phi_s^\dagger\Phi_s\right)\nonumber\\
    +&\lambda_7\left[\left(\Phi_s^\dagger\frac{\varphi_1+\varphi_2}{\sqrt{2}}\right)\left(\frac{\varphi_1^\dagger+\varphi_2^\dagger}{\sqrt{2}}\Phi_s\right)+\left(\Phi_s^\dagger\frac{\varphi_1-\varphi_2}{\sqrt{2}}\right)\left(\frac{\varphi_1^\dagger-\varphi_2^\dagger}{\sqrt{2}}\Phi_s\right)\right]\\
    +&\left[\lambda_{8}\left[\left(\Phi_s^\dagger\frac{\varphi_1+\varphi_2}{\sqrt{2}}\right)\left(\Phi_s^\dagger\frac{\varphi_1+\varphi_2}{\sqrt{2}}\right)+\left(\Phi_s^\dagger\frac{\varphi_1-\varphi_2}{\sqrt{2}}\right)\left(\Phi_s^\dagger\frac{\varphi_1-\varphi_2}{\sqrt{2}}\right)\right]+h.c\right]\nonumber\\
    +&\lambda_9\left(\varphi_1^\dagger\varphi_1^\dagger+\varphi_2^\dagger\varphi_2^\dagger\right)\left(\varphi_1\varphi_1+\varphi_2\varphi_2\right)+\lambda_{10}\left(\varphi_1^\dagger\varphi_2^\dagger+\varphi_2^\dagger\varphi_1^\dagger\right)\left(\varphi_1\varphi_2+\varphi_2\varphi_1\right)\nonumber\\
    +&\lambda_{11}\left(\varphi_1^\dagger\varphi_1^\dagger-\varphi_2^\dagger\varphi_2^\dagger\right)\left(\varphi_1\varphi_1-\varphi_2\varphi_2\right)+\lambda_{12}\left(\varphi_2^\dagger\varphi_1^\dagger-\varphi_1^\dagger\varphi_2^\dagger\right)\left(\varphi_2\varphi_1-\varphi_1\varphi_2\right)\nonumber
\end{align*}

After the spontaneous breaking of the electroweak and flavor symmetries, parameterized by the VEVs in Eq. \eqref{eq:vevsfin}, the mass matrices for the neutral scalar sector take the following forms,
\begin{align*}
    M^2_\rho=\left(
\begin{array}{ccc}
 2 \lambda _1 v_s^2 & \left(\lambda _6+\lambda _7+2 \lambda _8\right) v_d
   v_s & 0 \\
 \left(\lambda _6+\lambda _7+2 \lambda _8\right) v_d v_s & 2 \left(\lambda
   _2+\lambda _4+\lambda _9+\lambda _{11}\right) v_d^2 & 0 \\
 0 & 0 & 2 \left(\lambda _3-\lambda _4+\lambda _{10}-\lambda _{11}\right)
   v_d^2 \\
\end{array}
\right),
\end{align*}

\begin{align*}
    M_\eta^2=\left(
\begin{array}{ccc}
 -4 \lambda _8 v_d^2 & 2 \lambda _8 v_d v_s & 0 \\
 2 \lambda _8 v_d v_s & -\lambda _8 v_s^2 & 0 \\
 0 & 0 & -2 \left(\lambda _4+\lambda _5+\lambda _9-\lambda _{10}\right)
   v_d^2-\lambda _8 v_s^2 \\
\end{array}
\right)~,
\end{align*}
whereas, the mass matrix for the charged scalar sector is given by

\begin{align*}
    M^2_c=\left(
\begin{array}{ccc}
 \left(\lambda _7+2 \lambda _8\right) \left(-v_d^2\right) & \frac{1}{2}
   \left(\lambda _7+2 \lambda _8\right) v_d v_s & 0 \\
 \frac{1}{2} \left(\lambda _7+2 \lambda _8\right) v_d v_s & -\frac{1}{4}
   \left(\lambda _7+2 \lambda _8\right) v_s^2 & 0 \\
 0 & 0 & -\left(\left(2 \lambda _4+\lambda _9-\lambda _{10}+\lambda
   _{11}+\lambda _{12}\right) v_d^2\right)-\frac{1}{4} \left(\lambda _7+2
   \lambda _8\right) v_s^2 \\
\end{array}
\right).
\end{align*}
\section{Other Scenarios} \label{appB}

In the results presented in the previous sections, we consider the $D_4$ representation assignments for the lepton dublets as\footnote{See Tab.~\ref{tab:tab1}, in the section \ref{sec:TheModel}. }:
\begin{align}
    \label{eq:assignLept}
    L_1&\sim1_1,  & L_2&\sim1_3, & L_3&\sim1_3, & \ell_1&\sim1_1, & \ell_2&\sim1_3,  &\text{and}&& \ell_3&\sim1_3,
\end{align}
with three right-handed neutrino as $N_D\sim2$ and $N_s\sim1_3$. 

Nevertheless, we also analyzed alternative $D_4$ configurations, as well as other minimal non-abelian discrete groups such as $S_3$, $Q_6$, and $D_6$, which correspond to the lowest order possible. We begin by discussing the alternative $D_4$ scenario, followed by discussing the other groups mentioned above.
\subsection{$D_4$ implementations}
We also investigate the additional scenario above $D_4$ realizations: 3HDM$+N_D$ and 3HDM$+H+N_D$, where 3HDM refers to the scalar fields used in our main minimal model, and $H$ is an extra scalar. Note that the 3HDM$+H+N_D$ setup constitutes a 4HDM.
\subsubsection{3HDM$+N_D$}
In this case, the scalar potential remains identical to Eq.~\eqref{eq:Vtermsinv}, since the scalar field content is identical to the one presented in section \ref{sec:TheModel}. However, we explored different $D_4$ charge assignments for the charged leptons.

First, we consider the following quantum numbers for the fields, which are summarized in Tab.~\ref{tab:tabconf1}
\begin{table}[H]
\centering
\begin{tabular}{|l|l|l|l|l|l|l|l|l|l|}
\hline
        & $L_1$ & $L_2$ & $L_3$ & $\ell_1$ & $\ell_2$ & $\ell_3$ & $\phi_s$ & $\phi_D$ & $N_D$ \\ \hline
$D_4$   & $1_3$ & $1_3$ & $1_1$ & $1_3$    & $1_3$    & $1_1$    & $1_1$    & $2$      & $2$   \\ \hline
$SU(2)$ & $2$   & $2$   & $2$   & $1$      & $1$      & $1$      & $2$      & $2$      & $1$   \\ \hline
\end{tabular}
\caption{Quantum numbers for the first configuration.}
\label{tab:tabconf1}
\end{table}

For this configuration, we have the following structure for the neutrino mass matrix
\begin{align}
    \label{eq:mnuconf1}
    m_\nu= m_\nu^{active}+m_\nu^{dark}\simeq\begin{pmatrix}
        (a+d)y_1^\nu y_1^\nu & (a+d)y_1^\nu y_2^\nu & (a-d)y_1^\nu y_3^\nu\\
        (a+d)y_1^\nu y_2^\nu & (a+d)y_2^\nu y_2^\nu & (a-d)y_2^\nu y_3^\nu\\
        (a-d)y_1^\nu y_3^\nu & (a-d)y_2^\nu y_3^\nu & (a+d)y_3^\nu y_3^\nu
    \end{pmatrix}.
\end{align}

This matrix is a rank-2 and thus yields a single vanishing eigenvalue. Its corresponding eigenvector $\left(A,B,0\right)^T$ leads to a neutrino mixing matrix, $U_\nu$, which takes the following texture
\begin{equation}
    \label{eq:Uconf1}
    U_\nu=\begin{pmatrix}
        \times & \times & \times\\
        \times & \times & \times\\
        0 & \times & \times
    \end{pmatrix}.
\end{equation}

 We find that this configuration is phenomenologically excluded, since the final leptonic mixing matrix is defined as $U_\ell^\dagger U_\nu$ and, therefore, the zero element in $U_\nu$ propagates to the final observable mixing if not compensated by $U_\ell$. However, in this scenario, the $U_\ell$ matrix arising from the assignments under $D_4$ does not exhibit the required structure to remove this zero. Consequently, the vanishing element persists in the final mixing matrix, leading to predictions inconsistent with the experimental data.

The second exploration is summarized in Table \ref{tab:tabconf2}.
\begin{table}[H]
\centering
\begin{tabular}{|l|l|l|l|l|l|l|l|l|l|}
\hline
        & $L_1$ & $L_2$ & $L_3$ & $\ell_1$ & $\ell_2$ & $\ell_3$ & $\phi_s$ & $\phi_D$ & $N_D$ \\ \hline
$D_4$   & $1_1$ & $1_3$ & $1_3$ & $1_1$    & $1_3$    & $1_3$    & $1_1$    & $2$      & $2$   \\ \hline
$SU(2)$ & $2$   & $2$   & $2$   & $1$      & $1$      & $1$      & $2$      & $2$      & $1$   \\ \hline
\end{tabular}
\caption{Quantum numbers for the second configuration.}
\label{tab:tabconf2}
\end{table}

Similarly to the previous case, the neutrino mass matrix is
\begin{align}
    \label{eq:mnuconf2}
    m_\nu=m_\nu^{active}+m_\nu^{dark}\simeq\begin{pmatrix}
        (a+d)y_1^\nu y_1^\nu & (a-d)y_1^\nu y_2^\nu & (a-d)y_1^\nu y_3^\nu\\
        (a-d)y_1^\nu y_2^\nu & (a+d)y_2^\nu y_2^\nu & (a+d)y_2^\nu y_3^\nu\\
        (a-d)y_1^\nu y_3^\nu & (a+d)y_2^\nu y_3^\nu & (a+d)y_3^\nu y_3^\nu
    \end{pmatrix}.
\end{align}

The eigenvector associated with the vanishing eigenvalue now takes the form: $\left(0,A,B\right)^T$, providing a neutrino mixing matrix with the texture
\begin{align}
    \label{eq:Uconf2}
    U_\nu=\begin{pmatrix}
           0   \times \times\\
        \times \times \times\\
        \times \times \times\\
    \end{pmatrix}
\end{align}
However, as before, this zero cannot be removed by the $U_\ell$ matrix. 

Finally, we consider the third exploration, summarized in the Table \ref{tab:tabconf3}.
\begin{table}[H]
\centering
\begin{tabular}{|l|l|l|l|l|l|l|l|l|l|}
\hline
        & $L_1$ & $L_2$ & $L_3$ & $\ell_1$ & $\ell_2$ & $\ell_3$ & $\phi_s$ & $\phi_D$ & $N_D$ \\ \hline
$D_4$   & $1_3$ & $1_1$ & $1_3$ & $1_3$    & $1_1$    & $1_3$    & $1_1$    & $2$      & $2$   \\ \hline
$SU(2)$ & $2$   & $2$   & $2$   & $1$      & $1$      & $1$      & $2$      & $2$      & $1$   \\ \hline
\end{tabular}
\caption{Quantum numbers for the third configuration.}
\label{tab:tabconf3}
\end{table}

In this case, we find that the neutrino mixing matrix has the form

\begin{align}
    \label{eq:Uconf3}
    U_\nu=\begin{pmatrix}
        \times & \times & \times\\
        0 & \times & \times\\
        \times & \times & \times
    \end{pmatrix} ~,
\end{align}
but, once again, it remains impossible to remove the zero from the mixing matrix. 

Since none of these three configurations are capable of resolving this structural deficit, they are phenomenologically unviable. This conclusion is consistent with previous analyses in \cite{Grimus_2003, Grimus_2004}.

\subsubsection{4HDM$+N_D$}

For this implementation, we have added a new scalar field, $H\sim1_3$. In this case, the relevant particle content and quantum numbers are summarized in Table \ref{tab:3hdm_H}.
\begin{table}[H]
\centering
\begin{tabular}{|l|l|l|l|l|l|l|l|l|l|l|}
\hline
        & $L_1$ & $L_2$ & $L_3$ & $\ell_1$ & $\ell_2$ & $\ell_3$ & $\phi_s$ & $\phi_D$ & $N_D$ & $H$\\ \hline
$D_4$   & $1_1$ & $1_3$ & $1_3$ & $1_1$    & $1_3$    & $1_3$    & $1_1$    & $2$      & $2$   & $1_3$\\ \hline
$SU(2)$ & $2$   & $2$   & $2$   & $1$      & $1$      & $1$      & $2$      & $2$      & $1$   & $2$\\ \hline
\end{tabular}
\caption{Quantum numbers for the 4HDM model.}
\label{tab:3hdm_H}
\end{table}
The addition of the extra scalar $H$ introduces new terms to the potential, which now reads
\begin{align}
    \label{eq:VwithH}
     V&=\mu_s^2\Phi_s^\dagger\Phi_s+\mu_{d}^2\left[\Phi_D^\dagger\Phi_D\right]_{1_1}+\lambda_1\left(\Phi_s^\dagger\Phi_s\right)^2+\sum_{i=1}^4\lambda_{i+1}\left(\left[\Phi_D^\dagger\Phi_D\right]_{1_i}\right)^2\nonumber\\
    +&\lambda_6\left[\Phi_D^\dagger\Phi_D\right]_{1_1}\left(\Phi_s^\dagger\Phi_s\right)+\lambda_7\left[\left[\Phi_s^\dagger\Phi_D\right]_2\left[\Phi_D^\dagger\Phi_s\right]\right]_{1_1}+\left[\lambda_8\left(\left[\Phi_s^\dagger\Phi_D\right]_2\left[\Phi_s^\dagger\Phi_D\right]_2\right)_{1_1}+h.c\right]\nonumber\\
    +&\lambda_9\left[\Phi_D^\dagger\Phi_D^\dagger\right]_{1_1}\left[\Phi_D\Phi_D\right]_{1_1}+\lambda_{10}\left[\Phi_D^\dagger\Phi_D^\dagger\right]_{1_2}\left[\Phi_D\Phi_D\right]_{1_2}+\lambda_{11}\left[\Phi_D^\dagger\Phi_D^\dagger\right]_{1_3}\left[\Phi_D\Phi_D\right]_{1_3}\nonumber\\
    +&\lambda_{12}\left[\Phi_D^\dagger\Phi_D^\dagger\right]_{1_4}\left[\Phi_D\Phi_D\right]_{1_4}+\mu_H^2 \left(H^\dagger H\right)+\lambda_1^\prime\left(H^\dagger H\right)^2+\lambda_6^\prime\left[\Phi_D^\dagger\Phi_D\right]_{1_1}\left(H^\dagger H\right)\nonumber\\
    +&\lambda_7^\prime\left[\left[H^\dagger \Phi_D\right]_{2}\left[\Phi_D^\dagger H\right]\right]_{1_1}+\left[\lambda_8^\prime\left(\left[H^\dagger\Phi_D\right]_2\left[H^\dagger\Phi_D\right]_2\right)_{1_1}+\text{h.c}\right]\nonumber\\
    +&\lambda_1^{\prime\prime}\left(H^\dagger H\right)\left(\Phi_s^\dagger\Phi_s\right)+\left[\lambda_3^\prime\left(H^\dagger\Phi_s\right)\left(H^\dagger\Phi_s\right)+\text{h.c}\right]+\lambda_4^\prime\left(\Phi_s^\dagger H\right)\left(H^\dagger\Phi_s\right)\nonumber\\
    +&\lambda_{6}^{\prime\prime}\left[\left[\Phi_D^\dagger\Phi_D\right]_{1_3}\left(\Phi_s^\dagger H\right)+\text{h.c}\right]+\lambda_7^{\prime\prime}\left[\left[\left[H^\dagger\Phi_D\right]_2\left[\Phi_D^\dagger\Phi_s\right]_2\right]_{1_1}+\text{h.c}\right]\nonumber\\
    +&\lambda_8^{\prime\prime}\left(\left[\left[\Phi_s^\dagger\Phi_D\right]_2\left[H^\dagger\Phi_D\right]_2\right]_{1_1}+\text{h.c}\right).
\end{align}

Therefore, the Yukawa Lagrangian is given by
\begin{align}
    \label{eq:lyukawaleptonicoH}
    \mathcal{L}_{Y}=&y_1^\ell\overline{L_1}\Phi_s\ell_1+y_2^\ell\overline{L_2}\Phi_s\ell_2+y_3^\ell\overline{L_2}\Phi_s\ell_3+y_4^\ell\overline{L_3}\Phi_s\ell_2+y_5^\ell\overline{L_3}\Phi_s\ell_3+\nonumber\\
     +&y_6^\ell\overline{L_1}H\ell_2+y_7^\ell\overline{L_1}H\ell_3+y_8^\ell\overline{L_2}H\ell_1+y_9^\ell\overline{L_3}H\ell_1\nonumber\\
     +&y_1^\nu\overline{L_1}\left(\Tilde{\Phi}_DN_D\right)_{1_1}+y_2^\nu\overline{L_2}\left(\Tilde{\Phi}_DN_D\right)_{1_3}+y_3^\nu\overline{L_3}\left(\Tilde{\Phi}_DN_D\right)_{1_3}+M\left(\overline{N_D^C}N_D\right)_{1_1}+h.c.
\end{align}

In this construction, we successfully reproduce all observable neutrino masses and mixing, satisfying theoretical and phenomenological constraints, which makes the model phenomenologically viable. Nevertheless, the parameter space is sufficiently extensive to accommodate the experimental data independently, without the correlation characteristic of the minimal scenario. On the other hand, the extended scalar potential permits the introduction of a CP phase that is not present in the minimal model.

\subsection{$S_3$ implementation}
The $S_3$ group has three irreps: one doublet, $2$, and two singlets, $1$ and $1^\prime$. The product rules for these irreps are given by \cite{Ishimori:2010au}:
\begin{align}
    \label{eq:irrepsS3}
    2\otimes2=&1\oplus1^\prime\oplus2, & 1^\prime\otimes1^\prime=&1, & 2\otimes1=&2.
\end{align}
In the real representation, this product can be decomposed as:
\begin{align}
    \label{eq:decompositionproductS3}
    \begin{pmatrix}
    x_1\\
    x_2
    \end{pmatrix}_2\otimes\begin{pmatrix}
    y_1\\
    y_2
    \end{pmatrix}_2=&\left(x_1y_1+x_2y_2\right)_{1}+\left(x_1y_2-x_2y_1\right)_{1^\prime}+\begin{pmatrix}
        x_1y_2+x_2y_1\\
        x_1y_1-x_2y_2
    \end{pmatrix}_{2},\nonumber\\ 
     \newline
    \begin{pmatrix}
        x_1\\
        x_2
    \end{pmatrix}_2\otimes\begin{pmatrix}
        y^\prime
    \end{pmatrix}_{1^\prime}=&\begin{pmatrix}
        -x_2y^\prime\\
        x_1y^\prime   
    \end{pmatrix}_2,\\ 
    \newline
    \begin{pmatrix}
        x^\prime
    \end{pmatrix}_{1^\prime}\otimes\begin{pmatrix}
        y^\prime
    \end{pmatrix}_{1^\prime}=&\begin{pmatrix}
        x^\prime y^\prime
    \end{pmatrix}_1. \nonumber
\end{align}

For this realization, the matter fields are assigned to the following $S_3$ irreducible representations:
\begin{align}
    \label{eq:chargunderS3}
    L_1,\text{ }L_2,\text{ }L_3\sim1,\text{ } \Phi_s\sim1, \text{ }\Phi_D\sim2,\text{ } N_D\sim2.
\end{align}
The scalar potential for this assignment is
\begin{align}
    \label{eq:Vs3}
    V=&\mu_s^2\left(\Phi_s^\dagger\Phi_s\right)+\mu_d^2\left(\left(\Phi_1^\dagger\Phi_1\right)+\left(\Phi_2^\dagger\Phi_2\right)\right)\nonumber\\
    +&\lambda_1\left(\left(\Phi_1^\dagger\Phi_1\right)+\left(\Phi_2^\dagger\Phi_2\right)\right)^2+\lambda_2\left(\left(\Phi_1^\dagger\Phi_2\right)-\left(\Phi_2^\dagger\Phi_1\right)\right)^2 \nonumber\\ 
    +&\lambda_3\left[\left(\left(\Phi_1^\dagger\Phi_2\right)+\left(\Phi_2^\dagger\Phi_1\right)\right)^2+\left(\left(\Phi_1^\dagger\Phi_1\right)-\left(\Phi_2^\dagger\Phi_2\right)\right)^2\right]+\lambda_4\left(\Phi_s^\dagger\Phi_s\right)^2\nonumber\\
    +&\lambda_5\left(\left(\Phi_1^\dagger\Phi_1\right)+\left(\Phi_2^\dagger\Phi_2\right)\right)\left(\Phi_s^\dagger\Phi_s\right)\\
    +&\left[\lambda_6\left[\left(\Phi_s^\dagger\Phi_2\right)\left(\Phi_s^\dagger\Phi_2\right)+\left(\Phi_s^\dagger\Phi_1\right)\left(\Phi_s^\dagger\Phi_1\right)\right]+h.c\right]\nonumber\\
    +&\lambda_7\left[\left(\Phi_s^\dagger\Phi_1\right)\left(\Phi_1^\dagger\Phi_s\right)+\left(\Phi_s^\dagger\Phi_2\right)\left(\Phi_2^\dagger\Phi_s\right)\right]\nonumber\\
    +&\left[\lambda_8\left\lbrace\left(\Phi_s^\dagger\Phi_1\right)\left(\left(\Phi_1^\dagger\Phi_2\right)+\left(\Phi_2^\dagger\Phi_1\right)\right)+\left(\Phi_s^\dagger\Phi_2\right)\left(\left(\Phi_1^\dagger\Phi_1\right)-\left(\Phi_2^\dagger\Phi_2\right)\right)\right\rbrace+h.c\right].\nonumber
\end{align}

Note that the potential exhibits a $\mathbb{Z}_2$ invariance under the transformation $\phi_1\to-\phi_1$ \cite{Khater_2022,Kun_inas_2022}. This symmetry naturally restricts the  allowed terms in the Yukawa Lagrangian. For the neutrino sector, the general Lagrangian is given by

\begin{align}
    \label{eq:YukawaNuS3}
    \mathcal{L}^{Y-\nu}=y_1^\nu L_1\left(\Tilde{\Phi}_DN_D\right)+y_2^\nu L_2\left(\Tilde{\Phi}_DN_D\right)+y_3^\nu L_3\left(\Tilde{\Phi}_DN_D\right)+M \overline{N_D}^C N_D,
\end{align}
where the contraction of the two doublets takes the explicit form
\begin{align}
    \label{eq:Yukawa2x2S3}
     \left(\Tilde{\Phi}_DN_D\right)=\left(\Tilde{\Phi}_1N_1+\Tilde{\Phi}_2N_2\right)_1+\left(\Tilde{\Phi}_1N_2-\Tilde{\Phi}_2N_1\right)_{1^\prime}.
\end{align}

Since the leptons $L_i$ are all assigned under $S_3$ as $1$, the contraction $\left(\Tilde{\Phi}_DN_D\right)$ must also transform as $1$ to ensure an invariant Lagrangian. Given that the $1$ contraction is $\Tilde{\Phi}_1N_1+\Tilde{\Phi}_2N_2$, the Yukawa structure for the active and dark sectors is identical. This implies $m_\nu^{active}\propto m_{\nu}^{dark}$. Given that both matrices are of rank 1 and are proportional to one another, the total mass matrix $m_\nu=m_\nu^{active}+m_\nu^{dark}$ is also of rank 1. This implies the existence of only a single massive neutrino, a result that is phenomenologically excluded.

On the other hand, assigning $L_i\sim1^\prime$ leads to a $1^\prime$ contraction of $\left(\Tilde{\Phi}_DN_D\right)$ that is not $\mathbb{Z}_2$-invariant and is therefore forbidden.

\subsection{$Q_6$ implementation}
The $Q_6$ group has 12 elements of the form $a^mb^k$, for $m=0,1\dots,5$ and $k=0,1$, where the generators $a$ and $b$ satisfy \cite{Ishimori:2010au,VIEN2020115015}:
\begin{align}
    \label{eq:generatorQ6}
    a^6=&e, & b^2=&a^3, & b^{-1}ab&=a^{-1}.
\end{align}

The group irreps consist of four singlets, $1,1^\prime, 1^{\prime\prime},1^{\prime\prime\prime}$ and two doublets, $2,2^\prime$. The matrix representations of the generators for each irrep are \cite{Araki_2012}:
\begin{align}
    \label{eq:repgeneratorsQ6}
    1&: & a=&1 & b=&1,\\
    1&^\prime: &a=&1 & b=&-1,\\
    1&^{\prime\prime}:  &a=&-1 & b=&-i,\\
    1&^{\prime\prime\prime}: & a=&-1 & b=&i,\\
    2&: & a=&\begin{pmatrix}
                \omega_6 & 0\\
                0 & \omega_6^{-1}
             \end{pmatrix} & b=&\begin{pmatrix}
                                    0 & i\\
                                    i & 0
                                \end{pmatrix},\\
    2^\prime&:  &a=&\begin{pmatrix}
                \omega_6^2 & 0\\
                0 & \omega_6^{-2}
             \end{pmatrix} & b=&\begin{pmatrix}
                                    0 & 1\\
                                    1 & 0
                                \end{pmatrix},
\end{align}
where $\omega_6=\textbf{exp}\left(2i\pi/6\right)$. Note that, $1,\text{ }1^\prime$ and $2^\prime$ are the real representations. The tensor product between the doublets are defined as follows:
\begin{align}
    \label{eq:tensorproducQ6}
    2(x_1,x_2)\otimes2(y_1,y_2)=[x_1y_2-x_2y_1]_1+[x_1y_2+x_2y_1]_{1^\prime}+\begin{pmatrix}
        x_1y_1\\
        -x_2y_2
    \end{pmatrix}_{2^\prime},\\
    2(x_1,x_2)\otimes2^\prime(y_1,y_2)=[x_1y_1-x_2y_2]_{1^{\prime\prime}}+[x_1y_1+x_2y_2]_{1^{\prime\prime\prime}}+\begin{pmatrix}
        x_2y_1\\
        x_1y_2
    \end{pmatrix}_{2},\\
    2^\prime(x_1,x_2)\otimes2^\prime(y_1,y_2)=[x_1y_2+x_2y_1]_1+[x_1y_2-x_2y_1]_{1^\prime}+\begin{pmatrix}
        x_2y_2\\
        x_1y_1
    \end{pmatrix}_{2^\prime}.
\end{align}

We consider a 3HDM framework in which the scalar doublets transform as $\Phi_s\sim 1$ and $\Phi_D\sim2^\prime$, and the right-handed neutrinos as $N_D\sim 2^\prime$.

Since the generator $b$ for the $2^\prime$ representation has the same structure as the $\mathbb{Z}_2$ generator given in Eq.\eqref{eq:Bgenerator}, we can apply the same transformation $U$, defined in Eq. \eqref{eq:Utransformation}, to the $\Phi_D$ and $N_D$ doublets. In this basis, the product $\varphi_D\chi_D$ decomposes into two singlet representations, $1$ and $1^\prime$, according to
\begin{align}
    \label{eq:YukawaContracQ6}
    \left(\Tilde{\varphi}_D\chi_D\right)=\left(\Tilde{\varphi}_1\chi_1-\Tilde{\varphi}_2\chi_2\right)_{1}+\left(\Tilde{\varphi}_1\chi_2-\Tilde{\varphi}_2\chi_1\right)_{1^\prime}+2^\prime.
\end{align}

Assuming that the left-handed lepton doublets transform as $L_i \sim 1$, only these singlet contractions can enter the Yukawa sector. However, the $1^\prime$ contraction is odd under the $\mathbb{Z}_2$ symmetry due to $\varphi_2\to-\varphi_2$ and $\chi_2\to-\chi_2$, and is therefore forbidden. Consequently, the most general Yukawa Lagrangian that remains invariant under the $U$ transformation reads:
\begin{align}
    \label{eq:YukawaQ6}
    \mathcal{L}^{Y-\nu}=&y_1^{\nu}L_1\left(\Tilde{\varphi}_D\chi_D\right)_1+y_2^{\nu}L_2\left(\Tilde{\varphi}_D\chi_D\right)_1+y_3^{\nu}L_3\left(\Tilde{\varphi}_D\chi_D\right)_1+M \left(\overline{\chi_D}^C\chi_D\right)_1.
\end{align}
This Yukawa structure leads to a dark mass matrix $m_\nu^{dark}$ with the same structure as the active one $m_\nu^{active}$, implying
\begin{align}
    \label{eq:activeproptodark}
    m_\nu^{active}\propto m_\nu^{dark}.
\end{align}

Since both matrices are individually rank-1 and proportional, the total neutrino mass matrix
\begin{align}
    \label{eq:totmnuQ6}
    m_\nu=m_\nu^{active}+m_\nu^{dark},
\end{align}
is also rank-1. Consequently, this setup predicts only one massive neutrino, and is therefore not phenomenologically viable.

\subsection{$D_6$ implementation}
We also analyzed the $D_6$ group, which has two two-dimensional irreducible representations, $2$ and $2^\prime$, and four one-dimensional ones: $1,1^\prime,1^{\prime\prime},1^{\prime\prime\prime}$ \cite{Ishimori:2010au,Kajiyama_2007,Blum_2008}. The generators of this group satisfy the relations $a^6=\mathbb{I},\text{ }b^2=\mathbb{I},\text{ },aba=b$. For this case, we assign the scalar and fermionic doublets as $\Phi_D\sim2$ and $N_D\sim2$, respectively, while the charged-lepton doublets transform as singlets, $L_i\sim1$.

Although the generator representations differ from those of $Q_6$, the tensor products relevant for our model exhibit an analogous structure:
\begin{align}
    \label{eq:tensorproductD6}
    2\times2=&\begin{pmatrix}
        x_1\\
        x_2
    \end{pmatrix}\otimes\begin{pmatrix}
        y_1\\
        y_2
    \end{pmatrix}=\left(x_1y_2-x_2y_1\right)_{1^\prime}+\left(x_1y_1+x_2y_2\right)_{1}+2^\prime,\\
    2^\prime\times2^\prime=&\begin{pmatrix}
        a_1\\
        a_2
    \end{pmatrix}\otimes\begin{pmatrix}
        b_1\\
        b_2
    \end{pmatrix}=\left(a_1b_2-a_2b_1\right)_{1^\prime}+\left(a_1b_1+a_2b_2\right)_{1}+2^\prime,\\
    2\times2^\prime=&\begin{pmatrix}
        x_1\\
        x_2
    \end{pmatrix}\otimes\begin{pmatrix}
        a_1\\
        a_2
    \end{pmatrix}=\left(x_1a_2-x_2a_1\right)_{1^{\prime\prime\prime}}+\left(x_1a_1-x_2a_2\right)_{1^{\prime\prime}}+2.    
\end{align}

Since the doublet tensor products yielding singlet contractions have the same structure as in the $Q_6$ case, the Yukawa Lagrangian analysis leads to the same conclusion: the active and dark mass matrices are proportional, $m_\nu^{active}\propto m_\nu^{dark}$. Consequently, the total neutrino mass matrix $m_\nu$ is rank-1, which implies that this scenario is phenomenologically non-viable.
\bibliographystyle{JHEP}
\bibliography{references} 

@article{tanimoto1999neutrinomassesmixingsflavor,
      title={Neutrino Masses and Mixings with Flavor Symmetries}, 
      author={Morimitsu Tanimoto},
      year={1999},
      eprint={hep-ph/9910261},
      archivePrefix={arXiv},
      primaryClass={hep-ph},
      url={https://arxiv.org/abs/hep-ph/9910261}, 
}

@article{Morisi_2012,
   title={Neutrino masses and mixing: a flavour symmetry roadmap},
   volume={61},
   ISSN={1521-3978},
   url={http://dx.doi.org/10.1002/prop.201200125},
   DOI={10.1002/prop.201200125},
   number={4–5},
   journal={Fortschritte der Physik},
   publisher={Wiley},
   author={Morisi, S. and Valle, J.W.F.},
   year={2012},
   month=oct, pages={466–492} }

@article{Redigolo:2024ztw,
    author = "Redigolo, Diego and Tammaro, Michele and Tesi, Andrea",
    title = "{Large CP violation in flavor violating muon decays}",
    eprint = "2408.00847",
    archivePrefix = "arXiv",
    primaryClass = "hep-ph",
    doi = "10.1140/epjc/s10052-025-13742-9",
    journal = "Eur. Phys. J. C",
    volume = "85",
    number = "1",
    pages = "103",
    year = "2025"
}

@article{Sun:2025jmx,
    author = "Sun, Rong-Zhi and Zhao, Shu-Min and Liu, Ming-Yue and Han, Xing-Yu and Gao, Song and Dong, Xing-Xing",
    title = "{Lepton flavor violating decays $l_j\rightarrow l_i\gamma ,l_j \rightarrow 3l_i$ and $\mu \rightarrow e+ q\bar{q}$ in the N-B-LSSM}",
    eprint = "2502.18130",
    archivePrefix = "arXiv",
    primaryClass = "hep-ph",
    doi = "10.1140/epjc/s10052-025-14762-1",
    journal = "Eur. Phys. J. C",
    volume = "85",
    number = "9",
    pages = "1038",
    year = "2025"
}

@article{_vila_2022,
   title={Revisiting the scotogenic model with scalar dark matter},
   volume={49},
   ISSN={1361-6471},
   url={http://dx.doi.org/10.1088/1361-6471/ac5fb4},
   DOI={10.1088/1361-6471/ac5fb4},
   number={6},
   journal={Journal of Physics G: Nuclear and Particle Physics},
   publisher={IOP Publishing},
   author={Ávila, Ivania M and Cottin, Giovanna and Díaz, Marco A},
   year={2022},
   month=apr, pages={065001} }

@article{RevModPhys.59.671,
  title = {Massive neutrinos and neutrino oscillations},
  author = {Bilenky, S. M. and Petcov, S. T.},
  journal = {Rev. Mod. Phys.},
  volume = {59},
  issue = {3},
  pages = {671--754},
  numpages = {0},
  year = {1987},
  month = {Jul},
  publisher = {American Physical Society},
  doi = {10.1103/RevModPhys.59.671},
  url = {https://link.aps.org/doi/10.1103/RevModPhys.59.671}
}

@article{BILENKY1978225,
title = {Lepton mixing and neutrino oscillations},
journal = {Physics Reports},
volume = {41},
number = {4},
pages = {225-261},
year = {1978},
issn = {0370-1573},
doi = {https://doi.org/10.1016/0370-1573(78)90095-9},
url = {https://www.sciencedirect.com/science/article/pii/0370157378900959},
author = {S.M. Bilenky and B. Pontecorvo}
}

@article{2020,
   title={Planck 2018 results: VI. Cosmological parameters},
   volume={641},
   ISSN={1432-0746},
   url={http://dx.doi.org/10.1051/0004-6361/201833910},
   DOI={10.1051/0004-6361/201833910},
   journal={Astronomy \& Astrophysics},
   publisher={EDP Sciences},
   author={Aghanim, N. and Akrami, Y. and Ashdown, M. and Aumont, J. and Baccigalupi, C. and Ballardini, M. and Banday, A. J. and Barreiro, R. B. and Bartolo, N. and Basak, S. and Battye, R. and Benabed, K. and Bernard, J.-P. and Bersanelli, M. and Bielewicz, P. and Bock, J. J. and Bond, J. R. and Borrill, J. and Bouchet, F. R. and Boulanger, F. and Bucher, M. and Burigana, C. and Butler, R. C. and Calabrese, E. and Cardoso, J.-F. and Carron, J. and Challinor, A. and Chiang, H. C. and Chluba, J. and Colombo, L. P. L. and Combet, C. and Contreras, D. and Crill, B. P. and Cuttaia, F. and de Bernardis, P. and de Zotti, G. and Delabrouille, J. and Delouis, J.-M. and Di Valentino, E. and Diego, J. M. and Doré, O. and Douspis, M. and Ducout, A. and Dupac, X. and Dusini, S. and Efstathiou, G. and Elsner, F. and Enßlin, T. A. and Eriksen, H. K. and Fantaye, Y. and Farhang, M. and Fergusson, J. and Fernandez-Cobos, R. and Finelli, F. and Forastieri, F. and Frailis, M. and Fraisse, A. A. and Franceschi, E. and Frolov, A. and Galeotta, S. and Galli, S. and Ganga, K. and Génova-Santos, R. T. and Gerbino, M. and Ghosh, T. and González-Nuevo, J. and Górski, K. M. and Gratton, S. and Gruppuso, A. and Gudmundsson, J. E. and Hamann, J. and Handley, W. and Hansen, F. K. and Herranz, D. and Hildebrandt, S. R. and Hivon, E. and Huang, Z. and Jaffe, A. H. and Jones, W. C. and Karakci, A. and Keihänen, E. and Keskitalo, R. and Kiiveri, K. and Kim, J. and Kisner, T. S. and Knox, L. and Krachmalnicoff, N. and Kunz, M. and Kurki-Suonio, H. and Lagache, G. and Lamarre, J.-M. and Lasenby, A. and Lattanzi, M. and Lawrence, C. R. and Le Jeune, M. and Lemos, P. and Lesgourgues, J. and Levrier, F. and Lewis, A. and Liguori, M. and Lilje, P. B. and Lilley, M. and Lindholm, V. and López-Caniego, M. and Lubin, P. M. and Ma, Y.-Z. and Macías-Pérez, J. F. and Maggio, G. and Maino, D. and Mandolesi, N. and Mangilli, A. and Marcos-Caballero, A. and Maris, M. and Martin, P. G. and Martinelli, M. and Martínez-González, E. and Matarrese, S. and Mauri, N. and McEwen, J. D. and Meinhold, P. R. and Melchiorri, A. and Mennella, A. and Migliaccio, M. and Millea, M. and Mitra, S. and Miville-Deschênes, M.-A. and Molinari, D. and Montier, L. and Morgante, G. and Moss, A. and Natoli, P. and Nørgaard-Nielsen, H. U. and Pagano, L. and Paoletti, D. and Partridge, B. and Patanchon, G. and Peiris, H. V. and Perrotta, F. and Pettorino, V. and Piacentini, F. and Polastri, L. and Polenta, G. and Puget, J.-L. and Rachen, J. P. and Reinecke, M. and Remazeilles, M. and Renzi, A. and Rocha, G. and Rosset, C. and Roudier, G. and Rubiño-Martín, J. A. and Ruiz-Granados, B. and Salvati, L. and Sandri, M. and Savelainen, M. and Scott, D. and Shellard, E. P. S. and Sirignano, C. and Sirri, G. and Spencer, L. D. and Sunyaev, R. and Suur-Uski, A.-S. and Tauber, J. A. and Tavagnacco, D. and Tenti, M. and Toffolatti, L. and Tomasi, M. and Trombetti, T. and Valenziano, L. and Valiviita, J. and Van Tent, B. and Vibert, L. and Vielva, P. and Villa, F. and Vittorio, N. and Wandelt, B. D. and Wehus, I. K. and White, M. and White, S. D. M. and Zacchei, A. and Zonca, A.},
   year={2020},
   month=sep, pages={A6} }

@article{Bonilla_2023,
   title={Discrete dark matter mechanism as the source of neutrino mass scales},
   volume={2023},
   ISSN={1029-8479},
   url={http://dx.doi.org/10.1007/JHEP06(2023)078},
   DOI={10.1007/jhep06(2023)078},
   number={6},
   journal={Journal of High Energy Physics},
   publisher={Springer Science and Business Media LLC},
   author={Bonilla, Cesar and Herms, Johannes and Medina, Omar and Peinado, Eduardo},
   year={2023},
   month=jun }

@article{Grimus_2004,
   title={Lepton mixing angle {$ \theta_{13}= 0$} with a horizontal symmetry {$D_4$} },
   volume={2004},
   ISSN={1029-8479},
   url={http://dx.doi.org/10.1088/1126-6708/2004/07/078},
   DOI={10.1088/1126-6708/2004/07/078},
   number={07},
   journal={Journal of High Energy Physics},
   publisher={Springer Science and Business Media LLC},
   author={Grimus, W and Joshipura, A.S and Kaneko, S and Lavoura, L and Tanimoto, M},
   year={2004},
   month=jul, pages={078–078} }

@article{Grimus_2003,
   title={A discrete symmetry group for maximal atmospheric neutrino mixing},
   volume={572},
   ISSN={0370-2693},
   url={http://dx.doi.org/10.1016/j.physletb.2003.08.032},
   DOI={10.1016/j.physletb.2003.08.032},
   number={3–4},
   journal={Physics Letters B},
   publisher={Elsevier BV},
   author={Grimus, Walter and Lavoura, Luı́s},
   year={2003},
   month=oct, pages={189–195} }

@article{Adulpravitchai_2009,
   title={A supersymmetric {$D_4$} model for $\mu$--$\tau$ symmetry},
   volume={2009},
   ISSN={1029-8479},
   url={http://dx.doi.org/10.1088/1126-6708/2009/03/046},
   DOI={10.1088/1126-6708/2009/03/046},
   number={03},
   journal={Journal of High Energy Physics},
   publisher={Springer Science and Business Media LLC},
   author={Adulpravitchai, A and Blum, A and Hagedorn, C},
   year={2009},
   month=mar,
   pages={046--046}
}

@article{Aranda_2019,
   title={Dynamical generation of neutrino mass scales},
   volume={792},
   ISSN={0370-2693},
   url={http://dx.doi.org/10.1016/j.physletb.2019.01.068},
   DOI={10.1016/j.physletb.2019.01.068},
   journal={Physics Letters B},
   publisher={Elsevier BV},
   author={Aranda, Alfredo and Bonilla, Cesar and Peinado, Eduardo},
   year={2019},
   month=may, pages={40–42} }

@article{PhysRevD.82.116003,
  title = {Discrete dark matter},
  author = {Hirsch, M. and Morisi, S. and Peinado, E. and Valle, J. W. F.},
  journal = {Phys. Rev. D},
  volume = {82},
  issue = {11},
  pages = {116003},
  numpages = {5},
  year = {2010},
  month = {Dec},
  publisher = {American Physical Society},
  doi = {10.1103/PhysRevD.82.116003},
  url = {https://link.aps.org/doi/10.1103/PhysRevD.82.116003}
}

@article{Mandal_2021,
   title={The simplest scoto-seesaw model: WIMP dark matter phenomenology and Higgs vacuum stability},
   volume={819},
   ISSN={0370-2693},
   url={http://dx.doi.org/10.1016/j.physletb.2021.136458},
   DOI={10.1016/j.physletb.2021.136458},
   journal={Physics Letters B},
   publisher={Elsevier BV},
   author={Mandal, Sanjoy and Srivastava, Rahul and Valle, José W.F.},
   year={2021},
   month=aug, pages={136458} }

@article{Barreiros_2022,
   title={Flavour and dark matter in a scoto/type-II seesaw model},
   volume={2022},
   ISSN={1029-8479},
   url={http://dx.doi.org/10.1007/JHEP08(2022)030},
   DOI={10.1007/jhep08(2022)030},
   number={8},
   journal={Journal of High Energy Physics},
   publisher={Springer Science and Business Media LLC},
   author={Barreiros, D. M. and Câmara, H. B. and Joaquim, F. R.},
   year={2022},
   month=aug }

@article{Barreiros_2021,
   title={Minimal scoto-seesaw mechanism with spontaneous CP violation},
   volume={2021},
   ISSN={1029-8479},
   url={http://dx.doi.org/10.1007/JHEP04(2021)249},
   DOI={10.1007/jhep04(2021)249},
   number={4},
   journal={Journal of High Energy Physics},
   publisher={Springer Science and Business Media LLC},
   author={Barreiros, D. M. and Joaquim, F. R. and Srivastava, R. and Valle, J. W. F.},
   year={2021},
   month=apr }

@article{Rojas_2019,
   title={Simplest scoto-seesaw mechanism},
   volume={789},
   ISSN={0370-2693},
   url={http://dx.doi.org/10.1016/j.physletb.2018.12.014},
   DOI={10.1016/j.physletb.2018.12.014},
   journal={Physics Letters B},
   publisher={Elsevier BV},
   author={Rojas, Nicolás and Srivastava, Rahul and Valle, José W.F.},
   year={2019},
   month=feb, pages={132–136} }

@article{Esteban_2020,
   title={The fate of hints: updated global analysis of three-flavor neutrino oscillations},
   volume={2020},
   ISSN={1029-8479},
   url={http://dx.doi.org/10.1007/JHEP09(2020)178},
   DOI={10.1007/jhep09(2020)178},
   number={9},
   journal={Journal of High Energy Physics},
   publisher={Springer Science and Business Media LLC},
   author={Esteban, Ivan and Gonzalez-Garcia, M.C. and Maltoni, Michele and Schwetz, Thomas and Zhou, Albert},
   year={2020},
   month=sep }

@article{de_Salas_2021,
   title={2020 global reassessment of the neutrino oscillation picture},
   volume={2021},
   ISSN={1029-8479},
   url={http://dx.doi.org/10.1007/JHEP02(2021)071},
   DOI={10.1007/jhep02(2021)071},
   number={2},
   journal={Journal of High Energy Physics},
   publisher={Springer Science and Business Media LLC},
   author={de Salas, P. F. and Forero, D. V. and Gariazzo, S. and Martínez-Miravé, P. and Mena, O. and Ternes, C. A. and Tórtola, M. and Valle, J. W. F.},
   year={2021},
   month=feb }

@article{Aristizabal_Sierra_2011,
   title={On the importance of the 1-loop finite corrections to seesaw neutrino masses},
   volume={2011},
   ISSN={1029-8479},
   url={http://dx.doi.org/10.1007/JHEP08(2011)013},
   DOI={10.1007/jhep08(2011)013},
   number={8},
   journal={Journal of High Energy Physics},
   publisher={Springer Science and Business Media LLC},
   author={Aristizabal Sierra, D. and Yaguna, Carlos E.},
   year={2011},
   month=aug }

@article{Ishimori:2010au,
    author = "Ishimori, Hajime and Kobayashi, Tatsuo and Ohki, Hiroshi and Shimizu, Yusuke and Okada, Hiroshi and Tanimoto, Morimitsu",
    title = "{Non-Abelian Discrete Symmetries in Particle Physics}",
    eprint = "1003.3552",
    archivePrefix = "arXiv",
    primaryClass = "hep-th",
    reportNumber = "KUNS-2260",
    doi = "10.1143/PTPS.183.1",
    journal = "Prog. Theor. Phys. Suppl.",
    volume = "183",
    pages = "1--163",
    year = "2010"
}

@article{Blum_2008,
   title={Fermion masses and mixings from dihedral flavor symmetries with preserved subgroups},
   volume={77},
   ISSN={1550-2368},
   url={http://dx.doi.org/10.1103/PhysRevD.77.076004},
   DOI={10.1103/physrevd.77.076004},
   number={7},
   journal={Physical Review D},
   publisher={American Physical Society (APS)},
   author={Blum, A. and Hagedorn, C. and Lindner, M.},
   year={2008},
   month=apr }

@article{MELONI2011281,
title = {Stability of dark matter from the {$D_4\times Z_2^f$} flavor group},
journal = {Physics Letters B},
volume = {703},
number = {3},
pages = {281-287},
year = {2011},
issn = {0370-2693},
doi = {https://doi.org/10.1016/j.physletb.2011.07.084},
url = {https://www.sciencedirect.com/science/article/pii/S0370269311009038},
author = {D. Meloni and S. Morisi and E. Peinado}
}

@article{Ma_2006,
   title={Verifiable radiative seesaw mechanism of neutrino mass and dark matter},
   volume={73},
   ISSN={1550-2368},
   url={http://dx.doi.org/10.1103/PhysRevD.73.077301},
   DOI={10.1103/physrevd.73.077301},
   number={7},
   journal={Physical Review D},
   publisher={American Physical Society (APS)},
   author={Ma, Ernest},
   year={2006},
   month=apr }

@article{Escribano:2020iqq,
    author = "Escribano, Pablo and Reig, Mario and Vicente, Avelino",
    title = "{Generalizing the Scotogenic model}",
    eprint = "2004.05172",
    archivePrefix = "arXiv",
    primaryClass = "hep-ph",
    reportNumber = "IFIC/20-13",
    doi = "10.1007/JHEP07(2020)097",
    journal = "JHEP",
    volume = "07",
    pages = "097",
    year = "2020"
}

@article{deSalas:2018bym,
      author        = "De Salas, P.F. and Gariazzo, S. and Mena, O. and Ternes, C.A. and T{\'o}rtola, M.",
      title         = "{Neutrino Mass Ordering from Oscillations and Beyond: 2018 Status and Future Prospects}",
      eprint        = "1806.11051",
      archivePrefix = "arXiv",
      primaryClass  = "hep-ph",
      doi           = "10.3389/fspas.2018.00036",
      journal       = "Front. Astron. Space Sci.",
      volume        = "5",
      pages         = "36",
      year          = "2018"
}

@article{Gariazzo:2018pei,
      author         = "Gariazzo, S. and Archidiacono, M. and de Salas, P. F. and Mena, O. and Ternes, C. A. and T{\'o}rtola, M.",
      title          = "{Neutrino masses and their ordering: Global Data, Priors and Models}",
      journal        = "JCAP",
      volume         = "1803",
      year           = "2018",
      number         = "03",
      pages          = "011",
      doi            = "10.1088/1475-7516/2018/03/011",
      eprint         = "1801.04946",
      archivePrefix  = "arXiv",
      primaryClass   = "hep-ph",
      SLACcitation   = "%%CITATION = ARXIV:1801.04946;%%"
}

@article{Belyaev_2018,
   title={Anatomy of the inert two-Higgs-doublet model in the light of the LHC and non-LHC dark matter searches},
   volume={97},
   ISSN={2470-0029},
   url={http://dx.doi.org/10.1103/PhysRevD.97.035011},
   DOI={10.1103/physrevd.97.035011},
   number={3},
   journal={Physical Review D},
   publisher={American Physical Society (APS)},
   author={Belyaev, Alexander and Cacciapaglia, Giacomo and Ivanov, Igor P. and Rojas-Abatte, Felipe and Thomas, Marc},
   year={2018},
   month=feb }

@article{Lundstr_m_2009,
   title={Inert doublet model and LEP II limits},
   volume={79},
   ISSN={1550-2368},
   url={http://dx.doi.org/10.1103/PhysRevD.79.035013},
   DOI={10.1103/physrevd.79.035013},
   number={3},
   journal={Physical Review D},
   publisher={American Physical Society (APS)},
   author={Lundström, Erik and Gustafsson, Michael and Edsjö, Joakim},
   year={2009},
   month=feb }

@article{Castro_Alvaredo_2004,
   title={Integrable scattering theories with unstable particles},
   volume={35},
   ISSN={1434-6052},
   url={http://dx.doi.org/10.1140/epjc/s2004-01780-x},
   DOI={10.1140/epjc/s2004-01780-x},
   number={3},
   journal={The European Physical Journal C},
   publisher={Springer Science and Business Media LLC},
   author={Castro-Alvaredo, O. A. and Dreißig, J. and Fring, A.},
   year={2004},
   month=jun, pages={393–411} }

@article{Pierce_2007,
   title={Natural Dark Matter from an unnatural Higgs boson and new colored particles at the TeV scale},
   volume={2007},
   ISSN={1029-8479},
   url={http://dx.doi.org/10.1088/1126-6708/2007/08/026},
   DOI={10.1088/1126-6708/2007/08/026},
   number={08},
   journal={Journal of High Energy Physics},
   publisher={Springer Science and Business Media LLC},
   author={Pierce, Aaron and Thaler, Jesse},
   year={2007},
   month=aug, pages={026–026} }

@article{Cui_2017,
   title={Dark Matter Results from 54-Ton-Day Exposure of PandaX-II Experiment},
   volume={119},
   ISSN={1079-7114},
   url={http://dx.doi.org/10.1103/PhysRevLett.119.181302},
   DOI={10.1103/physrevlett.119.181302},
   number={18},
   journal={Physical Review Letters},
   publisher={American Physical Society (APS)},
   author={Cui, Xiangyi and Abdukerim, Abdusalam and Chen, Wei and Chen, Xun and Chen, Yunhua and Dong, Binbin and Fang, Deqing and Fu, Changbo and Giboni, Karl and Giuliani, Franco and Gu, Linhui and Gu, Yikun and Guo, Xuyuan and Guo, Zhifan and Han, Ke and He, Changda and Huang, Di and He, Shengming and Huang, Xingtao and Huang, Zhou and Ji, Xiangdong and Ju, Yonglin and Li, Shaoli and Li, Yao and Lin, Heng and Liu, Huaxuan and Liu, Jianglai and Ma, Yugang and Mao, Yajun and Ni, Kaixiang and Ning, Jinhua and Ren, Xiangxiang and Shi, Fang and Tan, Andi and Wang, Cheng and Wang, Hongwei and Wang, Meng and Wang, Qiuhong and Wang, Siguang and Wang, Xiuli and Wang, Xuming and Wu, Qinyu and Wu, Shiyong and Xiao, Mengjiao and Xie, Pengwei and Yan, Binbin and Yang, Yong and Yue, Jianfeng and Zhang, Dan and Zhang, Hongguang and Zhang, Tao and Zhang, Tianqi and Zhao, Li and Zhou, Jifang and Zhou, Ning and Zhou, Xiaopeng},
   year={2017},
   month=oct }

@article{Aprile_2023,
   title={First Dark Matter Search with Nuclear Recoils from the XENONnT Experiment},
   volume={131},
   ISSN={1079-7114},
   url={http://dx.doi.org/10.1103/PhysRevLett.131.041003},
   DOI={10.1103/physrevlett.131.041003},
   number={4},
   journal={Physical Review Letters},
   publisher={American Physical Society (APS)},
   author={Aprile, E. and Abe, K. and Agostini, F. and Ahmed Maouloud, S. and Althueser, L. and Andrieu, B. and Angelino, E. and Angevaare, J. R. and Antochi, V. C. and Antón Martin, D. and Arneodo, F. and Baudis, L. and Baxter, A. L. and Bazyk, M. and Bellagamba, L. and Biondi, R. and Bismark, A. and Brookes, E. J. and Brown, A. and Bruenner, S. and Bruno, G. and Budnik, R. and Bui, T. K. and Cai, C. and Cardoso, J. M. R. and Cichon, D. and Cimental Chavez, A. P. and Colijn, A. P. and Conrad, J. and Cuenca-García, J. J. and Cussonneau, J. P. and D’Andrea, V. and Decowski, M. P. and Di Gangi, P. and Di Pede, S. and Diglio, S. and Eitel, K. and Elykov, A. and Farrell, S. and Ferella, A. D. and Ferrari, C. and Fischer, H. and Flierman, M. and Fulgione, W. and Fuselli, C. and Gaemers, P. and Gaior, R. and Gallo Rosso, A. and Galloway, M. and Gao, F. and Glade-Beucke, R. and Grandi, L. and Grigat, J. and Guan, H. and Guida, M. and Hammann, R. and Higuera, A. and Hils, C. and Hoetzsch, L. and Hood, N. F. and Howlett, J. and Iacovacci, M. and Itow, Y. and Jakob, J. and Joerg, F. and Joy, A. and Kato, N. and Kara, M. and Kavrigin, P. and Kazama, S. and Kobayashi, M. and Koltman, G. and Kopec, A. and Kuger, F. and Landsman, H. and Lang, R. F. and Levinson, L. and Li, I. and Li, S. and Liang, S. and Lindemann, S. and Lindner, M. and Liu, K. and Loizeau, J. and Lombardi, F. and Long, J. and Lopes, J. A. M. and Ma, Y. and Macolino, C. and Mahlstedt, J. and Mancuso, A. and Manenti, L. and Marignetti, F. and Marrodán Undagoitia, T. and Martens, K. and Masbou, J. and Masson, D. and Masson, E. and Mastroianni, S. and Messina, M. and Miuchi, K. and Mizukoshi, K. and Molinario, A. and Moriyama, S. and Morå, K. and Mosbacher, Y. and Murra, M. and Müller, J. and Ni, K. and Oberlack, U. and Paetsch, B. and Palacio, J. and Peres, R. and Peters, C. and Pienaar, J. and Pierre, M. and Pizzella, V. and Plante, G. and Qi, J. and Qin, J. and Ramírez García, D. and Singh, R. and Sanchez, L. and dos Santos, J. M. F. and Sarnoff, I. and Sartorelli, G. and Schreiner, J. and Schulte, D. and Schulte, P. and Schulze Eißing, H. and Schumann, M. and Scotto Lavina, L. and Selvi, M. and Semeria, F. and Shagin, P. and Shi, S. and Shockley, E. and Silva, M. and Simgen, H. and Takeda, A. and Tan, P.-L. and Terliuk, A. and Thers, D. and Toschi, F. and Trinchero, G. and Tunnell, C. and Tönnies, F. and Valerius, K. and Volta, G. and Weinheimer, C. and Weiss, M. and Wenz, D. and Wittweg, C. and Wolf, T. and Wu, V. H. S. and Xing, Y. and Xu, D. and Xu, Z. and Yamashita, M. and Yang, L. and Ye, J. and Yuan, L. and Zavattini, G. and Zhong, M. and Zhu, T.},
   year={2023},
   month=jul }

@article{Aalbers_2023,
   title={First Dark Matter Search Results from the {LUX-ZEPLIN (LZ)} Experiment},
   volume={131},
   ISSN={1079-7114},
   url={http://dx.doi.org/10.1103/PhysRevLett.131.041002},
   DOI={10.1103/physrevlett.131.041002},
   number={4},
   journal={Physical Review Letters},
   publisher={American Physical Society (APS)},
   author={Aalbers, J. and Akerib, D. S. and Akerlof, C. W. and Al Musalhi, A. K. and Alder, F. and Alqahtani, A. and Alsum, S. K. and Amarasinghe, C. S. and Ames, A. and Anderson, T. J. and Angelides, N. and Araújo, H. M. and Armstrong, J. E. and Arthurs, M. and Azadi, S. and Bailey, A. J. and Baker, A. and Balajthy, J. and Balashov, S. and Bang, J. and Bargemann, J. W. and Barry, M. J. and Barthel, J. and Bauer, D. and Baxter, A. and Beattie, K. and Belle, J. and Beltrame, P. and Bensinger, J. and Benson, T. and Bernard, E. P. and Bhatti, A. and Biekert, A. and Biesiadzinski, T. P. and Birch, H. J. and Birrittella, B. and Blockinger, G. M. and Boast, K. E. and Boxer, B. and Bramante, R. and Brew, C. A. J. and Brás, P. and Buckley, J. H. and Bugaev, V. V. and Burdin, S. and Busenitz, J. K. and Buuck, M. and Cabrita, R. and Carels, C. and Carlsmith, D. L. and Carlson, B. and Carmona-Benitez, M. C. and Cascella, M. and Chan, C. and Chawla, A. and Chen, H. and Cherwinka, J. J. and Chott, N. I. and Cole, A. and Coleman, J. and Converse, M. V. and Cottle, A. and Cox, G. and Craddock, W. W. and Creaner, O. and Curran, D. and Currie, A. and Cutter, J. E. and Dahl, C. E. and David, A. and Davis, J. and Davison, T. J. R. and Delgaudio, J. and Dey, S. and de Viveiros, L. and Dobi, A. and Dobson, J. E. Y. and Druszkiewicz, E. and Dushkin, A. and Edberg, T. K. and Edwards, W. R. and Elnimr, M. M. and Emmet, W. T. and Eriksen, S. R. and Faham, C. H. and Fan, A. and Fayer, S. and Fearon, N. M. and Fiorucci, S. and Flaecher, H. and Ford, P. and Francis, V. B. and Fraser, E. D. and Fruth, T. and Gaitskell, R. J. and Gantos, N. J. and Garcia, D. and Geffre, A. and Gehman, V. M. and Genovesi, J. and Ghag, C. and Gibbons, R. and Gibson, E. and Gilchriese, M. G. D. and Gokhale, S. and Gomber, B. and Green, J. and Greenall, A. and Greenwood, S. and van der Grinten, M. G. D. and Gwilliam, C. B. and Hall, C. R. and Hans, S. and Hanzel, K. and Harrison, A. and Hartigan-O’Connor, E. and Haselschwardt, S. J. and Hernandez, M. A. and Hertel, S. A. and Heuermann, G. and Hjemfelt, C. and Hoff, M. D. and Holtom, E. and Hor, J. Y-K. and Horn, M. and Huang, D. Q. and Hunt, D. and Ignarra, C. M. and Jacobsen, R. G. and Jahangir, O. and James, R. S. and Jeffery, S. N. and Ji, W. and Johnson, J. and Kaboth, A. C. and Kamaha, A. C. and Kamdin, K. and Kasey, V. and Kazkaz, K. and Keefner, J. and Khaitan, D. and Khaleeq, M. and Khazov, A. and Khurana, I. and Kim, Y. D. and Kocher, C. D. and Kodroff, D. and Korley, L. and Korolkova, E. V. and Kras, J. and Kraus, H. and Kravitz, S. and Krebs, H. J. and Kreczko, L. and Krikler, B. and Kudryavtsev, V. A. and Kyre, S. and Landerud, B. and Leason, E. A. and Lee, C. and Lee, J. and Leonard, D. S. and Leonard, R. and Lesko, K. T. and Levy, C. and Li, J. and Liao, F.-T. and Liao, J. and Lin, J. and Lindote, A. and Linehan, R. and Lippincott, W. H. and Liu, R. and Liu, X. and Liu, Y. and Loniewski, C. and Lopes, M. I. and Lopez Asamar, E. and López Paredes, B. and Lorenzon, W. and Lucero, D. and Luitz, S. and Lyle, J. M. and Majewski, P. A. and Makkinje, J. and Malling, D. C. and Manalaysay, A. and Manenti, L. and Mannino, R. L. and Marangou, N. and Marzioni, M. F. and Maupin, C. and McCarthy, M. E. and McConnell, C. T. and McKinsey, D. N. and McLaughlin, J. and Meng, Y. and Migneault, J. and Miller, E. H. and Mizrachi, E. and Mock, J. A. and Monte, A. and Monzani, M. E. and Morad, J. A. and Morales Mendoza, J. D. and Morrison, E. and Mount, B. J. and Murdy, M. and Murphy, A. St. J. and Naim, D. and Naylor, A. and Nedlik, C. and Nehrkorn, C. and Neves, F. and Nguyen, A. and Nikoleyczik, J. A. and Nilima, A. and O’Dell, J. and O’Neill, F. G. and O’Sullivan, K. and Olcina, I. and Olevitch, M. A. and Oliver-Mallory, K. C. and Orpwood, J. and Pagenkopf, D. and Pal, S. and Palladino, K. J. and Palmer, J. and Pangilinan, M. and Parveen, N. and Patton, S. J. and Pease, E. K. and Penning, B. and Pereira, C. and Pereira, G. and Perry, E. and Pershing, T. and Peterson, I. B. and Piepke, A. and Podczerwinski, J. and Porzio, D. and Powell, S. and Preece, R. M. and Pushkin, K. and Qie, Y. and Ratcliff, B. N. and Reichenbacher, J. and Reichhart, L. and Rhyne, C. A. and Richards, A. and Riffard, Q. and Rischbieter, G. R. C. and Rodrigues, J. P. and Rodriguez, A. and Rose, H. J. and Rosero, R. and Rossiter, P. and Rushton, T. and Rutherford, G. and Rynders, D. and Saba, J. S. and Santone, D. and Sazzad, A. B. M. R. and Schnee, R. W. and Scovell, P. R. and Seymour, D. and Shaw, S. and Shutt, T. and Silk, J. J. and Silva, C. and Sinev, G. and Skarpaas, K. and Skulski, W. and Smith, R. and Solmaz, M. and Solovov, V. N. and Sorensen, P. and Soria, J. and Stancu, I. and Stark, M. R. and Stevens, A. and Stiegler, T. M. and Stifter, K. and Studley, R. and Suerfu, B. and Sumner, T. J. and Sutcliffe, P. and Swanson, N. and Szydagis, M. and Tan, M. and Taylor, D. J. and Taylor, R. and Taylor, W. C. and Temples, D. J. and Tennyson, B. P. and Terman, P. A. and Thomas, K. J. and Tiedt, D. R. and Timalsina, M. and To, W. H. and Tomás, A. and Tong, Z. and Tovey, D. R. and Tranter, J. and Trask, M. and Tripathi, M. and Tronstad, D. R. and Tull, C. E. and Turner, W. and Tvrznikova, L. and Utku, U. and Va’vra, J. and Vacheret, A. and Vaitkus, A. C. and Verbus, J. R. and Voirin, E. and Waldron, W. L. and Wang, A. and Wang, B. and Wang, J. J. and Wang, W. and Wang, Y. and Watson, J. R. and Webb, R. C. and White, A. and White, D. T. and White, J. T. and White, R. G. and Whitis, T. J. and Williams, M. and Wisniewski, W. J. and Witherell, M. S. and Wolfs, F. L. H. and Wolfs, J. D. and Woodford, S. and Woodward, D. and Worm, S. D. and Wright, C. J. and Xia, Q. and Xiang, X. and Xiao, Q. and Xu, J. and Yeh, M. and Yin, J. and Young, I. and Zarzhitsky, P. and Zuckerman, A. and Zweig, E. A.},
   year={2023},
   month=jul }

@article{Kannike_2012,
   title={Vacuum stability conditions from copositivity criteria},
   volume={72},
   ISSN={1434-6052},
   url={http://dx.doi.org/10.1140/epjc/s10052-012-2093-z},
   DOI={10.1140/epjc/s10052-012-2093-z},
   number={7},
   journal={The European Physical Journal C},
   publisher={Springer Science and Business Media LLC},
   author={Kannike, Kristjan},
   year={2012},
   month=jul }

@article{Maniatis_2015,
    author = "Maniatis, M. and Nachtmann, O.",
    title = "{Stability and symmetry breaking in the general $n$-Higgs-doublet model}",
    eprint = "1407.7774",
    archivePrefix = "arXiv",
    primaryClass = "hep-ph",
    doi = "10.1007/JHEP02(2015)058",
    journal = "JHEP",
    volume = "02",
    pages = "058",
    year = "2015"
}

@article{Degee_2013,
   title={Geometric minimization of highly symmetric potentials},
   volume={2013},
   ISSN={1029-8479},
   url={http://dx.doi.org/10.1007/JHEP02(2013)125},
   DOI={10.1007/jhep02(2013)125},
   number={2},
   journal={Journal of High Energy Physics},
   publisher={Springer Science and Business Media LLC},
   author={Degee, A. and Ivanov, I. P. and Keus, V.},
   year={2013},
   month=feb }

@article{Grimus_2008,
   title={The oblique parameters in multi-Higgs-doublet models},
   volume={801},
   ISSN={0550-3213},
   url={http://dx.doi.org/10.1016/j.nuclphysb.2008.04.019},
   DOI={10.1016/j.nuclphysb.2008.04.019},
   number={1–2},
   journal={Nuclear Physics B},
   publisher={Elsevier BV},
   author={Grimus, W. and Lavoura, L. and Ogreid, O.M. and Osland, P.},
   year={2008},
   month=sep, pages={81–96} }

@article{Peskin:1991sw,
    author = "Peskin, Michael E. and Takeuchi, Tatsu",
    title = "{Estimation of oblique electroweak corrections}",
    reportNumber = "SLAC-PUB-5618",
    doi = "10.1103/PhysRevD.46.381",
    journal = "Phys. Rev. D",
    volume = "46",
    pages = "381--409",
    year = "1992"
}

@article{ParticleDataGroup:2024cfk,
    author = "Navas, S. and others",
    collaboration = "Particle Data Group",
    title = "{Review of particle physics}",
    doi = "10.1103/PhysRevD.110.030001",
    journal = "Phys. Rev. D",
    volume = "110",
    number = "3",
    pages = "030001",
    year = "2024"
}

@article{Hern_ndez_2024,
   title={Phenomenology of extended multiHiggs doublet models with $S_4$ family symmetry},
   volume={84},
   ISSN={1434-6052},
   url={http://dx.doi.org/10.1140/epjc/s10052-024-13633-5},
   DOI={10.1140/epjc/s10052-024-13633-5},
   number={11},
   journal={The European Physical Journal C},
   publisher={Springer Science and Business Media LLC},
   author={Hernández, A. E. Cárcamo and Espinoza, Catalina and Gómez-Izquierdo, Juan Carlos and Marchant González, Juan and Mondragón, Myriam},
   year={2024},
   month=nov }

@article{Hern_ndez_2015,
   title={Precision measurements constraints on the number of Higgs doublets},
   volume={91},
   ISSN={1550-2368},
   url={http://dx.doi.org/10.1103/PhysRevD.91.095014},
   DOI={10.1103/physrevd.91.095014},
   number={9},
   journal={Physical Review D},
   publisher={American Physical Society (APS)},
   author={Hernández, A. E. Cárcamo and Kovalenko, Sergey and Schmidt, Iván},
   year={2015},
   month=may }

@article{Grimus_2008_2,
   title={A precision constraint on multi-Higgs-doublet models},
   volume={35},
   ISSN={1361-6471},
   url={http://dx.doi.org/10.1088/0954-3899/35/7/075001},
   DOI={10.1088/0954-3899/35/7/075001},
   number={7},
   journal={Journal of Physics G: Nuclear and Particle Physics},
   publisher={IOP Publishing},
   author={Grimus, W and Lavoura, L and Ogreid, O M and Osland, P},
   year={2008},
   month=may, pages={075001} }

@article{Sher:1991km,
    author = "Sher, Marc and Yuan, Yao",
    title = "{Rare B decays, rare tau decays and grand unification}",
    reportNumber = "WM-TH-91-107",
    doi = "10.1103/PhysRevD.44.1461",
    journal = "Phys. Rev. D",
    volume = "44",
    pages = "1461--1472",
    year = "1991"
}

@article{Lavoura2003,
    author = {Lavoura, L.},
    title = {{General formulae for $f_1 \to f_2 \gamma$}},
    journal = {The European Physical Journal C - Particles and Fields},
    year = {2003},
    month = {jul},
    volume = {29},
    number = {2},
    pages = {191--195},
    issn = {1434-6052},
    doi = {10.1140/epjc/s2003-01212-7},
    url = {https://doi.org/10.1140/epjc/s2003-01212-7}
}

@article{Toma_2014,
   title={Lepton flavor violation in the scotogenic model},
   volume={2014},
   ISSN={1029-8479},
   url={http://dx.doi.org/10.1007/JHEP01(2014)160},
   DOI={10.1007/jhep01(2014)160},
   number={1},
   journal={Journal of High Energy Physics},
   publisher={Springer Science and Business Media LLC},
   author={Toma, Takashi and Vicente, Avelino},
   year={2014},
   month=jan }

@article{megiicollaboration2024searchmutoegammadataset,
      title={A search for $\mu^+\to e^+\gamma$ with the first dataset of the MEG II experiment}, 
      author={MEG II collaboration and K. Afanaciev and A. M. Baldini and S. Ban and V. Baranov and H. Benmansour and M. Biasotti and G. Boca and P. W. Cattaneo and G. Cavoto and F. Cei and M. Chiappini and G. Chiarello and A. Corvaglia and F. Cuna and G. Dal Maso and A. De Bari and M. De Gerone and L. Ferrari Barusso and M. Francesconi and L. Galli and G. Gallucci and F. Gatti and L. Gerritzen and F. Grancagnolo and E. G. Grandoni and M. Grassi and D. N. Grigoriev and M. Hildebrandt and K. Ieki and F. Ignatov and F. Ikeda and T. Iwamoto and S. Karpov and P. -R. Kettle and N. Khomutov and S. Kobayashi and A. Kolesnikov and N. Kravchuk and V. Krylov and N. Kuchinskiy and W. Kyle and T. Libeiro and V. Malyshev and A. Matsushita and M. Meucci and S. Mihara and W. Molzon and Toshinori Mori and M. Nakao and D. Nicolò and H. Nishiguchi and A. Ochi and S. Ogawa and R. Onda and W. Ootani and A. Oya and D. Palo and M. Panareo and A. Papa and V. Pettinacci and A. Popov and F. Renga and S. Ritt and M. Rossella and A. Rozhdestvensky and P. Schwendimann and K. Shimada and G. Signorelli and M. Takahashi and G. F. Tassielli and K. Toyoda and Y. Uchiyama and M. Usami and A. Venturini and B. Vitali and C. Voena and K. Yamamoto and K. Yanai and T. Yonemoto and K. Yoshida and Yu. V. Yudin},
      year={2024},
      eprint={2310.12614},
      archivePrefix={arXiv},
      primaryClass={hep-ex},
      url={https://arxiv.org/abs/2310.12614}, 
}

@article{Grimus_2002,
   title={One-loop corrections to the seesaw mechanism in the multi-Higgs-doublet standard model},
   volume={546},
   ISSN={0370-2693},
   url={http://dx.doi.org/10.1016/S0370-2693(02)02672-2},
   DOI={10.1016/s0370-2693(02)02672-2},
   number={1–2},
   journal={Physics Letters B},
   publisher={Elsevier BV},
   author={Grimus, Walter and Lavoura, Luı́s},
   year={2002},
   month=oct, pages={86–95} }

@article{Staub_2014,
   title={{SARAH 4:} A tool for (not only SUSY) model builders},
   volume={185},
   ISSN={0010-4655},
   url={http://dx.doi.org/10.1016/j.cpc.2014.02.018},
   DOI={10.1016/j.cpc.2014.02.018},
   number={6},
   journal={Computer Physics Communications},
   publisher={Elsevier BV},
   author={Staub, Florian},
   year={2014},
   month=jun, pages={1773–1790} }

@article{B_langer_2018,
    title={micrOMEGAs5.0: Freeze-in},
    volume={231},
    ISSN={0010-4655},
    url={http://dx.doi.org/10.1016/j.cpc.2018.04.027},
    DOI={10.1016/j.cpc.2018.04.027},
    journal={Computer Physics Communications},
    publisher={Elsevier BV},
    author={B{\'e}langer, G. and Boudjema, F. and Goudelis, A. and Pukhov, A. and Zald{\'i}var, B.},
    year={2018},
    month=oct, pages={173–186} }

@article{Vien_2013,
   title={The {$D_4$} flavor symmetry in 3-3-1 model with neutral leptons},
   author={Vien, V. V. and Long, H. N.},
   journal={Int. J. Mod. Phys. A},
   volume={28},
   pages={1350159},
   year={2013},
   doi={10.1142/S0217732313501599} 
}

@article{PhysRevLett.104.021802,
  title = {Searches for Lepton Flavor Violation in the Decays ${\ensuremath{\tau}}^{\ifmmode\pm\else\textpm\fi{}}\ensuremath{\rightarrow}{e}^{\ifmmode\pm\else\textpm\fi{}}\ensuremath{\gamma}$ and ${\ensuremath{\tau}}^{\ifmmode\pm\else\textpm\fi{}}\ensuremath{\rightarrow}{\ensuremath{\mu}}^{\ifmmode\pm\else\textpm\fi{}}\ensuremath{\gamma}$},
  author = {Aubert, B. and Karyotakis, Y. and Lees, J. P. and Poireau, V. and Prencipe, E. and Prudent, X. and Tisserand, V. and Garra Tico, J. and Grauges, E. and Martinelli, M. and Palano, A. and Pappagallo, M. and Eigen, G. and Stugu, B. and Sun, L. and Battaglia, M. and Brown, D. N. and Hooberman, B. and Kerth, L. T. and Kolomensky, Yu. G. and Lynch, G. and Osipenkov, I. L. and Tackmann, K. and Tanabe, T. and Hawkes, C. M. and Soni, N. and Watson, A. T. and Koch, H. and Schroeder, T. and Asgeirsson, D. J. and Hearty, C. and Mattison, T. S. and McKenna, J. A. and Barrett, M. and Khan, A. and Randle-Conde, A. and Blinov, V. E. and Bukin, A. D. and Buzykaev, A. R. and Druzhinin, V. P. and Golubev, V. B. and Onuchin, A. P. and Serednyakov, S. I. and Skovpen, Yu. I. and Solodov, E. P. and Todyshev, K. Yu. and Bondioli, M. and Curry, S. and Eschrich, I. and Kirkby, D. and Lankford, A. J. and Lund, P. and Mandelkern, M. and Martin, E. C. and Stoker, D. P. and Atmacan, H. and Gary, J. W. and Liu, F. and Long, O. and Vitug, G. M. and Yasin, Z. and Sharma, V. and Campagnari, C. and Hong, T. M. and Kovalskyi, D. and Mazur, M. A. and Richman, J. D. and Beck, T. W. and Eisner, A. M. and Heusch, C. A. and Kroseberg, J. and Lockman, W. S. and Martinez, A. J. and Schalk, T. and Schumm, B. A. and Seiden, A. and Wang, L. and Winstrom, L. O. and Cheng, C. H. and Doll, D. A. and Echenard, B. and Fang, F. and Hitlin, D. G. and Narsky, I. and Ongmongkolkul, P. and Piatenko, T. and Porter, F. C. and Andreassen, R. and Mancinelli, G. and Meadows, B. T. and Mishra, K. and Sokoloff, M. D. and Bloom, P. C. and Ford, W. T. and Gaz, A. and Hirschauer, J. F. and Nagel, M. and Nauenberg, U. and Smith, J. G. and Wagner, S. R. and Ayad, R. and Toki, W. H. and Feltresi, E. and Hauke, A. and Jasper, H. and Karbach, T. M. and Merkel, J. and Petzold, A. and Spaan, B. and Wacker, K. and Kobel, M. J. and Nogowski, R. and Schubert, K. R. and Schwierz, R. and Bernard, D. and Latour, E. and Verderi, M. and Clark, P. J. and Playfer, S. and Watson, J. E. and Andreotti, M. and Bettoni, D. and Bozzi, C. and Calabrese, R. and Cecchi, A. and Cibinetto, G. and Fioravanti, E. and Franchini, P. and Luppi, E. and Munerato, M. and Negrini, M. and Petrella, A. and Piemontese, L. and Santoro, V. and Baldini-Ferroli, R. and Calcaterra, A. and de Sangro, R. and Finocchiaro, G. and Pacetti, S. and Patteri, P. and Peruzzi, I. M. and Piccolo, M. and Rama, M. and Zallo, A. and Contri, R. and Guido, E. and Lo Vetere, M. and Monge, M. R. and Passaggio, S. and Patrignani, C. and Robutti, E. and Tosi, S. and Morii, M. and Adametz, A. and Marks, J. and Schenk, S. and Uwer, U. and Bernlochner, F. U. and Lacker, H. M. and Lueck, T. and Volk, A. and Dauncey, P. D. and Tibbetts, M. and Behera, P. K. and Charles, M. J. and Mallik, U. and Cochran, J. and Crawley, H. B. and Dong, L. and Eyges, V. and Meyer, W. T. and Prell, S. and Rosenberg, E. I. and Rubin, A. E. and Gao, Y. Y. and Gritsan, A. V. and Guo, Z. J. and Arnaud, N. and D'Orazio, A. and Davier, M. and Derkach, D. and Firmino da Costa, J. and Grosdidier, G. and Le Diberder, F. and Lepeltier, V. and Lutz, A. M. and Malaescu, B. and Roudeau, P. and Schune, M. H. and Serrano, J. and Sordini, V. and Stocchi, A. and Wormser, G. and Lange, D. J. and Wright, D. M. and Bingham, I. and Burke, J. P. and Chavez, C. A. and Fry, J. R. and Gabathuler, E. and Gamet, R. and Hutchcroft, D. E. and Payne, D. J. and Touramanis, C. and Bevan, A. J. and Clarke, C. K. and Di Lodovico, F. and Sacco, R. and Sigamani, M. and Cowan, G. and Paramesvaran, S. and Wren, A. C. and Brown, D. N. and Davis, C. L. and Denig, A. G. and Fritsch, M. and Gradl, W. and Hafner, A. and Alwyn, K. E. and Bailey, D. and Barlow, R. J. and Jackson, G. and Lafferty, G. D. and West, T. J. and Yi, J. I. and Anderson, J. and Chen, C. and Jawahery, A. and Roberts, D. A. and Simi, G. and Tuggle, J. M. and Dallapiccola, C. and Salvati, E. and Cowan, R. and Dujmic, D. and Fisher, P. H. and Henderson, S. W. and Sciolla, G. and Spitznagel, M. and Yamamoto, R. K. and Zhao, M. and Patel, P. M. and Robertson, S. H. and Schram, M. and Biassoni, P. and Lazzaro, A. and Lombardo, V. and Palombo, F. and Stracka, S. and Cremaldi, L. and Godang, R. and Kroeger, R. and Sonnek, P. and Summers, D. J. and Zhao, H. W. and Nguyen, X. and Simard, M. and Taras, P. and Nicholson, H. and De Nardo, G. and Lista, L. and Monorchio, D. and Onorato, G. and Sciacca, C. and Raven, G. and Snoek, H. L. and Jessop, C. P. and Knoepfel, K. J. and LoSecco, J. M. and Wang, W. F. and Corwin, L. A. and Honscheid, K. and Kagan, H. and Kass, R. and Morris, J. P. and Rahimi, A. M. and Sekula, S. J. and Blount, N. L. and Brau, J. and Frey, R. and Igonkina, O. and Kolb, J. A. and Lu, M. and Rahmat, R. and Sinev, N. B. and Strom, D. and Strube, J. and Torrence, E. and Castelli, G. and Gagliardi, N. and Margoni, M. and Morandin, M. and Posocco, M. and Rotondo, M. and Simonetto, F. and Stroili, R. and Voci, C. and del Amo Sanchez, P. and Ben-Haim, E. and Bonneaud, G. R. and Briand, H. and Chauveau, J. and Hamon, O. and Leruste, Ph. and Marchiori, G. and Ocariz, J. and Perez, A. and Prendki, J. and Sitt, S. and Gladney, L. and Biasini, M. and Manoni, E. and Angelini, C. and Batignani, G. and Bettarini, S. and Calderini, G. and Carpinelli, M. and Cervelli, A. and Forti, F. and Giorgi, M. A. and Lusiani, A. and Morganti, M. and Neri, N. and Paoloni, E. and Rizzo, G. and Walsh, J. J. and Lopes Pegna, D. and Lu, C. and Olsen, J. and Smith, A. J. S. and Telnov, A. V. and Anulli, F. and Baracchini, E. and Cavoto, G. and Faccini, R. and Ferrarotto, F. and Ferroni, F. and Gaspero, M. and Jackson, P. D. and Li Gioi, L. and Mazzoni, M. A. and Morganti, S. and Piredda, G. and Renga, F. and Voena, C. and Ebert, M. and Hartmann, T. and Schr\"oder, H. and Waldi, R. and Adye, T. and Franek, B. and Olaiya, E. O. and Wilson, F. F. and Emery, S. and Esteve, L. and Hamel de Monchenault, G. and Kozanecki, W. and Vasseur, G. and Y\`eche, Ch. and Zito, M. and Allen, M. T. and Aston, D. and Bard, D. J. and Bartoldus, R. and Benitez, J. F. and Cenci, R. and Coleman, J. P. and Convery, M. R. and Dingfelder, J. C. and Dorfan, J. and Dubois-Felsmann, G. P. and Dunwoodie, W. and Field, R. C. and Franco Sevilla, M. and Fulsom, B. G. and Gabareen, A. M. and Graham, M. T. and Grenier, P. and Hast, C. and Innes, W. R. and Kaminski, J. and Kelsey, M. H. and Kim, H. and Kim, P. and Kocian, M. L. and Leith, D. W. G. S. and Li, S. and Lindquist, B. and Luitz, S. and Luth, V. and Lynch, H. L. and MacFarlane, D. B. and Marsiske, H. and Messner, R. and Muller, D. R. and Neal, H. and Nelson, S. and O'Grady, C. P. and Ofte, I. and Perl, M. and Ratcliff, B. N. and Roodman, A. and Salnikov, A. A. and Schindler, R. H. and Schwiening, J. and Snyder, A. and Su, D. and Sullivan, M. K. and Suzuki, K. and Swain, S. K. and Thompson, J. M. and Va'vra, J. and Wagner, A. P. and Weaver, M. and West, C. A. and Wisniewski, W. J. and Wittgen, M. and Wright, D. H. and Wulsin, H. W. and Yarritu, A. K. and Young, C. C. and Ziegler, V. and Chen, X. R. and Liu, H. and Park, W. and Purohit, M. V. and White, R. M. and Wilson, J. R. and Bellis, M. and Burchat, P. R. and Edwards, A. J. and Miyashita, T. S. and Ahmed, S. and Alam, M. S. and Ernst, J. A. and Pan, B. and Saeed, M. A. and Zain, S. B. and Soffer, A. and Spanier, S. M. and Wogsland, B. J. and Eckmann, R. and Ritchie, J. L. and Ruland, A. M. and Schilling, C. J. and Schwitters, R. F. and Wray, B. C. and Drummond, B. W. and Izen, J. M. and Lou, X. C. and Bianchi, F. and Gamba, D. and Pelliccioni, M. and Bomben, M. and Bosisio, L. and Cartaro, C. and Della Ricca, G. and Lanceri, L. and Vitale, L. and Azzolini, V. and Lopez-March, N. and Martinez-Vidal, F. and Milanes, D. A. and Oyanguren, A. and Albert, J. and Banerjee, Sw. and Bhuyan, B. and Choi, H. H. F. and Hamano, K. and King, G. J. and Kowalewski, R. and Lewczuk, M. J. and Lindsay, C. D. and Locke, C. B. and Nugent, I. M. and Roney, J. M. and Sobie, R. J. and Gershon, T. J. and Harrison, P. F. and Ilic, J. and Latham, T. E. and Mohanty, G. B. and Puccio, E. M. T. and Band, H. R. and Chen, X. and Dasu, S. and Flood, K. T. and Pan, Y. and Prepost, R. and Vuosalo, C. O. and Wu, S. L.},
  collaboration = {BABAR Collaboration},
  journal = {Phys. Rev. Lett.},
  volume = {104},
  issue = {2},
  pages = {021802},
  numpages = {7},
  year = {2010},
  month = {Jan},
  publisher = {American Physical Society},
  doi = {10.1103/PhysRevLett.104.021802},
  url = {https://link.aps.org/doi/10.1103/PhysRevLett.104.021802}
}

@article{2025,
   title={Joint neutrino oscillation analysis from the {T2K} and {NOvA} experiments},
   volume={646},
   ISSN={1476-4687},
   url={http://dx.doi.org/10.1038/s41586-025-09599-3},
   DOI={10.1038/s41586-025-09599-3},
   number={8086},
   journal={Nature},
   publisher={Springer Science and Business Media LLC},
   author={Abubakar, S. and Acero, M. A. and Acharya, B. and Adamson, P. and Anfimov, N. and Antoshkin, A. and Arrieta-Diaz, E. and Asquith, L. and Aurisano, A. and Azevedo, D. and Back, A. and Balashov, N. and Baldi, P. and Bambah, B. A. and Bannister, E. F. and Barros, A. and Bat, A. and Bays, K. and Bernstein, R. and Bezerra, T. J. C. and Bhatnagar, V. and Bhuyan, B. and Bian, J. and Booth, A. C. and Bowles, R. and Brahma, B. and Bromberg, C. and Buchanan, N. and Butkevich, A. and Calvez, S. and Carceller, J. M. and Carroll, T. J. and Catano-Mur, E. and Cesar, J. P. and Chirco, R. and Choudhary, B. C. and Christensen, A. and Cicala, M. F. and Coan, T. E. and Contreras, T. and Cooleybeck, A. and Coveyou, D. and Cremonesi, L. and Davies, G. S. and Derwent, P. F. and Ding, P. and Djurcic, Z. and Dobbs, K. and Dolce, M. and Dueñas Tonguino, D. and Dukes, E. C. and Dye, A. and Ehrlich, R. and Ewart, E. and Filip, P. and Frank, M. J. and Gallagher, H. R. and Giri, A. and Gomes, R. A. and Goodman, M. C. and Group, R. and Habig, A. and Hakl, F. and Hartnell, J. and Hatcher, R. and Hays, J. M. and He, M. and Heller, K. and Hewes, V. and Himmel, A. and Horoho, T. and Ivanova, A. and Jargowsky, B. and Kakorin, I. and Kalitkina, A. and Kaplan, D. M. and Khanam, A. and Kirezli, B. and Kleykamp, J. and Klimov, O. and Koerner, L. W. and Kolupaeva, L. and Kralik, R. and Kumar, A. and Kuruppu, C. D. and Kus, V. and Lackey, T. and Lang, K. and Lasorak, P. and Lesmeister, J. and Lister, A. and Liu, J. and Lock, J. A. and MacMahon, M. and Magill, S. and Mann, W. A. and Manoharan, M. T. and Manrique Plata, M. and Marshak, M. L. and Martinez-Casales, M. and Matveev, V. and Mehta, B. and Messier, M. D. and Meyer, H. and Miao, T. and Miller, W. H. and Mishra, S. R. and Mohanta, R. and Moren, A. and Morozova, A. and Mu, W. and Mualem, L. and Muether, M. and Mulder, K. and Myers, D. and Naples, D. and Nelleri, S. and Nelson, J. K. and Nichol, R. and Niner, E. and Norman, A. and Norrick, A. and Oh, H. and Olshevskiy, A. and Olson, T. and Ozkaynak, M. and Pal, A. and Paley, J. and Panda, L. and Patterson, R. B. and Pawloski, G. and Petti, R. and Plunkett, R. K. and Porter, J. C. C. and Prais, L. R. and Rafique, A. and Raj, V. and Rajaoalisoa, M. and Ramson, B. and Rebel, B. and Robles, E. and Roy, P. and Samoylov, O. and Sanchez, M. C. and Sánchez Falero, S. and Shanahan, P. and Sharma, P. and Sheshukov, A. and Shivam and Shmakov, A. and Shorrock, W. and Shukla, S. and Singh, I. and Singh, P. and Singh, V. and Singh Chhibra, S. and Singha, D. K. and Smith, A. and Smolik, J. and Snopok, P. and Solomey, N. and Sousa, A. and Soustruznik, K. and Strait, M. and Suter, L. and Sutton, A. and Swain, S. and Sweeney, C. and Sztuc, A. and Talukdar, N. and Tas, P. and Thakore, T. and Thomas, J. and Tiras, E. and Titus, M. and Torun, Y. and Tran, D. and Trokan-Tenorio, J. and Urheim, J. and Vahle, P. and Vallari, Z. and Vockerodt, K. J. and Waldron, A. V. and Wallbank, M. and Warburton, T. K. and Weber, C. and Wetstein, M. and Whittington, D. and Wickremasinghe, D. A. and Wolcott, J. and Wu, S. and Wu, W. and Wu, W. and Xiao, Y. and Yaeggy, B. and Yahaya, A. and Yankelevich, A. and Yonehara, K. and Zadorozhnyy, S. and Zalesak, J. and Zwaska, R. and Abe, K. and Abe, S. and Adhkary, H. and Akutsu, R. and Alarakia-Charles, H. and Hakim, Y. I. Alj and Monsalve, S. Alonso and Anthony, L. and Aoki, S. and Apte, K. A. and Arai, T. and Arihara, T. and Arimoto, S. and Ashida, Y. and Atkin, E. T. and Babu, N. and Baranov, V. and Barker, G. J. and Barr, G. and Barrow, D. and Bates, P. and Bathe-Peters, L. and Batkiewicz-Kwasniak, M. and Baudis, N. and Berardi, V. and Berns, L. and Bhattacharjee, S. and Blanchet, A. and Blondel, A. and Boistier, P. M. M. and Bolognesi, S. and Bordoni, S. and Boyd, S. B. and Bronner, C. and Bubak, A. and Buizza Avanzini, M. and Caballero, J. A. and Cadoux, F. and Calabria, N. F. and Cao, S. and Cap, S. and Carabadjac, D. and Cartwright, S. L. and Casado, M. P. and Catanesi, M. G. and Chakrani, J. and Chalumeau, A. and Cherdack, D. and Chvirova, A. and Coleman, J. and Collazuol, G. and Cormier, F. and Craplet, A. A. L. and Cudd, A. and D’ago, D. and Dalmazzone, C. and Daret, T. and Dasgupta, P. and Davis, C. and Davydov, Yu. I. and de Perio, P. and De Rosa, G. and Dealtry, T. and Densham, C. and Dergacheva, A. and Dharmapal Banerjee, R. and Di Lodovico, F. and Diaz Lopez, G. and Dolan, S. and Douqa, D. and Doyle, T. A. and Drapier, O. and Duffy, K. E. and Dumarchez, J. and Dunne, P. and Dygnarowicz, K. and Eguchi, A. and Elias, J. and Emery-Schrenk, S. and Erofeev, G. and Ershova, A. and Eurin, G. and Fedorova, D. and Fedotov, S. and Feltre, M. and Feng, L. and Ferlewicz, D. and Finch, A. J. and Fitton, M. D. and Forza, C. and Friend, M. and Fujii, Y. and Fukuda, Y. and Furui, Y. and García-Marcos, J. and Germer, A. C. and Giannessi, L. and Giganti, C. and Girgus, M. and Glagolev, V. and Gonin, M. and González Jiménez, R. and González Rosa, J. and Goodman, E. A. G. and Gorshanov, K. and Govindaraj, P. and Grassi, M. and Guigue, M. and Guo, F. Y. and Hadley, D. R. and Han, S. and Harris, D. A. and Harris, R. J. and Hasegawa, T. and Hasnip, C. M. and Hassani, S. and Hastings, N. C. and Hayato, Y. and Heitkamp, I. and Henaff, D. and Hino, Y. and Holeczek, J. and Holin, A. and Holvey, T. and Hong Van, N. T. and Honjo, T. and Hooft, M. C. F. and Hosokawa, K. and Hu, J. and Ichikawa, A. K. and Ieki, K. and Ikeda, M. and Ishida, T. and Ishitsuka, M. and Izmaylov, A. and Jachowicz, N. and Jenkins, S. J. and Jesús-Valls, C. and Jia, M. and Jiang, J. J. and Ji, J. Y. and Jones, T. P. and Jonsson, P. and Joshi, S. and Jung, C. K. and Kabirnezhad, M. and Kaboth, A. C. and Kakuno, H. and Kameda, J. and Karpova, S. and Kasturi, V. S. and Kataoka, Y. and Katori, T. and Kawamura, Y. and Kawaue, M. and Kearns, E. and Khabibullin, M. and Khotjantsev, A. and Kikawa, T. and King, S. and Kiseeva, V. and Kisiel, J. and Klustová, A. and Kneale, L. and Kobayashi, H. and Koch, L. and Kodama, S. and Kolupanova, M. and Konaka, A. and Kormos, L. L. and Koshio, Y. and Kowalik, K. and Kudenko, Y. and Kudo, Y. and Kumar Jha, A. and Kurjata, R. and Kurochka, V. and Kutter, T. and Labarga, L. and Lachat, M. and Lachner, K. and Lagoda, J. and Lakshmi, S. M. and Lamers James, M. and Langella, A. and Langridge, D. H. and Laporte, J.-F. and Last, D. and Latham, N. and Laveder, M. and Lavitola, L. and Lawe, M. and Leon Silverio, D. and Levorato, S. and Lewis, S. V. and Li, B. and Lin, C. and Litchfield, R. P. and Liu, S. L. and Li, W. and Longhin, A. and Lopez Moreno, A. and Ludovici, L. and Lu, X. and Lux, T. and Machado, L. N. and Magaletti, L. and Mahn, K. and Mahtani, K. K. and Mandal, M. and Manly, S. and Marino, A. D. and Martin, D. G. R. and Martinez Caicedo, D. A. and Martinez, L. and Martini, M. and Matsubara, T. and Matsumoto, R. and Matveev, V. and Mauger, C. and Mavrokoridis, K. and McCauley, N. and McFarland, K. S. and McGrew, C. and McKean, J. and Mefodiev, A. and Megias, G. D. and Mellet, L. and Metelko, C. and Mezzetto, M. and Miki, S. and Mikola, V. and Miller, E. W. and Minamino, A. and Mineev, O. and Mine, S. and Mirabito, J. and Miura, M. and Moriyama, S. and Moriyama, S. and Morrison, P. and Mueller, Th. A. and Munford, D. and Muñoz, A. and Munteanu, L. and Nagai, Y. and Nakadaira, T. and Nakagiri, K. and Nakahata, M. and Nakajima, Y. and Nakamura, K. D. and Nakano, Y. and Nakayama, S. and Nakaya, T. and Nakayoshi, K. and Naseby, C. E. R. and Nguyen, D. T. and Nguyen, V. Q. and Niewczas, K. and Nishimori, S. and Nishimura, Y. and Noguchi, Y. and Nosek, T. and Nova, F. and Nugent, J. C. and O’Keeffe, H. M. and O’Sullivan, L. and Okazaki, R. and Okinaga, W. and Okumura, K. and Okusawa, T. and Onda, N. and Ospina, N. and Osu, L. and Oyama, Y. and Paolone, V. and Pasternak, J. and Payne, D. and Peacock, T. and Pfaff, M. and Pickering, L. and Popov, B. and Portocarrero Yrey, A. J. and Posiadala-Zezula, M. and Prabhu, Y. S. and Prasad, H. and Pupilli, F. and Quilain, B. and Quyen, P. T. and Radicioni, E. and Radics, B. and Ramirez, M. A. and Ramsden, R. and Ratoff, P. N. and Reh, M. and Reina, G. and Riccio, C. and Riley, D. W. and Rondio, E. and Roth, S. and Roy, N. and Rubbia, A. and Russo, L. and Rychter, A. and Saenz, W. and Sakashita, K. and Samani, S. and Sánchez, F. and Sandford, E. M. and Sato, Y. and Schefke, T. and Schloesser, C. M. and Scholberg, K. and Scott, M. and Seiya, Y. and Sekiguchi, T. and Sekiya, H. and Sekiya, T. and Seppala, D. and Sgalaberna, D. and Shaikhiev, A. and Shiozawa, M. and Shiraishi, Y. and Shvartsman, A. and Skrobova, N. and Skwarczynski, K. and Smyczek, D. and Smy, M. and Sobczyk, J. T. and Sobel, H. and Soler, F. J. P. and Speers, A. J. and Spina, R. and Srivastava, A. and Stowell, P. and Stroke, Y. and Suslov, I. A. and Suzuki, A. and Suzuki, S. Y. and Tada, M. and Tairafune, S. and Takeda, A. and Teklu, A. and Takeuchi, Y. and Tanaka, H. K. and Tanigawa, H. and Tereshchenko, V. V. and Thamm, N. and Touramanis, C. and Tran, N. and Tsukamoto, T. and Tzanov, M. and Uchida, Y. and Vagins, M. and Varghese, M. and Vasilyev, I. and Vasseur, G. and Villa, E. and Virginet, U. and Vladisavljevic, T. and Wachala, T. and Wakabayashi, D. and Wallace, H. T. and Walsh, J. G. and Wan, L. and Wark, D. and Wascko, M. O. and Weber, A. and Wendell, R. and Wilking, M. J. and Wilkinson, C. and Wilson, J. R. and Wood, K. and Wret, C. and Xia, J. and Yamamoto, K. and Yamamoto, T. and Yanagisawa, C. and Yang, Y. and Yano, T. and Yershov, N. and Yevarouskaya, U. and Yokoyama, M. and Yoshimoto, Y. and Yoshimura, N. and Zaki, R. and Zalewska, A. and Zalipska, J. and Zarnecki, G. and Zhang, J. and Zhao, X. Y. and Zheng, H. and Zhong, H. and Zhu, T. and Ziembicki, M. and Zimmerman, E. D. and Zito, M. and Zsoldos, S.},
   year={2025},
   month=oct, pages={818–824} }

@article{Abe_2023,
  title = {Search for Majorana Neutrinos with the Complete KamLAND-Zen Dataset},
  author = {Abe, S. and Araki, T. and Chiba, K. and Eda, T. and Eizuka, M. and Funahashi, Y. and Furuto, A. and Gando, A. and Gando, Y. and Goto, S. and Hachiya, T. and Hata, K. and Ichimura, K. and Ieki, S. and Ikeda, H. and Inoue, K. and Ishidoshiro, K. and Kamei, Y. and Kawada, N. and Kishimoto, Y. and Koga, M. and Marthe, A. and Matsumoto, Y. and Mitsui, T. and Miyake, H. and Morita, D. and Nakajima, R. and Nakamura, K. and Nakamura, R. and Nakamura, R. and Nakane, J. and Ono, T. and Ozaki, H. and Saito, K. and Sakai, T. and Shimizu, I. and Shirai, J. and Shiraishi, K. and Suzuki, A. and Tachibana, K. and Tamae, K. and Watanabe, H. and Watanabe, K. and Yoshida, S. and Umehara, S. and Fushimi, K. and Kotera, K. and Urano, Y. and Berger, B. E. and Fujikawa, B. K. and Learned, J. G. and Maricic, J. and Fu, Z. and Ghosh, S. and Smolsky, J. and Winslow, L. A. and Efremenko, Y. and Karwowski, H. J. and Markoff, D. M. and Tornow, W. and Dell'Oro, S. and O'Donnell, T. and Detwiler, J. A. and Enomoto, S. and Decowski, M. P. and Weerman, K. M. and Grant, C. and Penek, \"O. and Song, H. and Li, A. and Axani, S. N. and Garcia, M. and Sarfraz, M.},
  collaboration = {KamLAND-Zen Collaboration},
  journal = {Phys. Rev. Lett.},
  volume = {135},
  issue = {26},
  pages = {262501},
  numpages = {7},
  year = {2025},
  month = {Dec},
  publisher = {American Physical Society},
  doi = {10.1103/jkf6-48j8},
  url = {https://link.aps.org/doi/10.1103/jkf6-48j8}
}

@article{Bonilla_2020,
   title={Neutrino phenomenology in a left-right {$D_4$} symmetric model},
   volume={102},
   ISSN={2470-0029},
   url={http://dx.doi.org/10.1103/PhysRevD.102.036006},
   DOI={10.1103/physrevd.102.036006},
   number={3},
   journal={Physical Review D},
   publisher={American Physical Society (APS)},
   author={Bonilla, Cesar and de la Vega, Leon M. G. and Ferro-Hernandez, R. and Nath, Newton and Peinado, Eduardo},
   year={2020},
   month=aug }

@article{Esteban_2024,
   title={NuFit-6.0: updated global analysis of three-flavor neutrino oscillations},
   volume={2024},
   ISSN={1029-8479},
   url={http://dx.doi.org/10.1007/JHEP12(2024)216},
   DOI={10.1007/jhep12(2024)216},
   number={12},
   journal={Journal of High Energy Physics},
   publisher={Springer Science and Business Media LLC},
   author={Esteban, Ivan and Gonzalez-Garcia, M. C. and Maltoni, Michele and Martinez-Soler, Ivan and Pinheiro, João Paulo and Schwetz, Thomas},
   year={2024},
   month=dec }

@article{Khater_2022,

   title={Dark matter in three-Higgs-doublet models with S3 symmetry},

   volume={2022},

   ISSN={1029-8479},

   url={http://dx.doi.org/10.1007/JHEP01(2022)120},

   DOI={10.1007/jhep01(2022)120},

   number={1},

   journal={Journal of High Energy Physics},

   publisher={Springer Science and Business Media LLC},

   author={Khater, W. and Kunčinas, A. and Ogreid, O. M. and Osland, P. and Rebelo, M. N.},

   year={2022},

   month=jan }

@article{Kun_inas_2022,

   title={Dark matter in a {CP-}violating three-Higgs-doublet model with {$S_3$} symmetry},

   volume={106},

   ISSN={2470-0029},

   url={http://dx.doi.org/10.1103/PhysRevD.106.075002},

   DOI={10.1103/physrevd.106.075002},

   number={7},

   journal={Physical Review D},

   publisher={American Physical Society (APS)},

   author={Kunčinas, A. and Ogreid, O. M. and Osland, P. and Rebelo, M. N.},

   year={2022},

   month=oct }

@article{Araki_2012,

   title={{$Q_6$} flavor symmetry model for the extension of the minimal standard model by three right-handed sterile neutrinos},

   volume={85},

   ISSN={1550-2368},

   url={http://dx.doi.org/10.1103/PhysRevD.85.065016},

   DOI={10.1103/physrevd.85.065016},

   number={6},

   journal={Physical Review D},

   publisher={American Physical Society (APS)},

   author={Araki, Takeshi and Li, Y. F.},

   year={2012},

   month=mar }

@article{VIEN2020115015,

title = {{$B-L$ extension of the {Standard Model} with $Q_6$ symmetry}},

journal = {Nuclear Physics B},

volume = {956},

pages = {115015},

year = {2020},

issn = {0550-3213},

doi = {https://doi.org/10.1016/j.nuclphysb.2020.115015},

url = {https://www.sciencedirect.com/science/article/pii/S0550321320301012},

author = {V.V. Vien}

}

@article{Kajiyama_2007,

   title={{$D_6$} family symmetry and cold dark matter at CERN LHC},

   volume={75},

   ISSN={1550-2368},

   url={http://dx.doi.org/10.1103/PhysRevD.75.033001},

   DOI={10.1103/physrevd.75.033001},

   number={3},

   journal={Physical Review D},

   publisher={American Physical Society (APS)},

   author={Kajiyama, Yuji and Kubo, Jisuke and Okada, Hiroshi},

   year={2007},

   month=feb }

@article{legendcollaboration2021legend1000preconceptualdesignreport,
      title={{LEGEND-1000} Preconceptual Design Report}, 
      author={LEGEND Collaboration and N. Abgrall and I. Abt and M. Agostini and A. Alexander and C. Andreoiu and G. R. Araujo and F. T. Avignone III and W. Bae and A. Bakalyarov and M. Balata and M. Bantel and I. Barabanov and A. S. Barabash and P. S. Barbeau and C. J. Barton and P. J. Barton and L. Baudis and C. Bauer and E. Bernieri and L. Bezrukov and K. H. Bhimani and V. Biancacci and E. Blalock and A. Bolozdynya and S. Borden and B. Bos and E. Bossio and A. Boston and V. Bothe and R. Bouabid and S. Boyd and R. Brugnera and N. Burlac and M. Busch and A. Caldwell and T. S. Caldwell and R. Carney and C. Cattadori and Y. -D. Chan and A. Chernogorov and C. D. Christofferson and P. -H. Chu and M. Clark and T. Cohen and D. Combs and T. Comellato and R. J. Cooper and I. A. Costa and V. D'Andrea and J. A. Detwiler and A. Di Giacinto and N. Di Marco and J. Dobson and A. Drobizhev and M. R. Durand and F. Edzards and Yu. Efremenko and S. R. Elliott and A. Engelhardt and L. Fajt and N. Faud and M. T. Febbraro and F. Ferella and D. E. Fields and F. Fischer and M. Fomina and H. Fox and J. Franchi and R. Gala and A. Galindo-Uribarri and A. Gangapshev and A. Garfagnini and A. Geraci and C. Gilbert and M. Gold and C. Gooch and K. P. Gradwohl and M. P. Green and G. F. Grinyer and A. Grobov and J. Gruszko and I. Guinn and V. E. Guiseppe and V. Gurentsov and Y. Gurov and K. Gusev and B. Hacket and F. Hagemann and J. Hakenmüeller and M. Haranczyk and L. Hauertmann and C. R. Haufe and C. Hayward and B. Heffron and F. Henkes and R. Henning and D. Hervas Aguilar and J. Hinton and R. Hodak and H. Hoffmann and W. Hofmann and A. Hostiuc and J. Huang and M. Hult and M. Ibrahim Mirza and J. Jochum and R. Jones and D. Judson and M. Junker and J. Kaizer and V. Kazalov and Y. Kermaïdic and H. Khushbakht and M. Kidd and T. Kihm and K. Kilgus and I. Kim and A. Klimenko and K. T. Knöpfle and O. Kochetov and S. I. Konovalov and I. Kontul and K. Kool and L. L. Kormos and V. N. Kornoukhov and M. Korosec and P. Krause and V. V. Kuzminov and J. M. López-Castaño and K. Lang and M. Laubenstein and E. León and B. Lehnert and A. Leonhardt and A. Li and M. Lindner and I. Lippi and X. Liu and J. Liu and D. Loomba and A. Lubashevskiy and B. Lubsandorzhiev and N. Lusardi and Y. Müller and M. Macko and C. Macolino and B. Majorovits and F. Mamedov and W. Maneschg and L. Manzanillas and G. Marshall and R. D. Martin and E. L. Martin and R. Massarczyk and D. Mei and S. J. Meijer and S. Mertens and M. Misiaszek and E. Mondragon and M. Morella and B. Morgan and T. Mroz and D. Muenstermann and C. J. Nave and I. Nemchenok and M. Neuberger and T. K. Oli and G. Orebi Gann and G. Othman and V. Palušova and R. Panth and L. Papp and L. S. Paudel and K. Pelczar and J. Perez Perez and L. Pertoldi and W. Pettus and P. Piseri and A. W. P. Poon and P. Povinec and A. Pullia and D. C. Radford and Y. A. Ramachers and C. Ransom and L. Rauscher and M. Redchuk and A. L. Reine and S. Riboldi and K. Rielage and S. Rozov and E. Rukhadze and N. Rumyantseva and J. Runge and N. W. Ruof and R. Saakyan and S. Sailer and G. Salamanna and F. Salamida and D. J. Salvat and V. Sandukovsky and S. Schönert and A. Schültz and M. Schütt and D. C. Schaper and J. Schreiner and O. Schulz and M. Schuster and M. Schwarz and B. Schwingenheuer and O. Selivanenko and M. Shafiee and E. Shevchik and M. Shirchenko and Y. Shitov and H. Simgen and F. Simkovic and M. Skorokhvatov and M. Slavickova and K. Smolek and A. Smolnikov and J. A. Solomon and G. Song and K. Starosta and I. Stekl and M. Stommel and D. Stukov and R. R. Sumathi and D. A. Sweigart and K. Szczepaniec and L. Taffarello and D. Tagnani and R. Tayloe and D. Tedeschi and M. Turqueti and R. L. Varner and S. Vasilyev and A. Veresnikova and K. Vetter and C. Vignoli and C. Vogl and K. von Sturm and D. Waters and J. C. Waters and W. Wei and C. Wiesinger and J. F. Wilkerson and M. Willers and C. Wiseman and M. Wojcik and V. H. -S. Wu and W. Xu and E. Yakushev and T. Ye and C. -H. Yu and V. Yumatov and N. Zaretski and J. Zeman and I. Zhitnikov and D. Zinatulina and A. -K. Zschocke and A. J. Zsigmond and K. Zuber and G. Zuzel},
      year={2021},
      eprint={2107.11462},
      archivePrefix={arXiv},
      primaryClass={physics.ins-det},
      url={https://arxiv.org/abs/2107.11462}, 
}

@article{Andringa_2016,
   title={Current Status and Future Prospects of the {SNO+} Experiment},
   volume={2016},
   ISSN={1687-7365},
   url={http://dx.doi.org/10.1155/2016/6194250},
   DOI={10.1155/2016/6194250},
   journal={Advances in High Energy Physics},
   publisher={Wiley},
   author={Andringa, S. and Arushanova, E. and Asahi, S. and Askins, M. and Auty, D. J. and Back, A. R. and Barnard, Z. and Barros, N. and Beier, E. W. and Bialek, A. and Biller, S. D. and Blucher, E. and Bonventre, R. and Braid, D. and Caden, E. and Callaghan, E. and Caravaca, J. and Carvalho, J. and Cavalli, L. and Chauhan, D. and Chen, M. and Chkvorets, O. and Clark, K. and Cleveland, B. and Coulter, I. T. and Cressy, D. and Dai, X. and Darrach, C. and Davis-Purcell, B. and Deen, R. and Depatie, M. M. and Descamps, F. and Di Lodovico, F. and Duhaime, N. and Duncan, F. and Dunger, J. and Falk, E. and Fatemighomi, N. and Ford, R. and Gorel, P. and Grant, C. and Grullon, S. and Guillian, E. and Hallin, A. L. and Hallman, D. and Hans, S. and Hartnell, J. and Harvey, P. and Hedayatipour, M. and Heintzelman, W. J. and Helmer, R. L. and Hreljac, B. and Hu, J. and Iida, T. and Jackson, C. M. and Jelley, N. A. and Jillings, C. and Jones, C. and Jones, P. G. and Kamdin, K. and Kaptanoglu, T. and Kaspar, J. and Keener, P. and Khaghani, P. and Kippenbrock, L. and Klein, J. R. and Knapik, R. and Kofron, J. N. and Kormos, L. L. and Korte, S. and Kraus, C. and Krauss, C. B. and Labe, K. and Lam, I. and Lan, C. and Land, B. J. and Langrock, S. and LaTorre, A. and Lawson, I. and Lefeuvre, G. M. and Leming, E. J. and Lidgard, J. and Liu, X. and Liu, Y. and Lozza, V. and Maguire, S. and Maio, A. and Majumdar, K. and Manecki, S. and Maneira, J. and Marzec, E. and Mastbaum, A. and McCauley, N. and McDonald, A. B. and McMillan, J. E. and Mekarski, P. and Miller, C. and Mohan, Y. and Mony, E. and Mottram, M. J. and Novikov, V. and O’Keeffe, H. M. and O’Sullivan, E. and Orebi Gann, G. D. and Parnell, M. J. and Peeters, S. J. M. and Pershing, T. and Petriw, Z. and Prior, G. and Prouty, J. C. and Quirk, S. and Reichold, A. and Robertson, A. and Rose, J. and Rosero, R. and Rost, P. M. and Rumleskie, J. and Schumaker, M. A. and Schwendener, M. H. and Scislowski, D. and Secrest, J. and Seddighin, M. and Segui, L. and Seibert, S. and Shantz, T. and Shokair, T. M. and Sibley, L. and Sinclair, J. R. and Singh, K. and Skensved, P. and Sörensen, A. and Sonley, T. and Stainforth, R. and Strait, M. and Stringer, M. I. and Svoboda, R. and Tatar, J. and Tian, L. and Tolich, N. and Tseng, J. and Tseung, H. W. C. and Van Berg, R. and Vázquez-Jáuregui, E. and Virtue, C. and von Krosigk, B. and Walker, J. M. G. and Walker, M. and Wasalski, O. and Waterfield, J. and White, R. F. and Wilson, J. R. and Winchester, T. J. and Wright, A. and Yeh, M. and Zhao, T. and Zuber, K.},
   year={2016},
   pages={1–21} }

@article{Adhikari_2021,
   title={{nEXO:} neutrinoless double beta decay search beyond 1028 year half-life sensitivity},
   volume={49},
   ISSN={1361-6471},
   url={http://dx.doi.org/10.1088/1361-6471/ac3631},
   DOI={10.1088/1361-6471/ac3631},
   number={1},
   journal={Journal of Physics G: Nuclear and Particle Physics},
   publisher={IOP Publishing},
   author={Adhikari, G and Al Kharusi, S and Angelico, E and Anton, G and Arnquist, I J and Badhrees, I and Bane, J and Belov, V and Bernard, E P and Bhatta, T and Bolotnikov, A and Breur, P A and Brodsky, J P and Brown, E and Brunner, T and Caden, E and Cao, G F and Cao, L and Chambers, C and Chana, B and Charlebois, S A and Chernyak, D and Chiu, M and Cleveland, B and Collister, R and Czyz, S A and Dalmasson, J and Daniels, T and Darroch, L and DeVoe, R and Di Vacri, M L and Dilling, J and Ding, Y Y and Dolgolenko, A and Dolinski, M J and Dragone, A and Echevers, J and Elbeltagi, M and Fabris, L and Fairbank, D and Fairbank, W and Farine, J and Ferrara, S and Feyzbakhsh, S and Fu, Y S and Gallina, G and Gautam, P and Giacomini, G and Gillis, W and Gingras, C and Goeldi, D and Gornea, R and Gratta, G and Hardy, C A and Harouaka, K and Heffner, M and Hoppe, E W and House, A and Iverson, A and Jamil, A and Jewell, M and Jiang, X S and Karelin, A and Kaufman, L J and Kotov, I and Krücken, R and Kuchenkov, A and Kumar, K S and Lan, Y and Larson, A and Leach, K G and Lenardo, B G and Leonard, D S and Li, G and Li, S and Li, Z and Licciardi, C and Lindsay, R and MacLellan, R and Mahtab, M and Martel-Dion, P and Masbou, J and Massacret, N and McElroy, T and McMichael, K and Peregrina, M Medina and Michel, T and Mong, B and Moore, D C and Murray, K and Nattress, J and Natzke, C R and Newby, R J and Ni, K and Nolet, F and Nusair, O and Ondze, J C Nzobadila and Odgers, K and Odian, A and Orrell, J L and Ortega, G S and Overman, C T and Parent, S and Perna, A and Piepke, A and Pocar, A and Pratte, J-F and Priel, N and Radeka, V and Raguzin, E and Ramonnye, G J and Rao, T and Rasiwala, H and Rescia, S and Retière, F and Ringuette, J and Riot, V and Rossignol, T and Rowson, P C and Roy, N and Saldanha, R and Sangiorgio, S and Shang, X and Soma, A K and Spadoni, F and Stekhanov, V and Sun, X L and Tarka, M and Thibado, S and Tidball, A and Todd, J and Totev, T and Triambak, S and Tsang, R H M and Tsang, T and Vachon, F and Veeraraghavan, V and Viel, S and Vivo-Vilches, C and Vogel, P and Vuilleumier, J-L and Wagenpfeil, M and Wager, T and Walent, M and Wamba, K and Wang, Q and Wei, W and Wen, L J and Wichoski, U and Wilde, S and Worcester, M and Wu, S X and Wu, W H and Wu, X and Xia, Q and Yan, W and Yang, H and Yang, L and Zeldovich, O and Zhao, J and Ziegler, T},
   year={2021},
   month=dec, pages={015104} }

@article{alfonso2025sensitivitycupidexperiment0nubetabeta,
      title={Sensitivity of the CUPID experiment to $0\nu\beta\beta$ decay of {$^{100}$Mo}}, 
      author={K. Alfonso and A. Armatol and C. Augier and F. T. Avignone III and O. Azzolini and A. S. Barabash and G. Bari and A. Barresi and D. Baudin and F. Bellini and G. Benato and L. Benussi and V. Berest and M. Beretta and L. Bergé and M. Bettelli and M. Biassoni and J. Billard and F. Boffelli and V. Boldrini and E. D. Brandani and C. Brofferio and C. Bucci and M. Buchynska and J. Camilleri and A. Campani and J. Cao and C. Capelli and S. Capelli and V. Caracciolo and L. Cardani and P. Carniti and N. Casali and E. Celi and C. Chang and M. Chapellier and H. Chen and D. Chiesa and D. Cintas and M. Clemenza and I. Colantoni and S. Copello and O. Cremonesi and R. J. Creswick and A. D'Addabbo and I. Dafinei and F. A. Danevich and F. DeDominicis and M. De Jesus and P. de Marcillac and S. Dell'Oro and S. Di Domizio and S. Di Lorenzo and T. Dixon and A. Drobizhev and L. Dumoulin and M. El Idrissi and M. Faverzani and E. Ferri and F. Ferri and F. Ferroni and E. Figueroa Feliciano and J. Formaggio and A. Franceschi and S. Fu and B. K. Fujikawa and J. Gascon and S. Ghislandi and A. Giachero and M. Girola and L. Gironi and A. Giuliani and P. Gorla and C. Gotti and C. Grant and P. Gras and P. V. Guillaumon and T. D. Gutierrez and K. Han and E. V. Hansen and K. M. Heeger and D. L. Helis and H. Z. Huang and M. T. Hurst and L. Imbert and A. Juillard and G. Karapetrov and G. Keppel and H. Khalife and V. V. Kobychev and Yu. G. Kolomensky and R. Kowalski and H. Lattaud and M. Lefevre and M. Lisovenko and R. Liu and Y. Liu and P. Loaiza and L. Ma and F. Mancarella and N. Manenti and A. Mariani and L. Marini and S. Marnieros and M. Martinez and R. H. Maruyama and Ph. Mas and D. Mayer and G. Mazzitelli and E. Mazzola and Y. Mei and M. N. Moore and S. Morganti and T. Napolitano and M. Nastasi and J. Nikkel and C. Nones and E. B. Norman and V. Novosad and I. Nutini and T. O'Donnell and E. Olivieri and M. Olmi and B. T. Oregui and S. Pagan and M. Pageot and L. Pagnanini and D. Pasciuto and L. Pattavina and M. Pavan and Ö. Penek and H. Peng and G. Pessina and V. Pettinacci and C. Pira and S. Pirro and O. Pochon and D. V. Poda and T. Polakovic and O. G. Polischuk and E. G. Pottebaum and S. Pozzi and E. Previtali and A. Puiu and S. Puranam and S. Quitadamo and A. Rappoldi and G. L. Raselli and A. Ressa and R. Rizzoli and C. Rosenfeld and P. Rosier and M. Rossella and J. A. Scarpaci and B. Schmidt and R. Serino and A. Shaikina and K. Shang and V. Sharma and V. N. Shlegel and V. Singh and M. Sisti and P. Slocum and D. Speller and P. T. Surukuchi and L. Taffarello and S. Tomassini and C. Tomei and A. Torres and J. A. Torres and D. Tozzi and V. I. Tretyak and D. Trotta and M. Velazquez and K. J. Vetter and S. L. Wagaarachchi and G. Wang and L. Wang and R. Wang and B. Welliver and J. Wilson and K. Wilson and L. A. Winslow and F. Xie and M. Xue and J. Yang and V. Yefremenko and V. I. Umatov and M. M. Zarytskyy and T. Zhu and A. Zolotarova and S. Zucchelli},
      year={2025},
      eprint={2504.14369},
      archivePrefix={arXiv},
      primaryClass={nucl-ex},
      url={https://arxiv.org/abs/2504.14369}, 
}

@article{feruglio2017neutrinomassesmodularforms,
      title={Are neutrino masses modular forms?}, 
      author={Ferruccio Feruglio},
      year={2017},
      eprint={1706.08749},
      archivePrefix={arXiv},
      primaryClass={hep-ph},
      url={https://arxiv.org/abs/1706.08749}, 
}

@article{nomura2024nonholomorphicmodularmathcala4symmetric,
      title={Non-Holomorphic Modular $\mathcal{A}_4$ Symmetric Scotogenic Model}, 
      author={Takaaki Nomura and Hiroshi Okada and Oleg Popov},
      year={2024},
      eprint={2409.12547},
      archivePrefix={arXiv},
      primaryClass={hep-ph},
      url={https://arxiv.org/abs/2409.12547}, 
}

@article{nomura2025radiativeneutrinomassmodel,
      title={A radiative neutrino mass model with leptoquarks under non-holomorphic modular $\mathcal{A}_4$ symmetry}, 
      author={Takaaki Nomura and Hiroshi Okada and Xing-Yu Wang},
      year={2025},
      eprint={2504.21404},
      archivePrefix={arXiv},
      primaryClass={hep-ph},
      url={https://arxiv.org/abs/2504.21404}, 
}

@article{PhysRevD.110.123537,
  title = {Critical look at the cosmological neutrino mass bound},
  author = {Naredo-Tuero, Daniel and Escudero, Miguel and Enrique Fernandez-Martinez and Marcano, Xabier and Poulin, Vivian},
  journal = {Phys. Rev. D},
  volume = {110},
  issue = {12},
  pages = {123537},
  numpages = {21},
  year = {2024},
  month = {Dec},
  publisher = {American Physical Society},
  doi = {10.1103/PhysRevD.110.123537},
  url = {https://link.aps.org/doi/10.1103/PhysRevD.110.123537}
}

@article{Bonilla:2026jzk,
    author = "Bonilla, Cesar and Layana-Ramirez, Andres",
    title = "{Neutrino Masses and Dark Matter Stability in 3HDMs with Minimal Non-Abelian Discrete Symmetries}",
    eprint = "2607.07853",
    archivePrefix = "arXiv",
    primaryClass = "hep-ph",
    month = "7",
    year = "2026"
}

@article{Garcia-Cely_2016,
doi = {10.1088/1475-7516/2016/02/043},
url = {https://doi.org/10.1088/1475-7516/2016/02/043},
year = {2016},
month = {feb},
publisher = {},
volume = {2016},
number = {02},
pages = {043},
author = {Garcia-Cely, Camilo and Gustafsson, Michael and Ibarra, Alejandro},
title = {Probing the inert doublet dark matter model with Cherenkov telescopes},
journal = {Journal of Cosmology and Astroparticle Physics}
}

@article{Boucenna:2011tj,
    author = "Boucenna, M. S. and Hirsch, M. and Morisi, S. and Peinado, E. and Taoso, M. and Valle, J. W. F.",
    title = "{Phenomenology of Dark Matter from $A_4$ Flavor Symmetry}",
    eprint = "1101.2874",
    archivePrefix = "arXiv",
    primaryClass = "hep-ph",
    doi = "10.1007/JHEP05(2011)037",
    journal = "JHEP",
    volume = "05",
    pages = "037",
    year = "2011"
}

@article{Boucenna_2012,
   title={Predictive discrete dark matter model and neutrino oscillations},
   volume={86},
   ISSN={1550-2368},
   url={http://dx.doi.org/10.1103/PhysRevD.86.073008},
   DOI={10.1103/physrevd.86.073008},
   number={7},
   journal={Physical Review D},
   publisher={American Physical Society (APS)},
   author={Boucenna, M. S. and Morisi, S. and Peinado, E. and Valle, J. W. F. and Shimizu, Yusuke},
   year={2012},
   month=Oct }

\end{document}